\documentclass[aps,pre,reprint,superscriptaddress,floatfix]{revtex4-2}

\usepackage{comment}
\usepackage{amsmath}
\usepackage{amssymb}
\usepackage{bm}
\usepackage{graphicx}
\usepackage{physics}
\usepackage{color}
\usepackage{hyperref}
\usepackage[caption=false]{subfig}
\usepackage{tikz}
\usetikzlibrary{positioning}
\graphicspath{{Figures/}{./}}

\begin{document}

\title{Anti-synchronized collective rotation in magnetic colloidal particles under rotating magnetic fields}

\author{Adithya Bharath}
\email[]{adithyabharath81@gmail.com}
\author{A. V. Anil Kumar}
\email[]{anil@niser.ac.in}
\affiliation{School of Physical Sciences, National Institute of Science Education and Research, Jatni, Bhubaneswar 752050, India}
\affiliation{Homi Bhabha National Institute, Anushakti Nagar, Mumbai, India}

\date{\today}

\begin{abstract}

Collective synchronization is a common response of interacting magnetic particles subjected to rotating external fields. Here we demonstrate a distinct nonequilibrium state in a two-dimensional cluster of magnetic colloidal particles in which the ordered crystalline core rotates, on average, opposite to the direction of the applied rotating magnetic field, while the defect-rich outer region continues to rotate with the drive. We term this state anti-synchronized collective rotation. By combining structural, spectral, bond-resolved, and layer-resolved torque analyses, we show that anti-synchronization is associated with a broader defect-rich interface, enhanced and dynamically heterogeneous bond deformation, and a reorganization of torque transmission between the surface and the crystalline interior. In the synchronized state, steric torques compensate the opposing dipolar contribution in the outer core, maintaining a positive net interaction torque. In the anti-synchronized state this compensation becomes insufficient, producing a negative net torque in the outer core and enabling core–surface rotational decoupling. Increasing the driving frequency restores coherent rotation through a transition characterized by competing positive- and negative-angular-velocity populations and hysteresis, consistent with a discontinuous, first-order-like dynamical transition. These results demonstrate how deformation and spatial redistribution of interaction torques can generate counter-rotating collective states in driven colloidal matter.

\end{abstract}

\maketitle

\section{Introduction}

Driven many-body systems provide a broad framework for understanding the emergence of collective behavior in interacting systems maintained away from thermodynamic equilibrium by external forcing. Periodic driving, in particular, can give rise to a rich variety of dynamical phenomena, including synchronization \cite{bruot,golastanian,juniper}, pattern formation \cite{cross,aranson}, dynamical self-assembly and collective motion \cite{palacci,grzybowski,snezhko}, and transitions between distinct nonequilibrium states \cite{tierno1,sides,korniss}. Soft-matter systems provide particularly suitable platforms for investigating such phenomena because their mesoscopic constituents and relatively weak interaction energies allow their structures and dynamics to be readily modified by external perturbations and thermal fluctuations. Colloidal suspensions, in particular, can be driven away from equilibrium through externally imposed mechanical forcing \cite{wagner} or particle-level self-propulsion \cite{tailleur}, leading to a wide range of collective responses. One prominent example is lane formation in mixtures of oppositely driven colloidal particles, which has been observed in Brownian dynamics simulations of both two- and three-dimensional systems \cite{dzubiella,reichhardt}. Similar behavior, including local lane formation, has been reported experimentally in oppositely charged colloidal suspensions subjected to steady \cite{leunissen,vissers} and alternating electric fields \cite{vissers1}. Beyond lane formation, Reichhardt {\it et al.} \cite{reichhardt1} demonstrated the emergence of moving clump and stripe phases in driven colloidal systems with competing short-range attractive and long-range repulsive interactions on disordered substrates. These examples illustrate how the interplay between external driving, interparticle interactions, and structural rearrangements can generate diverse collective states and interesting dynamical phenomena.

Among externally driven soft-matter systems, magnetic colloidal suspensions provide a versatile platform for investigating nonequilibrium collective dynamics, as their structural organization and dynamical response can be controlled through the strength, frequency, and orientation of applied magnetic fields. These systems have attracted sustained interest over several decades owing to their rich structural and dynamical behavior. A prominent class of magnetoresponsive soft materials is ferrofluids, consisting of magnetic nanoparticles dispersed in a liquid carrier \cite{torres,park}. The application of external magnetic fields can substantially modify their structural and rheological properties by promoting the formation and reorganization of particle aggregates. Owing to the anisotropic nature of dipole–dipole interactions, magnetic particles can assemble into a variety of structures, including linear chains, rings, and branched networks \cite{klokkenburg,furst,kantorovich}. Beyond static or quasistatic field-induced assembly, time-dependent magnetic fields provide an additional means of controlling particle organization and motion. In particular, rotating magnetic fields can induce dynamical self-assembly, structural reorganization, and collective rotational dynamics through the competition between magnetic interactions, viscous dissipation, and the timescale of external forcing. Early studies by Kopelman and coworkers demonstrated that individual magnetic colloids can rotate synchronously with a rotating field at sufficiently low driving frequencies, whereas above a characteristic threshold, viscous resistance prevents sustained frequency locking \cite{kopelman}.  Extending this picture to interacting particles, Tierno investigated the dynamics of dipolarly coupled colloidal rotators, demonstrating how magnetic interactions give rise to coupled rotational responses, including synchronization and beating phenomena \cite{tierno}. Jäger and coworkers showed through computer simulations that the loss of single-particle synchronization can be accompanied by a collective structural transition from a layered to a disordered state \cite{jager1}. The connection between synchronization and structural organization was further demonstrated experimentally by Yan and coworkers, who investigated the self-assembly of magnetic Janus colloids under rotating magnetic fields \cite{yan1}. Subsequently, Yan, Bae, and Granick demonstrated the formation of rotating colloidal crystals exhibiting structural rearrangements, dislocations, and mobile domain boundaries, highlighting the importance of internal deformation in the dynamics of field-driven colloidal assemblies \cite{yan}. The role of hydrodynamic interactions, particularly the coupling between rotational and translational motion mediated by the surrounding fluid, has also been investigated theoretically and computationally \cite{jager}.  These studies highlight the close connection between magnetic synchronization, collective structural organization, and internal particle rearrangements, raising the broader question of how the rotational response of a deformable colloidal assembly is determined by the interplay between external forcing and interparticle interactions.

While the synchronization of individual magnetic moments with an external rotating field has been extensively investigated\cite{jager1}, the collective rotational dynamics of interacting magnetic colloids involve additional degrees of freedom arising from their internal structure and interparticle interactions. In contrast to the rotation of individual magnetic moments, the orbital motion of particles within a colloidal assembly is governed by the interaction forces transmitted through the particle network, which may undergo continuous deformation and structural rearrangement under periodic forcing\cite{yan}. An important issue, therefore, is to understand whether the interplay between magnetic interactions, internal deformation, and spatially heterogeneous force transmission can give rise to collective rotational motion opposite to the direction of the applied field. 

Collective rotation opposite to the direction of an externally applied rotating field represents an unconventional dynamical response that challenges the usual picture of field-driven synchronization. Such counter-rotating collective motion has been reported in swarms of magnetotactic bacteria subjected to rotating magnetic fields, where hydrodynamic interactions play an important role in determining the direction of collective rotation \cite{belovs}. These observations demonstrate that the collective rotational response of an interacting system need not follow the direction of the applied magnetic forcing, even when the individual magnetic moments are driven by the same external field. However, the emergence of counter-rotation in structurally ordered magnetic colloidal assemblies, particularly the possibility of oppositely rotating regions within a single connected cluster, remains comparatively unexplored. Understanding such behaviour requires examining how microscopic structural rearrangements and the spatial transmission of interparticle forces influence the collective rotational dynamics. In this work, we investigate the collective rotational dynamics of a two-dimensional assembly of interacting magnetic colloidal particles subjected to a uniform rotating magnetic field. We identify a distinct nonequilibrium state, which we term anti-synchronized collective rotation, in which the structurally ordered crystalline core rotates, on average, opposite to the direction of the applied field, while the defect-rich outer region continues to rotate with the drive. Through a systematic analysis of structural ordering, defect dynamics, microscopic bond fluctuations, and spatially resolved interaction torques, we elucidate the mechanism underlying this unconventional rotational response. Our results reveal that anti-synchronization is accompanied by enhanced structural deformation and a reorganization of torque transmission between the outer region and the crystalline interior. In particular, the competition between dipolar and steric contributions produces a negative net interaction torque in the outer core, enabling rotational decoupling between the crystalline interior and the field-following boundary. Upon increasing the driving frequency, the system undergoes a hysteretic transition to synchronized collective rotation, characterized by competing positive- and negative-angular-velocity states near the transition. These findings demonstrate how the interplay between microscopic deformation and spatially heterogeneous interaction torques can give rise to counter-rotating collective states in periodically driven colloidal matter.

\section{Model and Methodology}

We consider a two-dimensional suspension of $N$ circular magnetic colloidal particles subjected to a spatially uniform rotating magnetic field. Each particle possesses a permanent magnetic dipole moment rigidly attached to its body axis and undergoes overdamped  Brownian motion. The dynamics arise from the interplay between steric repulsion, magnetic dipole--dipole interactions, thermal fluctuations, and the external magnetic torque.
The translational and rotational dynamics are described by overdamped Langevin equations. The position $\mathbf{r}_i$ of the $i$th particle evolves according to

\begin{equation}
\dot{\mathbf r}_i
=
\frac{\mathbf{F_i}}{\gamma_{\rm{t}}}
+
\sqrt{2D_t}\,
\boldsymbol{\eta}_i(t),
\label{eq:translation}
\end{equation}

\noindent where $\gamma_t$ is the translational viscous coefficient, and $D_t$ denotes the translational diffusion coefficient. The Gaussian white noise satisfies
\begin{align}
\langle \eta_{i\alpha}(t)\rangle &=0,\\
\langle \eta_{i\alpha}(t)\eta_{j\beta}(t')\rangle
&=
\delta_{ij}\delta_{\alpha\beta}\delta(t-t').
\end{align}

\noindent The particle orientation evolves according to
\begin{equation}
\dot{\phi}_i
=
\frac{\tau_i}{\gamma_{\rm{r}}}
+
\sqrt{2D_r}\,
\xi_i(t),
\label{eq:rotation}
\end{equation}
\noindent where $\gamma_r$ is the rotational viscous coefficient, $D_r$ is the rotational diffusion coefficient, and $\xi_i(t)$ is Gaussian white noise.
The total force acting on particle $i$ is
\begin{equation}
\mathbf F_i
=
\sum_{j\neq i}
\left(
\mathbf{F}_{ij}^{\rm{WCA}}
+
\mathbf{F}_{ij}^{\rm{dd}}
\right),
\label{eq:total_force}
\end{equation}
\noindent while the total torque consists of contributions from the dipolar interactions and the external magnetic field,
\begin{equation}
\tau_i
=
\tau_i^{\rm dip}
+
\tau_i^{\mathrm{ext}}.
\label{eq:total_torque}
\end{equation}

\noindent The short-range steric repulsion is modeled using the Weeks--Chandler--Andersen (WCA) potential \cite{Weeks1971},
\begin{equation}
U_{\mathrm{WCA}}(r)=
\begin{cases} 
4\epsilon  \left[ \left(\frac{\sigma}{r_{ij}}\right)^{12} - \frac{1}{2}\left(\frac{\sigma}{r_{ij}}\right)^6 \right] + \epsilon, & r_{ij} < 2^{1/6}\sigma, \\
0, & r_{ij} \ge 2^{1/6}\sigma,
\end{cases}
\label{eq:wca_potential}
\end{equation}
\noindent with the corresponding pairwise repulsive force \\
$\mathbf{F}_{ij}^{\rm{WCA}} = -\boldsymbol{\nabla}_i U_{\mathrm{WCA}}(r_{ij})$. Magnetic interactions between particles are described by the dipole--dipole interaction
\begin{equation}
\begin{split}
\mathbf{F}_{ij}^{\rm{dd}} =  & \frac{3\mu_0\mu^2}{4\pi r_{ij}^4} \left[ \hat{\mathbf{r}}_{ij} (\hat{\mathbf{e}}_i \cdot \hat{\mathbf{e}}_j) + \hat{\mathbf{e}}_i (\hat{\mathbf{r}}_{ij} \cdot \hat{\mathbf{e}}_j) \right.  \\ & \left. + \hat{\mathbf{e}}_j (\hat{\mathbf{r}}_{ij} \cdot \hat{\mathbf{e}}_i) - 5\hat{\mathbf{r}}_{ij} (\hat{\mathbf{r}}_{ij} \cdot \hat{\mathbf{e}}_i) (\hat{\mathbf{r}}_{ij} \cdot \hat{\mathbf{e}}_j) \right].
\end{split}
\label{eq:dipole-force}
\end{equation}

In addition to the WCA and dipole--dipole interactions, the particles are subjected to an external magnetic field rotating with constant angular frequency $\omega_{\rm f}$,

\begin{equation}
\mathbf B(t)
=
B
\left(
\cos\omega_{\rm f}t,
\sin\omega_{\rm f}t
\right),
\label{eq:b_field}
\end{equation}

\noindent where $B$ is the field amplitude. The magnetic torque exerted on each particle is

\begin{equation}
\tau_i^{\mathrm{ext}}
=
\mu B
\sin\!\left(
\omega_{\rm f}t-\phi_i
\right),
\label{eq:torque_ext}
\end{equation}

\noindent where $\mu$ denotes the magnitude of the particle dipole moment. Here \(\phi_i\) denotes the direction of the permanently attached magnetic dipole moment.

We employ reduced units throughout the simulations. Lengths are scaled by particle diameter $\sigma$, energy is scaled by $\epsilon$,time is scaled by $\gamma_{t}\sigma^2/\epsilon$, angular velocity is scaled by  $\epsilon/(\gamma_t \sigma^2)$  and the magnetic dipole moment  is scaled by $\sqrt{\frac{4\pi \epsilon \sigma^3}{\mu_0}}$, and the magnetic field is scaled by the $\sqrt{\frac{9\mu_{0}\epsilon}{4\pi\sigma^3}}$. We impose the constraint \(\gamma_r=3\gamma_t\). 

The equations of motion are integrated using the Euler-Maruyama scheme \cite{Kloeden1977} with a fixed timestep $\Delta t$ . Periodic boundary conditions are employed in both spatial directions.
Unless otherwise stated, the simulations are performed at fixed dipole moment $\mu=7$, temperature $T=0$ and number of particles $N=448$. The field amplitude and driving frequency are varied to construct the nonequilibrium phase diagram. Each simulation is evolved for $10^6$ integration steps, and statistical quantities are computed by averaging over the final $5\times10^5$ steps after the system has reached a steady state. To characterize the nonequilibrium steady states, we monitor several structural and dynamical observables.
The local crystalline order is characterized using the hexatic order parameter \cite{Nelson1979}

\begin{equation}
\psi_6(i)
=
\frac{1}{N_i}
\sum_{j=1}^{N_i}
e^{6\mathrm{i}\theta_{ij}},
\label{eq:psi6}
\end{equation}

\noindent from which the global hexatic order parameter $|\Psi_6|$ is calculated. Magnetic synchronization with the rotating field is quantified using the synchronization parameter $\alpha = \omega_{\mathrm{p}}/\omega_{\rm f}$, where $\omega_{\mathrm{p}}$ is the angular velocity at which the individual particle orientations follow the field.
We characterize the instantaneous rotational motion of the cluster by its angular velocity obtained from the particle positions and velocities relative to the cluster center of mass, given by 

\begin{equation}
\Omega_{\rm{cl}}(t)=
\frac{\sum_i [\mathbf{r}_i(t)\times\mathbf{v}_i(t)]_z}
{\sum_i \left|\mathbf{r}_i(t)\right|^2}.
\label{eq:omega_cluster}
\end{equation}

To characterize fluctuations in the collective rotational dynamics and investigate the transition between anti-synchronized and synchronized states, we analyze the probability distribution of the crystalline-core angular velocity, $P(\Omega_{\mathrm{core}})$.

\section{Results and discussion}

\subsection{Dynamical phases under rotating magnetic fields}

\begin{figure}[!htbp]
\centering
\includegraphics[width=\linewidth]{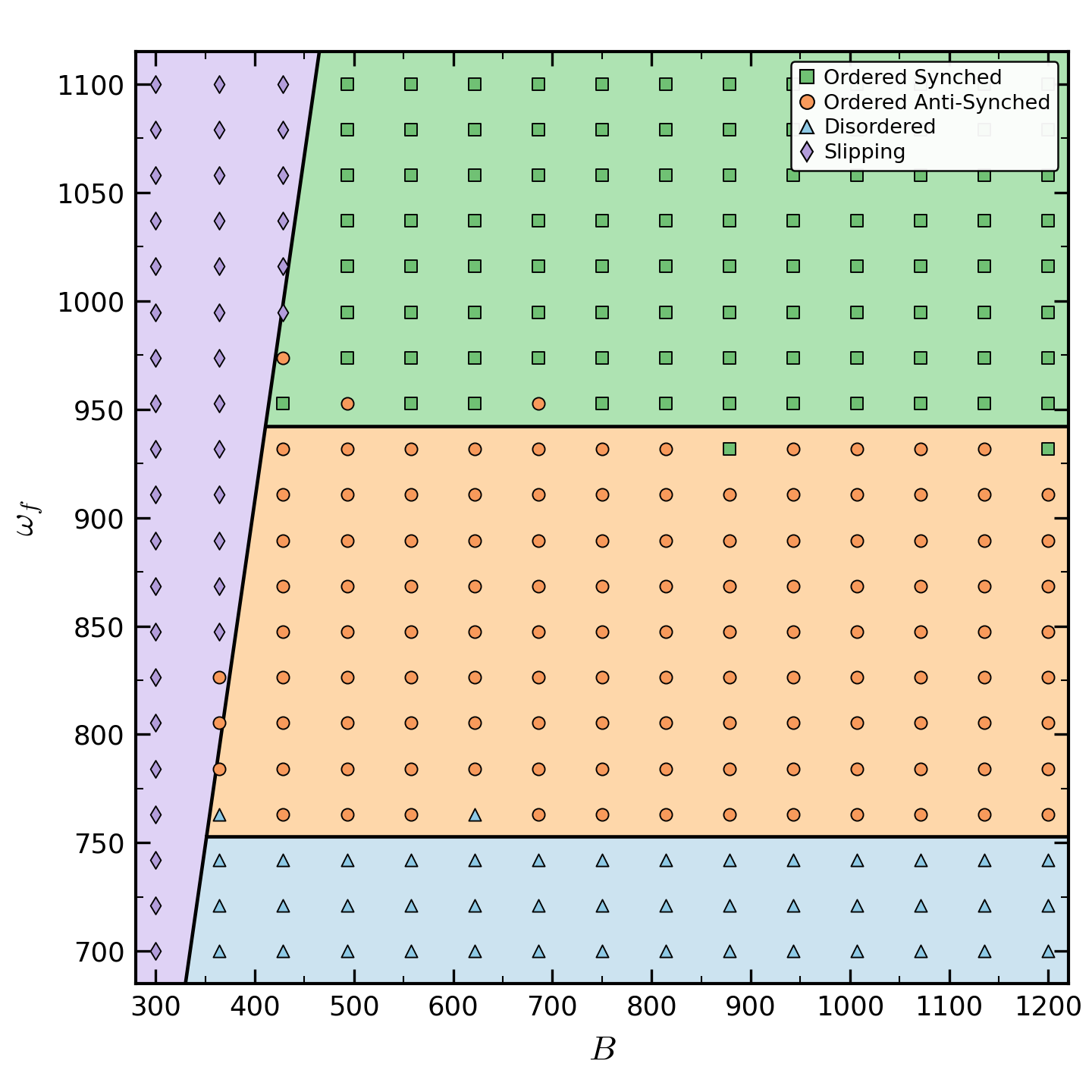}
\caption{
Phase diagram of magnetic colloids subjected to a rotating magnetic field in the $(B,\omega_{\rm f})$ plane. Four dynamical regimes are identified: disordered, anti-synchronized, synchronized, and slipping. Structural order is determined from the hexatic correlation length, while the rotational state is classified from the sign of the crystalline-core angular velocity, $\Omega_{\rm core}$ and the synchronization parameter, $\alpha$. Parameters: $N=448$, timestep $\Delta t=6\times10^{-6}$, $\mu=7$.
}
\label{fig:phase_diagram}
\end{figure}

We first investigate the collective response of the magnetic colloidal particle cluster by systematically varying the amplitude $B$ and angular frequency $\omega_{\rm f}$ of the externally applied rotating magnetic field. The resulting nonequilibrium states are characterized by both the structural order of the cluster and its collective rotational dynamics. The corresponding phase diagram is shown Fig.~\ref{fig:phase_diagram}.  To quantify structural organization, we calculate the spatial correlation function of the local hexatic order parameter,  $g_6(r)$.
\begin{equation}
g_6(r)=\frac{
\left\langle
\psi_6(\mathbf{r})\psi_6^*(\mathbf{0})
\right\rangle_{|\mathbf{r}|=r}
}{
\left\langle
|\psi_6(\mathbf{0})|^2
\right\rangle
}.
\label{eq:g6_corr}
\end{equation}
\noindent 
We determine the orientational correlation length, $\xi$, by performing a least-squares linear fit to $\log[g_6(r)]$ as a function of $r$.  We classify the cluster as ordered when the correlation length is comparable to or larger than the cluster size,  $\xi \geq L_s$,  and as disordered when $\xi < L_s$. Here $L_s$ denotes the system size, approximated by the diameter of the cluster.  The rotational state of an ordered cluster is then distinguished using the angular velocity of its crystalline core. A positive core angular velocity corresponds to rotation in the direction of the external field, whereas a negative value corresponds to counter-rotation. Magnetic synchronization with the external field is quantified by the ratio $\alpha = \omega_{\mathrm{p}}/\omega_{\rm f}$., with $\alpha<0.95$ used to identify the slipping regime.

\begin{figure*}[!t]
    \centering
    \subfloat[]{\includegraphics[width=0.48\textwidth]{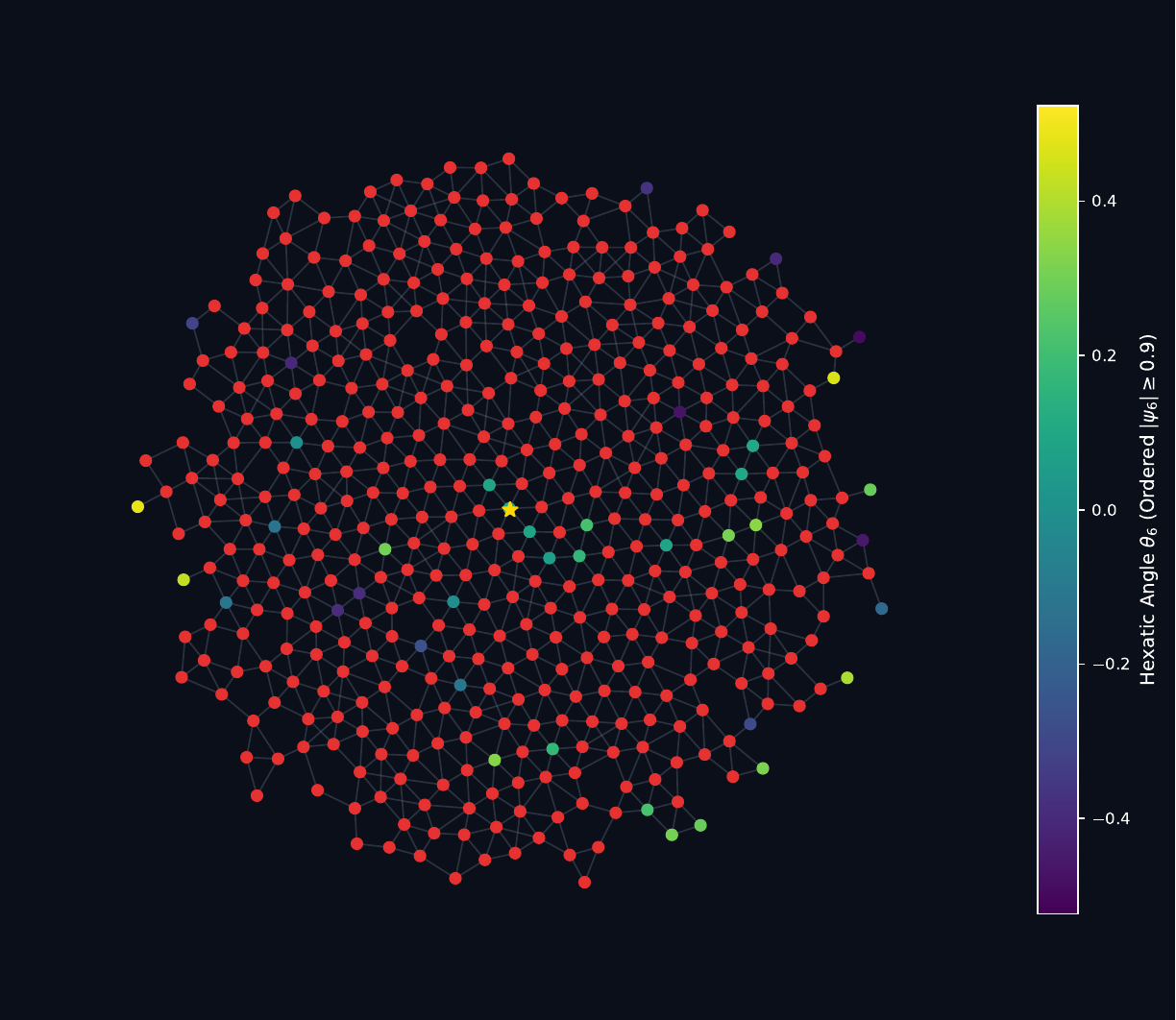}\label{fig:disordered}}
    \hfill
    \subfloat[]{\includegraphics[width=0.48\textwidth]{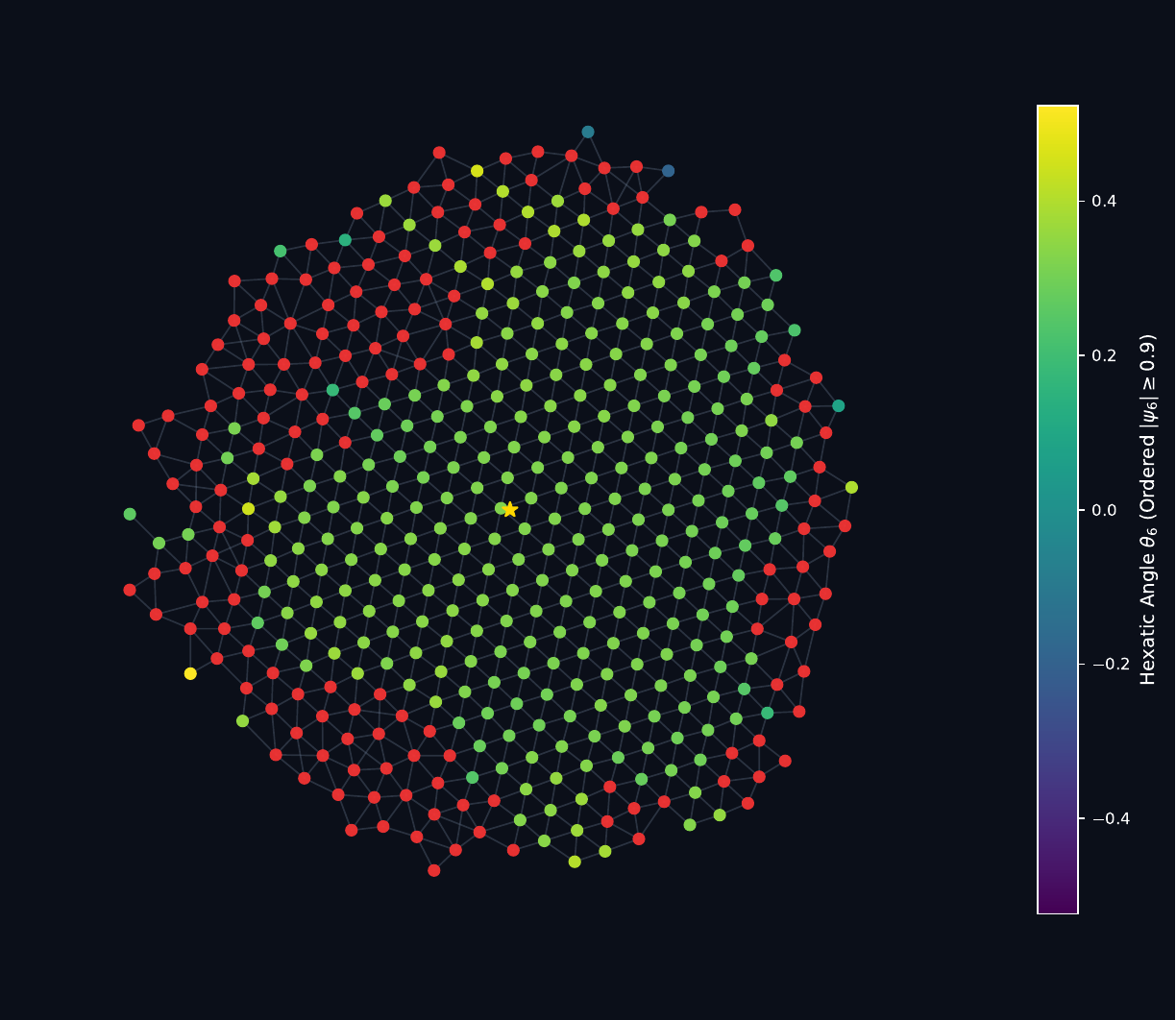}\label{fig:anti}}
    \\[-1ex]
    \subfloat[]{\includegraphics[width=0.48\textwidth]{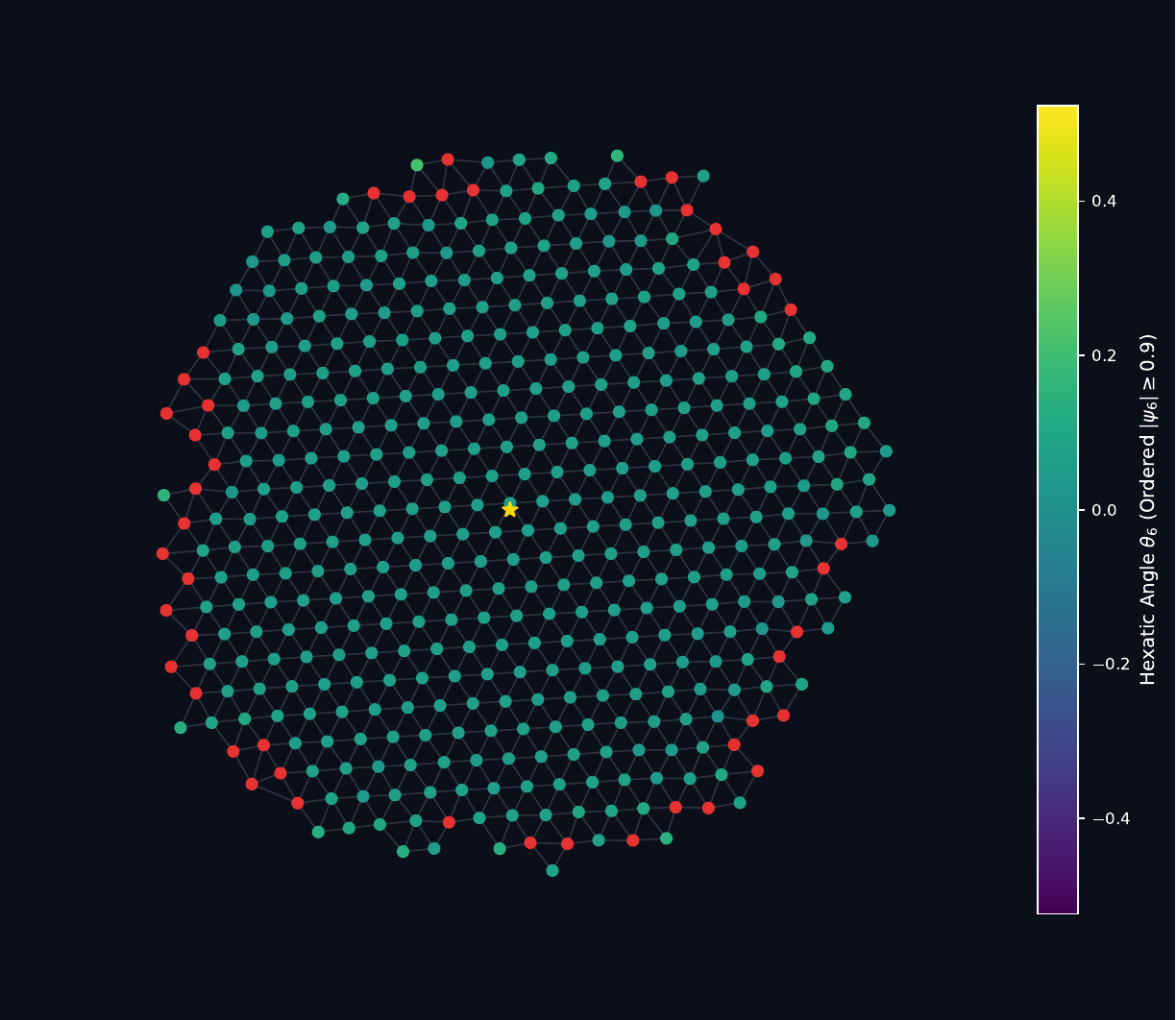}\label{fig:synched}}
    \hfill
    \subfloat[]{\includegraphics[width=0.48\textwidth]{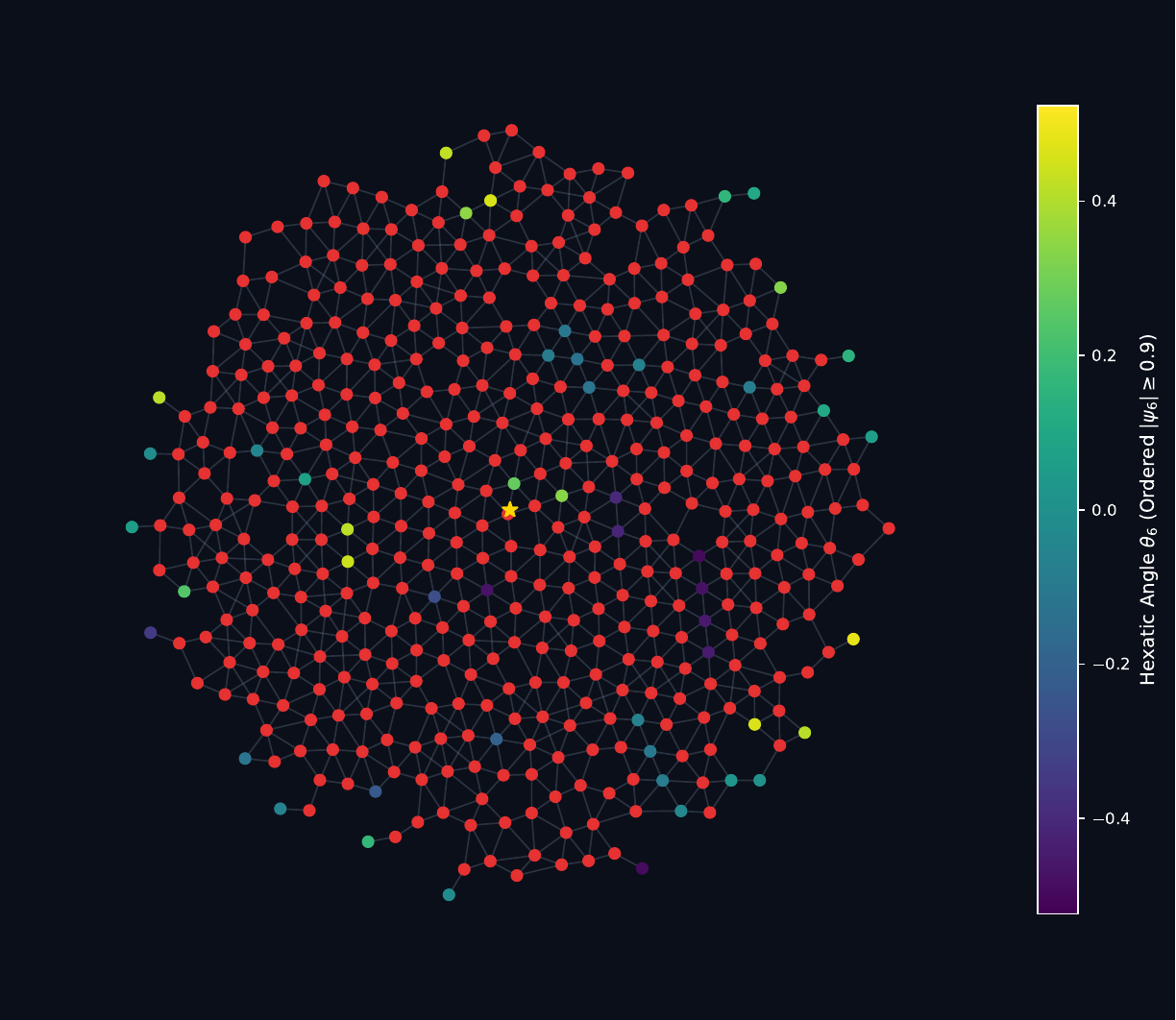}\label{fig:slipping}}
    \caption{Snapshots of the four dynamical regimes for the magnetic colloidal particles under a rotating magnetic field. (a) Disordered phase: \textbf{B}=1020, $\omega_{\rm f}=300$. (b) Anti-synchronized phase: \textbf{B}=1200, $\omega_{\rm f}=903$. (c) Synchronized phase: \textbf{B}=1200, $\omega_{\rm f}=1850$. (d) Slipping phase: \textbf{B}=480, $\omega_{\rm f}=2200$. The red particles have $|\psi_6| <0.90$, The colored particles are colored by their local $\psi_6$ orientation angle.}
    \label{fig:snapshots}
\end{figure*}

In our simulations, we observe four distinct dynamical regimes: a disordered state, an anti-synchronized state, a synchronized state, and a slipping state. Representative configurations are shown in Fig. ~\ref{fig:snapshots}. At low driving frequencies and field amplitudes, the particles do not develop long-range crystalline order and the cluster remains in a disordered state (Fig. ~\ref{fig:snapshots}(a)). Upon increasing the driving frequency, crystalline order develops, but the resulting ordered cluster does not immediately synchronize with the rotating field. Instead, a distinct anti-synchronized state emerges (Fig. ~\ref{fig:snapshots}(b)). In this regime, the crystalline core rotates opposite to the direction of the applied field, while the defect-rich outer boundary rotates in the same direction as the field. An example of this anti-synchronized state is shown in supplemental movie 1 for $\omega_{\rm f}$ =  850\cite{movie1}.  With a further increase in driving frequency, the rotational dynamics change qualitatively. The boundary becomes increasingly locked to the crystalline interior, and the cluster eventually rotates coherently in the direction of the applied field, defining the synchronized state (Fig. ~\ref{fig:snapshots}(c)). A visualization of this  synchronized rotation of the cluster is provided as supplemental movie 2 \cite{movie2}.  At still higher driving frequencies, the magnetic moments can no longer follow the rapidly rotating field continuously. The particles undergo repeated slipping events, accompanied by a progressive reduction of structural order, resulting in the slipping regime (Fig. ~\ref{fig:snapshots}(d)). 

The coupling between magnetic synchronization and structural ordering has also been observed in earlier studies of dipolar colloids under rotating magnetic fields. Jäger and Klapp demonstrated that changes in the synchronization of individual magnetic moments can be accompanied by transitions between ordered and disordered configurations \cite{jager1}. Experimental investigations of rotating magnetic Janus colloids have similarly revealed transitions between disordered and crystalline states, together with structural rearrangements within rotating assemblies \cite{yan}. In the present system, however, an additional dynamical distinction emerges within the structurally ordered regime: the crystalline core can rotate either along or opposite to the direction of the external field.

An important feature of the phase diagram is that structural ordering and synchronization do not occur simultaneously. In particular, the anti-synchronized regime occupies an intermediate range of driving conditions in which the cluster is structurally ordered but its collective rotational response is opposite to that of the applied field. Thus, the sign of the core angular velocity provides a dynamical distinction between two structurally similar ordered states.
The emergence of the anti-synchronized state is therefore not simply a loss of magnetic synchronization. Rather, it represents a distinct collective mode in which different regions of the same connected cluster exhibit different rotational responses. As discussed below, this separation between the crystalline core and the defect-rich boundary is central to the mechanism underlying counter-rotation.

\subsection{Microscopic origin of cluster rotation}
\label{subsec:cluster_rotation}

Before considering the collective cluster, it is useful to identify the microscopic mechanism that allows a deformable magnetic aggregate to rotate. We first consider the simplest composite structure, a pair of interacting particles, or dimer. Let $r_{ij}$ denote the separation between the particles and $\theta_{ij}$ the orientation of their bond. Let $\delta$ be the phase difference between particle orientation and the magnetic field direction, then the instantaneous angle entering the dipolar interaction can be written as $\psi_{ij}=\omega_{\rm f}t-\theta_{ij}-\delta$.  For the dimer, the relative radial and angular dynamics obtained from the dipolar interaction are (see Appendix A for derivation)

\begin{equation}
\begin{aligned}
\frac{\dot{r}_{ij}}{2} &= \frac{3\mu^2}{r_{ij}^{4}}\left(1-3\cos^2\psi_{ij}\right) + F_{r}^{\mathrm{WCA}}, \\[6pt]
\dot{\theta}_{ij} &= \frac{6\mu^2}{r_{ij}^{5}}\sin\left(2\psi_{ij}\right).
\end{aligned}
\label{eq:dimer_dynamics}
\end{equation}

\noindent Thus, the dipolar interaction possesses an explicitly tangential component proportional to $\sin(2\psi_{ij})$. This component produces a nonzero angular velocity of the bond and provides the microscopic origin of collective rotational motion.  For an extended cluster, the corresponding collective torque is obtained within the first approximation by summing the pairwise dipolar torques about the instantaneous cluster center of mass,

\begin{equation}
\tau_{\mathrm{cl}} = \sum_i \sum_{j \neq i} \tau_{ij},
\label{eq:torque_sum}
\end{equation}

\noindent where $\tau_{ij}$ is the dipolar torque exerted by particle $j$ on particle $i$ with respect to the center of mass. For a symmetric cluster, the radial forces do not directly contribute to the net torque about the center of mass, giving
\begin{equation}
\tau_{\mathrm{cl}}
=\sum_{i<j}
\frac{3\mu^2}{r_{ij}^{3}}
\sin\left(2\psi_{ij}\right).
\label{eq:cluster_torque}
\end{equation}
\begin{figure}[!t]
    \centering
    \includegraphics[width=0.48\textwidth]{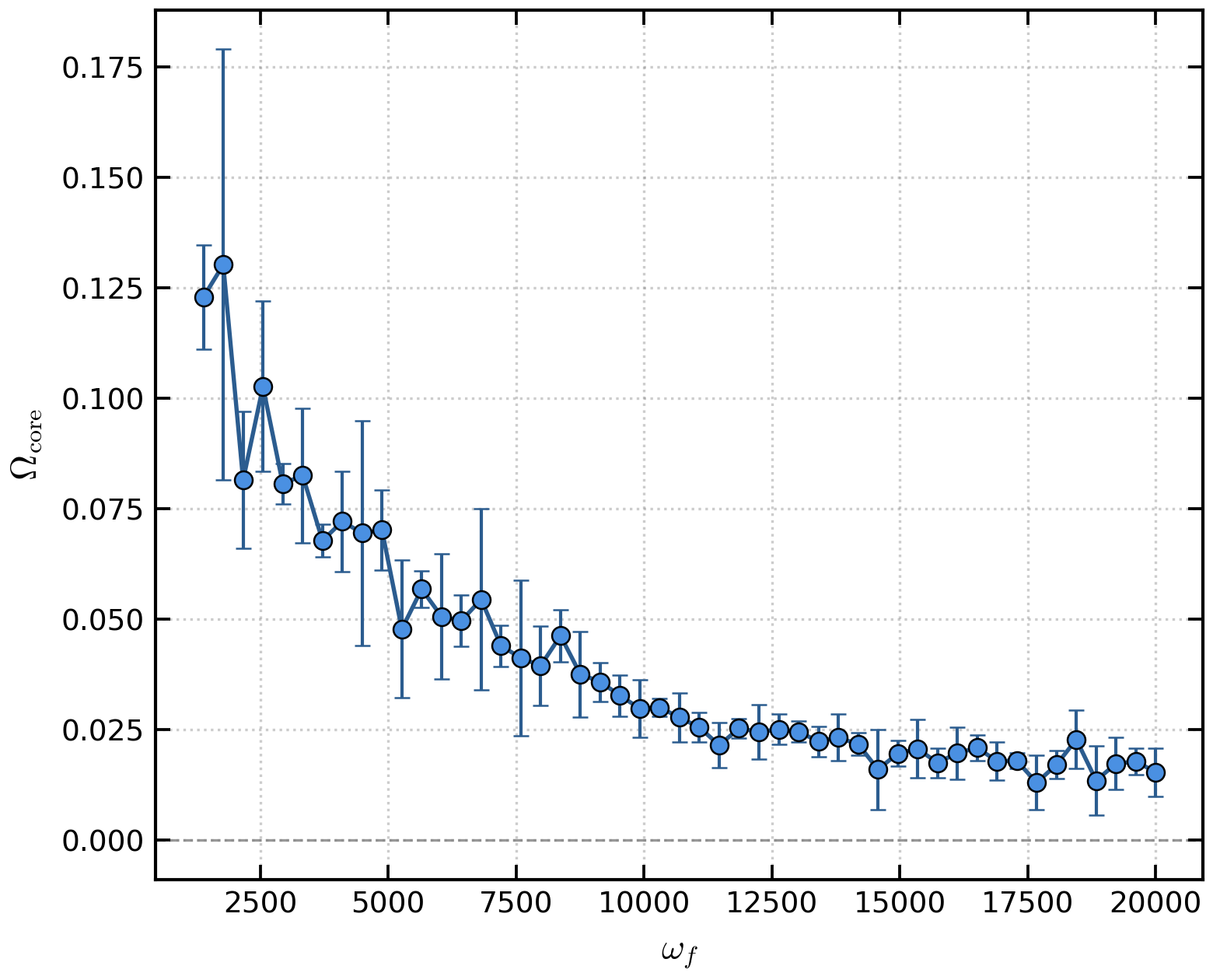}
    \caption{Decay of core angular velocity with frequency.Each data point is averaged over 5 simulations of random initial conditions. The error bars shows the standard deviation of the mean angular velocities from these 5 simulations.To account for high frequencies we used a smaller time step of dt=$2 \times 10^{-6}$.}
    \label{fig:omega_cl vs omega}
\end{figure}
\begin{figure*}[!t]
    \centering
    \subfloat[]{\includegraphics[width=0.48\textwidth]{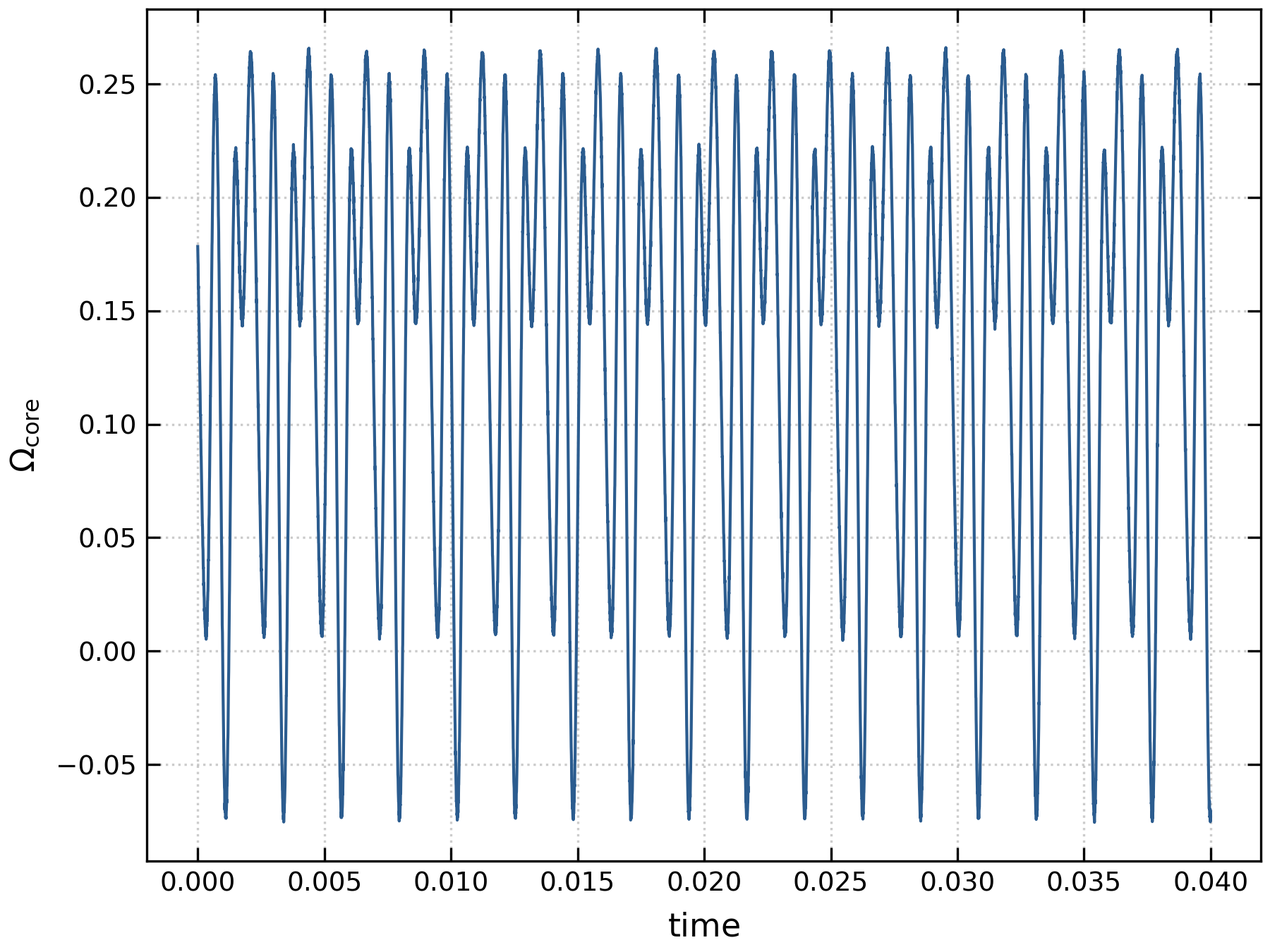}\label{fig:synch time}}
    \hfill
    \subfloat[]{\includegraphics[width=0.48\textwidth]{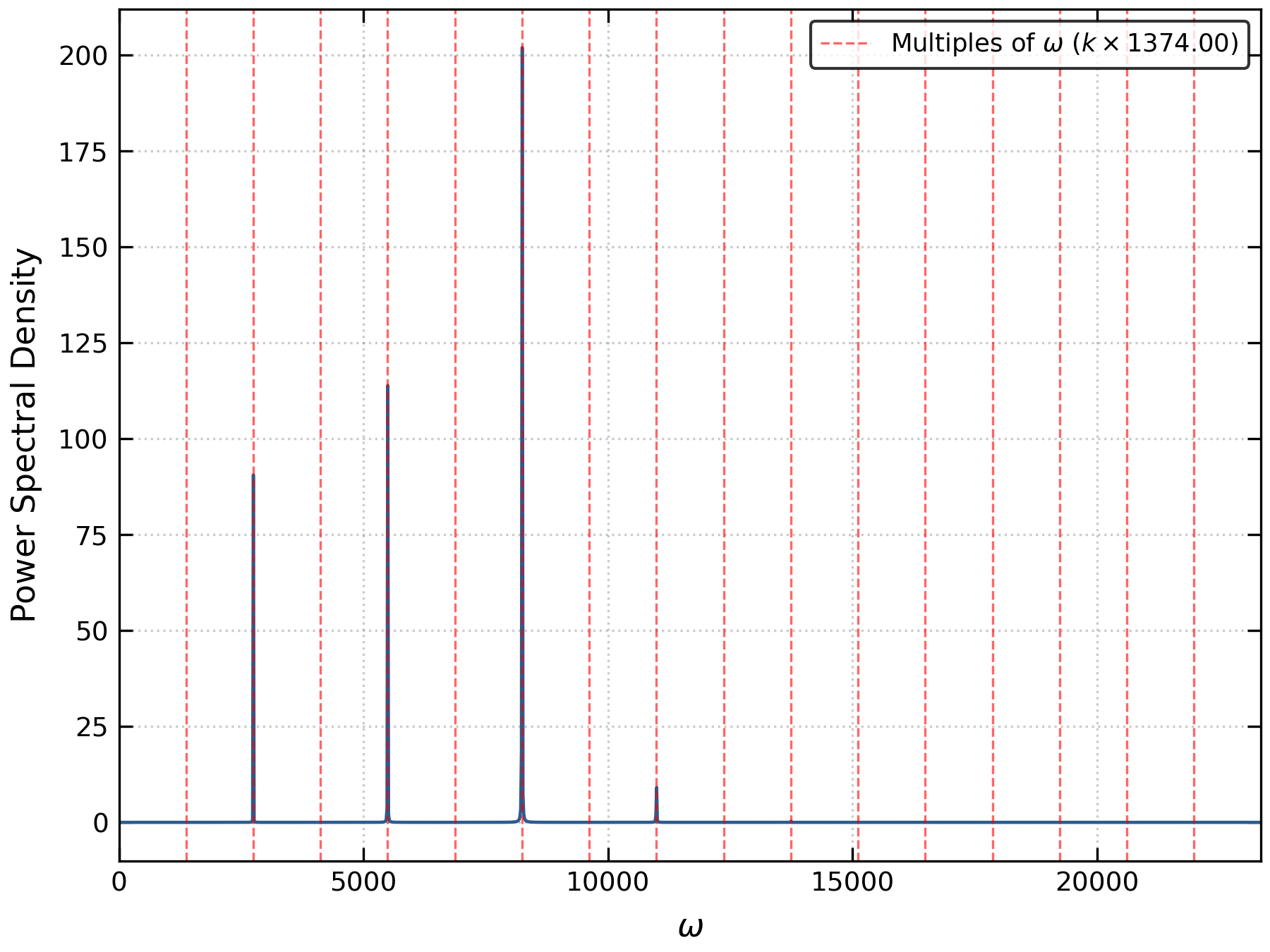}\label{fig:synch fft}}
    \\[-1ex]
    \subfloat[]{\includegraphics[width=0.48\textwidth]{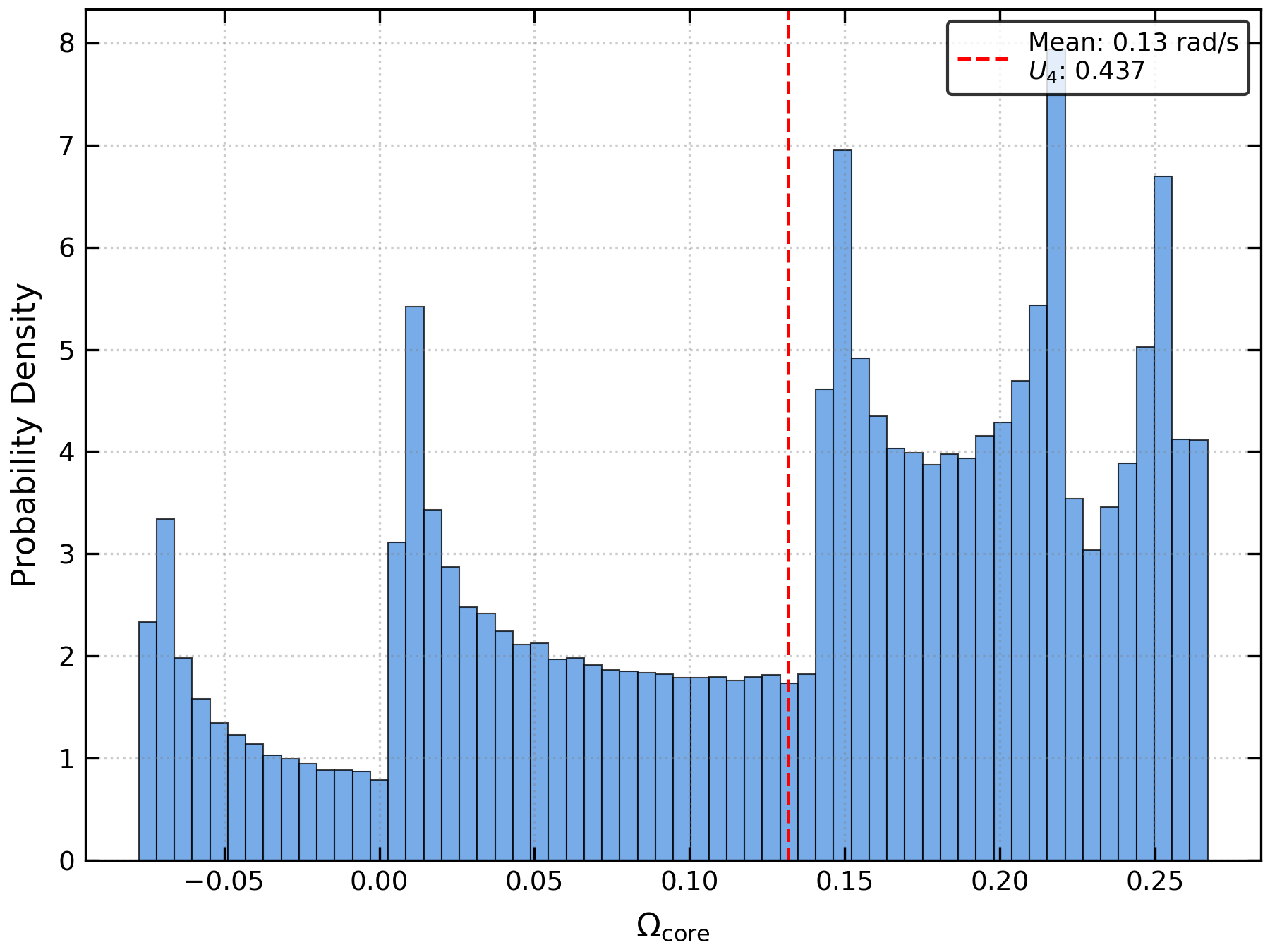}\label{fig:synch dist}}
    \hfil    \subfloat[]{\includegraphics[width=0.48\textwidth]{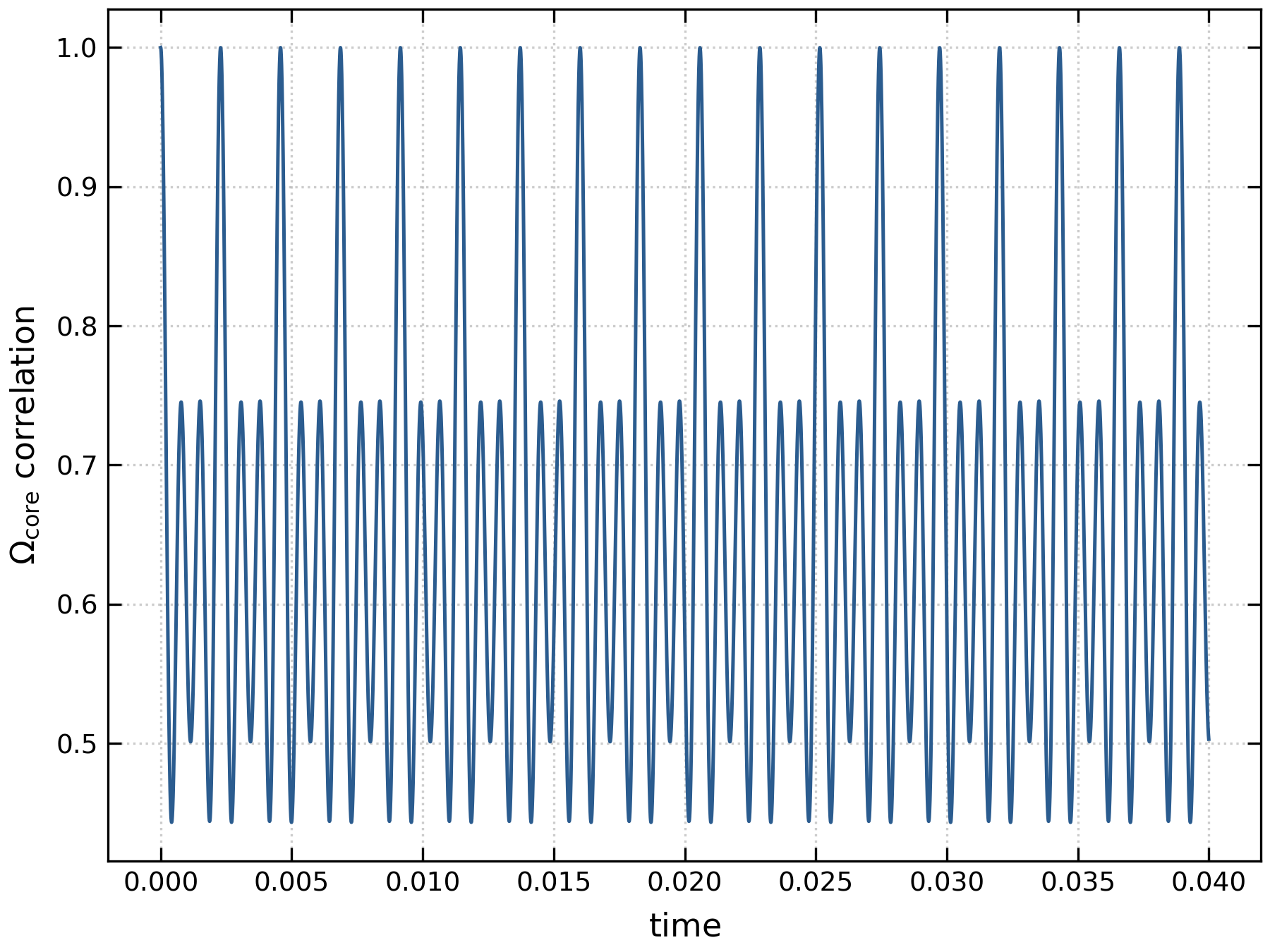}\label{fig:synch corr}}
    \caption{The results are from a cluster at $\omega_{\rm f}=1374, B=1200$ showing rotation in synchronized phase.(a) Instantaneous angular velocity of the core as a function of time. (b) Normalized frequency spectrum of the angular velocity. (c) Probability distribution function of the angular velocity. (d) Normalized time correlation function of the angular velocity. Note that all the plots are representative, The waveform of angular velocity can show diverse modes of rotation.}
\label{fig:synch_omega}    
\end{figure*}

\noindent  The collective angular velocity can therefore be expressed through the effective torque balance
\begin{equation}
\gamma_{\mathrm{eff}} \Omega_{\rm{cl}}
=
\tau_{\mathrm{cl}},
\label{eq:torque_balance}
\end{equation}

\noindent where $\gamma_{\mathrm{eff}} = \sum_i r_i^2$, and $\Omega_{\rm{cl}}(t)$ is the instantaneous collective angular velocity defined in Eq.~\eqref{eq:omega_cluster}. This relation makes explicit that collective rotation originates from the sum of the tangential components of the pairwise dipolar forces. The radial component of the dipolar interaction plays a complementary role. It maintains the integrity of the cluster and produces time-dependent bond deformations. For particle $i$, the net radial dipolar force is

\begin{equation}
\mathbf{F}_{\mathrm{r},i} = \sum_{j \neq i} \frac{3\mu^2}{r_{ij}^{4}}
\left(1-3\cos^2\psi_{ij}\right)\hat{\mathbf{r}}_{ij}.
\label{eq:radial_force}
\end{equation}

\noindent For a symmetric circular cluster, the unit vectors $\hat{\mathbf{r}}_{ij}$ sum to a net radial vector directed away from the center of mass. While the instantaneous radial force oscillates between attractive and repulsive states at frequency $2\omega_{\rm f}$, its time average over one field period yields a net attractive cohesive force toward the center of mass. Furthermore, the alternating attractive and repulsive radial force drives microscopic bond-length oscillations. From Appendix B, we obtain a characteristic bond-oscillation length scale that scales as $\omega_{\rm f}^{-1/5}$. In the high-frequency limit, this means the bond oscillations are strongly suppressed. Time averaging yields $\langle\frac{\cos^2\psi_{ij}}{r^4_{ij}}\rangle \approx \frac{\langle\cos^2\psi_{ij}\rangle}{r_{ij}^4} \approx \frac{1}{2 r^4_{ij}}$ assuming $\langle \Delta \theta_{ij} \rangle$ is small in a period, resulting in a time-averaged cohesive force
\begin{equation}
\langle\mathbf{F}_{\mathrm{r},i}\rangle = -\sum_{j \neq i} \frac{3\mu^2}{2r_{ij}^{4}}\hat{\mathbf{r}}_{ij}.
\label{eq:radial_force_avg}
\end{equation}
\noindent Time averaging of dipolar interactions under rotating magnetic fields is known to produce an effective attractive interaction between colloidal particles \cite{coughlan,martinez}. The present expression follows from the corresponding dipolar interaction in the simplified limit considered here. At high driving frequency as the bond oscillations get suppressed, $\langle\frac{\sin2\psi_{ij}}{r^4_{ij}}\rangle \approx \frac{\langle\sin2\psi_{ij}\rangle}{r^4_{ij}} \approx 0$, assuming $\langle \Delta \theta_{ij} \rangle$ is small in a period.  In this limit $\tau_{cl}\sim 0$ and $\Omega_{\rm{cl}}\sim 0$, consistent with the simulation results in Fig.~\ref{fig:omega_cl vs omega}.  This microscopic picture also clarifies an important point: collective rotation requires a deformable interaction network. In the absence of internal bond deformation, the time-dependent modulation of the dipolar forces cannot generate the same deformation-mediated rotational response. The collective rotation observed here therefore reflects the coupled evolution of bond lengths, bond orientations, and interaction torques within a deformable particle network.

\subsection{Synchronized collective rotation}

We next examine the synchronized state, in which the cluster undergoes coherent collective rotation in the same direction as the externally applied magnetic field. To characterize the temporal dynamics of this state, we calculate the instantaneous collective angular velocity, $\Omega_{\rm{cl}}(t)$, from the orbital motion of the particles about the instantaneous center of mass of the cluster. A representative time series for $\omega_{\rm f}=1374$ and $B=1200$ is shown in Fig.~\ref{fig:synch_omega}(a).  The angular velocity remains predominantly positive and exhibits a regular oscillatory modulation, demonstrating that the synchronized state is not characterized by a constant angular velocity but by a periodically modulated collective rotation. The periodic nature of this motion becomes particularly clear from the corresponding power spectrum shown in Fig.~\ref{fig:synch_omega}(b). A pronounced harmonic structure is observed, with peaks at $2\omega_{\rm f}$, $4\omega_{\rm f}$, $6\omega_{\rm f}$, and $8\omega_{\rm f}$. In the idealized limit of a rigid cluster, the interparticle separations remain fixed while the bond orientations rotate coherently with the cluster. The resulting dipolar torque therefore contains a leading oscillatory contribution of the form $ \tau_{\rm cl}(t)\sim \sin(2\omega_{\rm f}t+\phi)$, giving rise naturally to a response at twice the driving frequency. The presence of substantial spectral weight at $4\omega_{\rm f}$, $6\omega_{\rm f}$, $8\omega_{\rm f}$, and higher harmonics, however, shows that the synchronized cluster does not behave as a perfectly rigid rotating body. Instead, the particle network undergoes periodic microscopic deformations during its collective rotation. In particular, the interparticle separations and relative bond orientations vary with time, thereby modulating the magnitude and direction of the dipolar forces and introducing nonlinearities into the instantaneous torque acting on the cluster. These nonlinear bond-level dynamics generate the higher harmonics observed in the power spectrum.

\begin{figure*}[!t]
    \centering
    \subfloat[]{\includegraphics[width=0.48\textwidth]{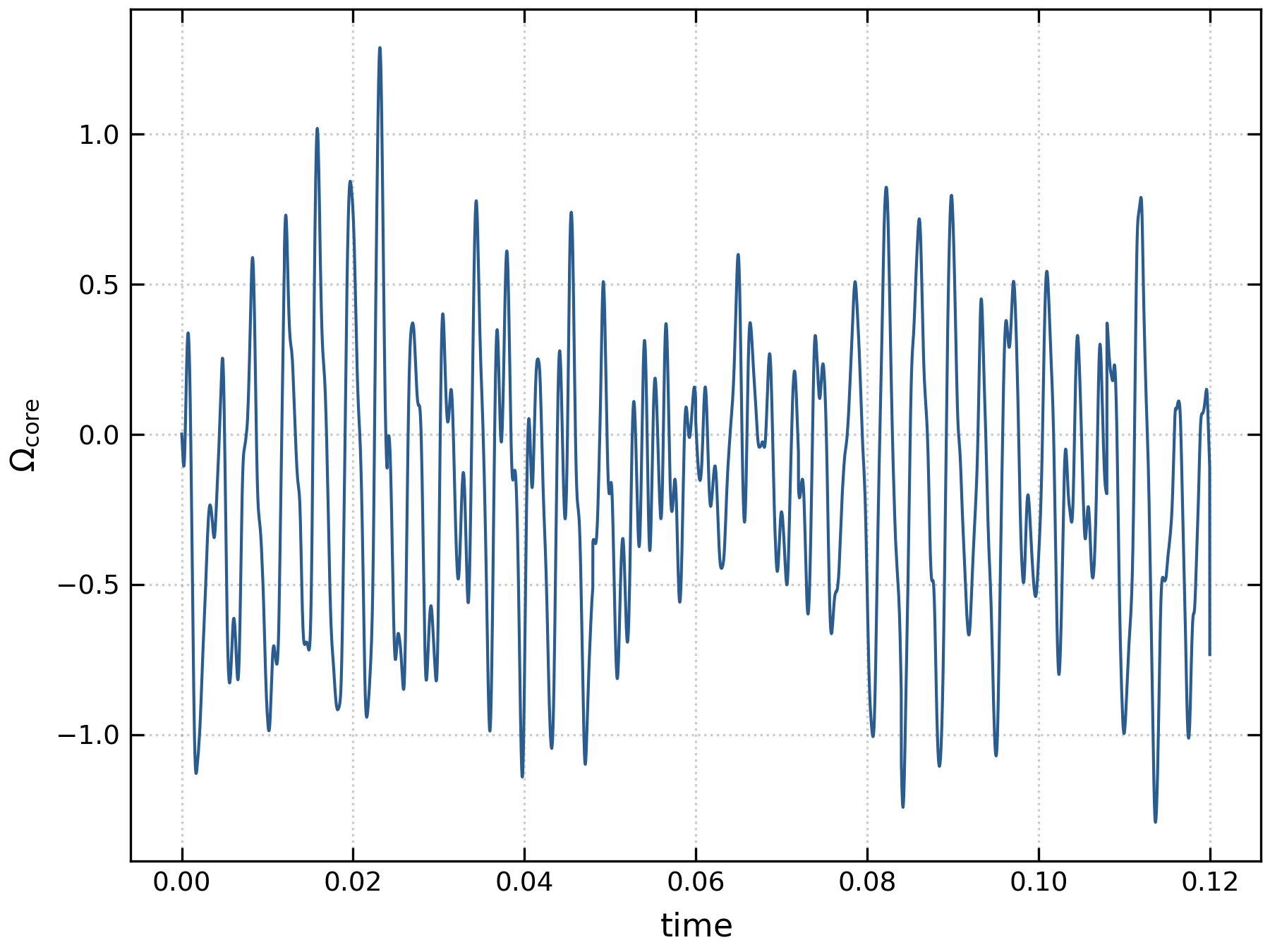}\label{fig:anti_time}}
    \hfill
    \subfloat[]{\includegraphics[width=0.48\textwidth]{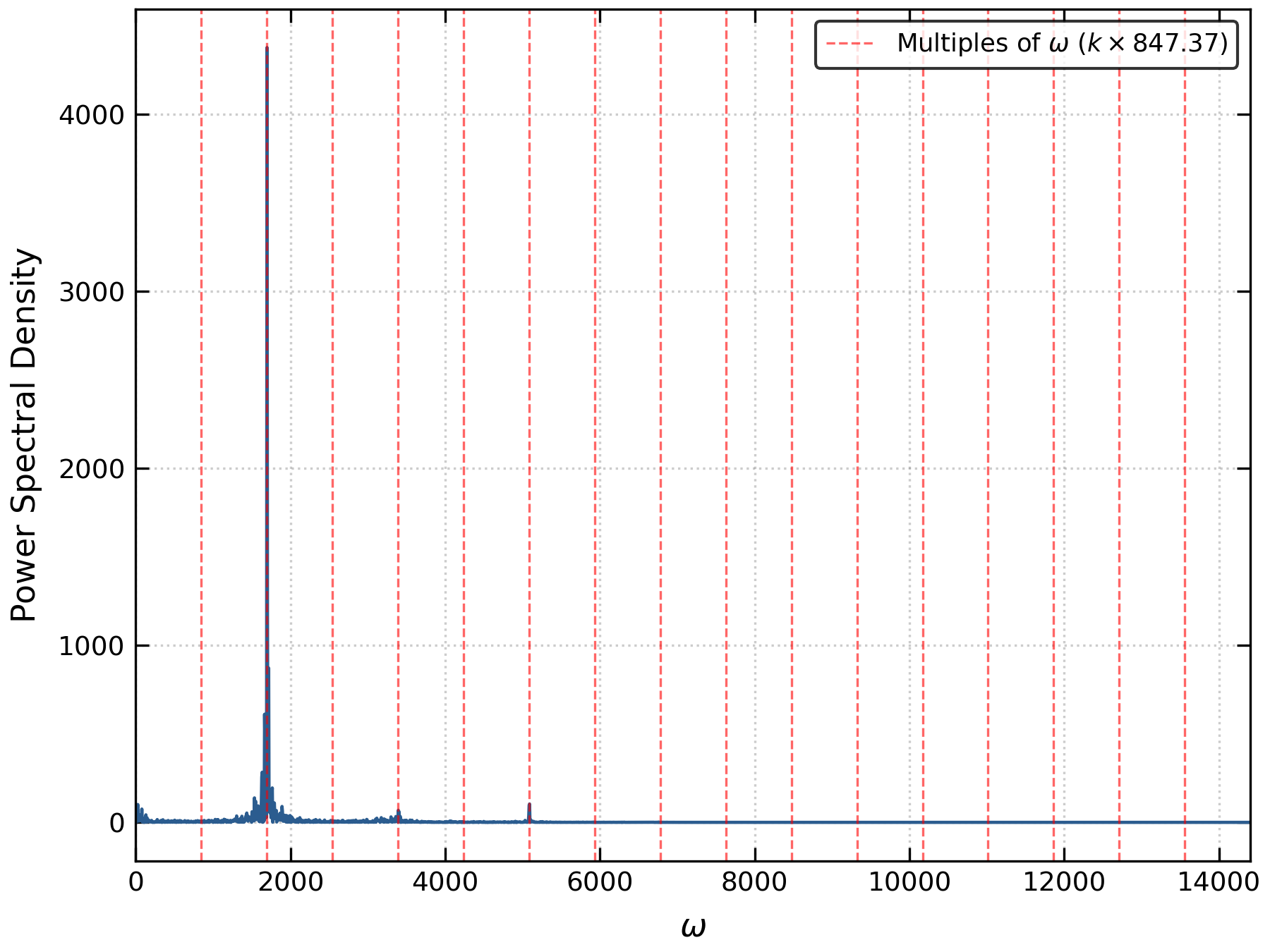}\label{fig:anti_fft}}
    \\[-1ex]
    \subfloat[]{\includegraphics[width=0.48\textwidth]{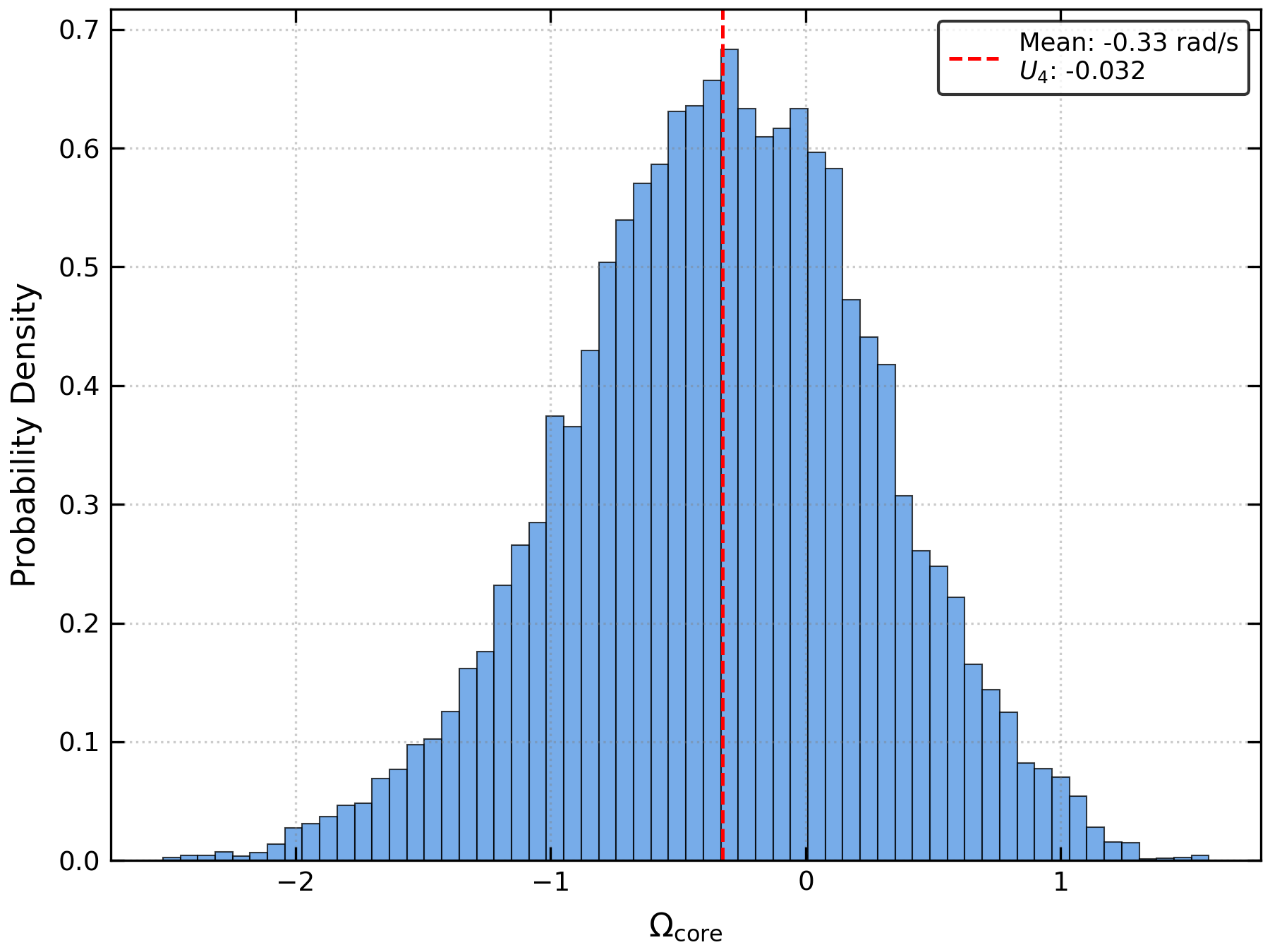}\label{fig:anti_dist}}
    \hfill
    \subfloat[]{\includegraphics[width=0.48\textwidth]{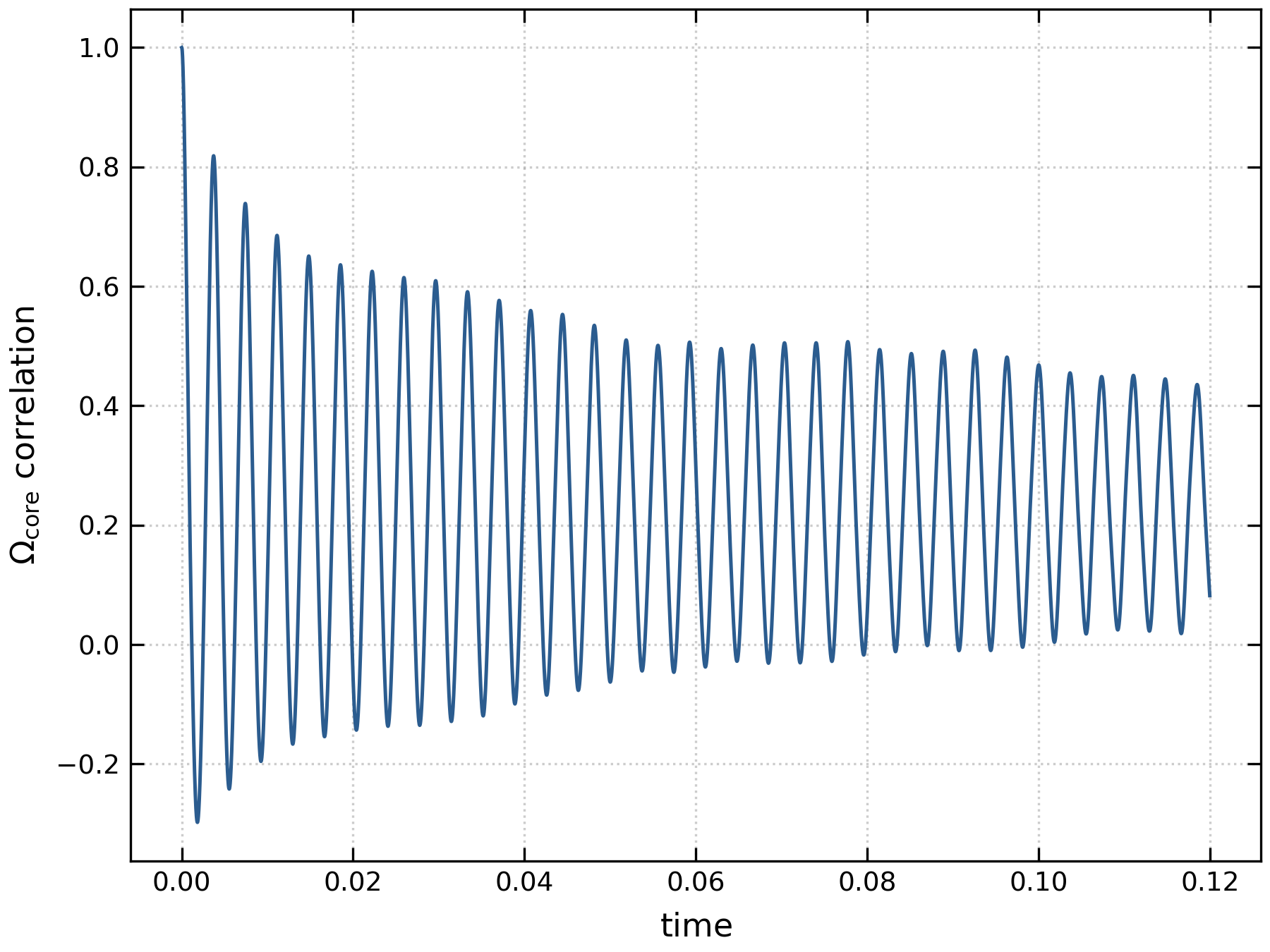}\label{fig:anti_corr}}
    \caption{The results are from simulations run at $\omega_{\rm f}=847, B=1200$ showing rotation in anti-synchronized phase.(a)Instantaneous angular velocity of the core as a function of time. (b) Normalized frequency spectrum of the angular velocity. (c) Probability distribution function of the angular velocity. (d) Normalized time correlation function of the angular velocity. Note that all the plots are representative, The waveform of angular velocity can show diverse modes of rotation.}
    \label{fig:anti_synch_omega}
\end{figure*}

Further information about the synchronized dynamics is provided by the probability distribution of the core angular velocity, $P(\Omega_{\rm core})$, shown in Fig.~\ref{fig:synch_omega}(c). The distribution is broad and structured rather than sharply localized around a single angular velocity, consistent with the strong temporal modulation visible in Fig.~\ref{fig:synch_omega}(a). Importantly, however, its mean remains positive, $\langle\Omega_{\rm core}\rangle\simeq0.13$, confirming that the net rotation of the crystalline core is in the same direction as the external magnetic field. Thus, synchronization in the present system should be understood in terms of the direction and long-time average of the collective rotation rather than as uniform rigid-body rotation at a constant instantaneous angular velocity. The broad distribution reflects the periodic variation of $\Omega_{\rm core}(t)$ during each driving cycle. The temporal coherence of this periodically modulated rotation is further demonstrated by the normalized autocorrelation function of \(\Omega_{\rm core}\), shown in Fig.~\ref{fig:synch_omega}(d). The correlation function exhibits persistent, nearly periodic oscillations over the entire observation window, with no appreciable systematic decay. This behavior indicates that the oscillations of the core angular velocity remain temporally correlated over many driving cycles. In other words, although the instantaneous angular velocity varies substantially with time, these variations are not random fluctuations; rather, they form a reproducible periodic response associated with the external driving. This long-lived oscillatory correlation provides an additional dynamical signature of the synchronized state and distinguishes it from the anti-synchronized regime, where the angular-velocity correlations decay substantially with time, as discussed in the next section.

The time series, power spectrum, angular-velocity distribution, and autocorrelation function collectively demonstrate that the synchronized state exhibits dynamics considerably richer than simple rigid-body rotation. The cluster possesses a positive mean angular velocity and retains long-lived temporal coherence with the external drive, while simultaneously exhibiting strong periodic modulation and a hierarchy of higher harmonics generated by internal deformation.
In this respect, the synchronized response is consistent with previous studies of magnetically driven colloidal systems, where rotating magnetic fields can induce synchronized particle dynamics and dynamically ordered structures \cite{jager,yan,han}. In driven dipolar colloids, synchronization with the rotating field generates an effective attractive interaction that promotes two-dimensional cluster formation \cite{jager}. Experiments on magnetic Janus colloids have likewise demonstrated the formation of hexagonally ordered rotating crystals, with hydrodynamic interactions playing a key role in the collective rotation and subsequent defect dynamics \cite{yan}. In contrast, we find that collective rotation of the cluster persists even in the absence of hydrodynamic interactions.
The present synchronized state nevertheless displays an important additional feature. Although the cluster is synchronized at the level of its mean collective rotation, the pronounced higher harmonics in Fig.~\ref{fig:synch_omega}(b), together with the broad distribution of instantaneous angular velocities in Fig.~\ref{fig:synch_omega}(c), reveal that synchronization does not imply rigid-body motion. Instead, the particle network remains dynamically deformable and undergoes periodic internal rearrangements while preserving coherent rotation. The persistent oscillations of the angular-velocity autocorrelation function in Fig.~\ref{fig:synch_omega}(d) further demonstrate that these variations are temporally organized rather than stochastic. Thus, the synchronized cluster may be viewed as a dynamically coherent but internally deformable structure, in which macroscopic phase locking coexists with nontrivial microscopic bond dynamics.

\subsection{Emergence of the Anti-Synchronized State}

The most remarkable feature of the phase diagram is the emergence of a collective rotational state in which the ordered interior of the cluster rotates in a direction opposite to that of the externally applied rotating magnetic field. This counter-rotating state is qualitatively different from the conventional synchronized response expected for magnetically driven particles, in which the entire cluster rotates coherently with the external drive. Instead, the cluster develops a pronounced spatial separation in its rotational dynamics, with the crystalline interior and the defect-rich boundary exhibiting opposite responses to the applied field. This behavior therefore represents a distinct nonequilibrium dynamical state of the colloidal crystal, in which the internal rotational response is not determined solely by the instantaneous direction of the external magnetic torque. The temporal dynamics of the counter-rotating core are illustrated in Fig.~\ref{fig:anti_synch_omega}(a) for $\omega_{\rm f}=847$ and $B=1200$. In contrast to the regular periodic modulation observed in the synchronized state, $\Omega_{\rm core}(t)$ displays considerably stronger and less regular fluctuations, repeatedly exploring both positive and negative instantaneous angular velocities. Nevertheless, the time-averaged angular velocity remains negative, $\langle\Omega_{\rm core}\rangle\simeq-0.33$, demonstrating that the net motion of the crystalline interior is opposite to the external drive. Thus, anti-synchronization does not imply that the core rotates uniformly in the reverse direction at every instant. Rather, it represents a statistically persistent counter-rotating state superimposed on substantial temporal fluctuations.

This distinction is also apparent in the power spectrum shown in Fig.~\ref{fig:anti_synch_omega}(b). The spectrum retains a pronounced contribution near $2\omega_{\rm f}$, demonstrating that the dynamics of the counter-rotating core remain coupled to the periodic magnetic forcing. However, compared with the synchronized state, the harmonic structure is considerably less sharply organized: higher harmonics are strongly reduced, while appreciable spectral weight appears at frequencies below the principal $2\omega_{\rm f}$ response. The anti-synchronized state therefore retains a field-driven periodic component but also contains slower dynamical modes. In the present system, these low-frequency contributions are consistent with slower structural rearrangements and enhanced deformation of the defect-rich surface and the adjoining particle network. We emphasize, however, that the power spectrum alone does not identify the microscopic origin of these low-frequency modes. The probability distribution $P(\Omega_{\rm core})$, shown in Fig.~\ref{fig:anti_synch_omega}(c), provides a complementary view of these fluctuations. The distribution is broad and extends over both positive and negative angular velocities, but its mean is shifted to negative values. This broad distribution is consistent with the strongly fluctuating time series in Fig.~\ref{fig:anti_synch_omega}(a): although the instantaneous core motion can transiently follow the field, the accumulated rotational response remains counter-directed. This observation is important for distinguishing anti-synchronization from a simple rigid-body rotation in the opposite direction. The anti-synchronized core instead exhibits a fluctuating nonequilibrium state whose statistically preferred direction of rotation opposes the external drive. A particularly clear distinction between the synchronized and anti-synchronized states is provided by the normalized angular-velocity autocorrelation function in Fig.~\ref{fig:anti_synch_omega}(d). The correlation initially exhibits oscillations associated with the periodic forcing, but their amplitude decays substantially with increasing time. This behavior contrasts with the synchronized state [Fig.~\ref{fig:synch_omega}(d)], where persistent oscillations survive throughout the observation window with little appreciable decay. The loss of correlation in the anti-synchronized phase demonstrates that the core does not maintain the same degree of long-time phase coherence as in the synchronized state. Taken together with the broadened power spectrum and angular-velocity distribution, this result indicates that anti-synchronization is a dynamically fluctuating state rather than simply the synchronized state with the sign of $\Omega_{\rm core}$ reversed.

\begin{figure}[!t]
    \centering
    \includegraphics[width=0.48\textwidth]{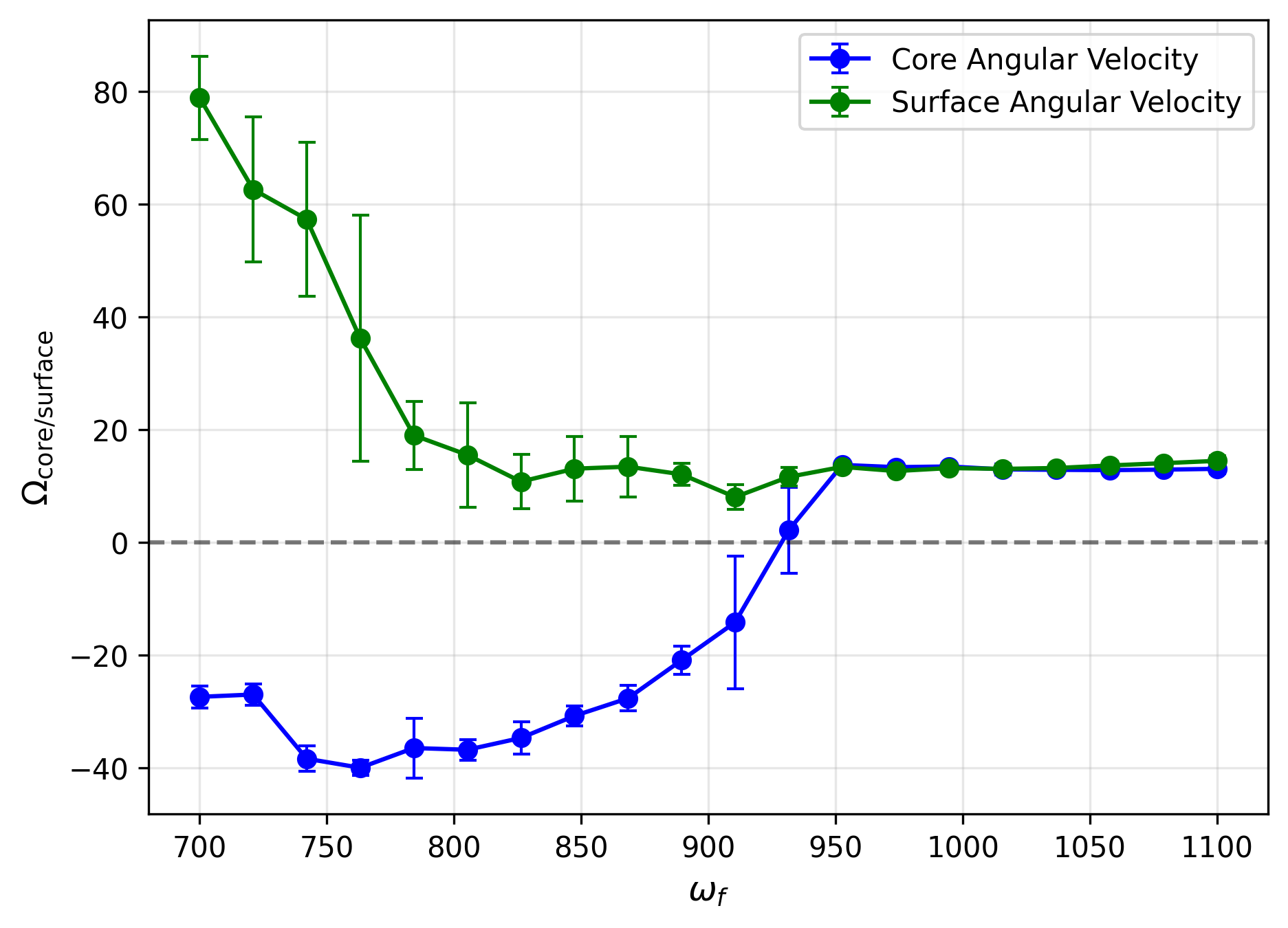}
    \caption{Comparison of the orbital angular velocities of the cluster core and boundary for a cluster B=1200.The particles are divided into layers with respect to the distance from the center of the cluster. The outer 3 layers are taken to be the surface, the rest is the core.Each
    data point is averaged over 5 simulations of random initial
    conditions. The error bars shows the standard deviation of the
    mean angular velocities from these 5 simulations}
    \label{fig:orbital-omega-comparison}
\end{figure}

The spatial separation of the rotational dynamics becomes particularly evident when the mean angular velocities of the core and surface are examined separately as a function of the driving frequency, as shown in Fig.~\ref{fig:orbital-omega-comparison}. Throughout the anti-synchronized regime, the surface angular velocity remains positive, indicating that the outer layers continue to rotate in the same direction as the magnetic field. The crystalline core, by contrast, possesses a pronounced negative angular velocity over a broad range of driving frequencies. At the lower frequencies shown, the magnitude of the counter-rotation increases before reaching its strongest response at intermediate frequencies, after which $\Omega_{\rm core}$ progressively increases toward zero as the driving frequency is raised. Approaching the anti-synchronized--synchronized transition, the magnitude of the negative core angular velocity decreases rapidly, whereas the surface continues to maintain a positive rotational response. Near the transition, \(\Omega_{\rm core}\) changes sign and subsequently approaches the surface angular velocity. At still higher frequencies, the two curves nearly coincide, indicating that the dynamical distinction between the surface and interior has largely disappeared and the cluster rotates coherently in the field direction. The transition to synchronization can therefore be viewed as a recovery of rotational coherence between two regions that are dynamically decoupled in the anti-synchronized state.

In the anti-synchronized phase, the defect-rich boundary layer rotates in the same direction as the externally applied magnetic field, indicating that the surface particles remain strongly coupled to, and dynamically influenced by, the rotating magnetic drive. In striking contrast, the ordered crystalline core rotates steadily in the opposite direction. Thus, despite being part of a single connected cluster, the system develops two dynamically distinct rotational responses: a field-following boundary and a counter-rotating crystalline interior. The coexistence of these oppositely rotating regions demonstrates that the rotational response of the cluster is strongly heterogeneous and cannot be described by a single collective angular velocity. Instead, the boundary and core become dynamically decoupled, with the intervening defect-rich region accommodating the mismatch between their rotational motions. This behavior contrasts sharply with the synchronized phase. As shown in Fig.~\ref{fig:orbital-omega-comparison}, increasing the driving frequency progressively reduces the difference between the angular velocities of the core and boundary. The counter-rotation of the core weakens, while its angular velocity increases toward that of the field-following boundary. At sufficiently high driving frequencies, both regions rotate in the same direction with nearly identical angular velocities, indicating the recovery of coherent collective rotation. In this regime, the distinction between the rotational dynamics of the interior and surface effectively disappears, and the entire cluster responds as a single dynamically synchronized object. The evolution from counter-rotation to synchronization therefore involves a progressive restoration of coherent rotational motion across the cluster rather than simply a change in the response of individual particles. 

Counter-rotation relative to a rotating magnetic field has previously been reported in magnetotactic bacterial swarms, where hydrodynamic interactions between swimming bacteria can generate collective rotation opposite to the applied field\cite{belovs}. The present state is qualitatively different: the counter-rotation is spatially heterogeneous within a single connected cluster, with the defect-rich surface following the external field while the crystalline interior rotates against it. 

The coexistence of opposite rotational directions within a single mechanically connected crystal raises two central questions. First, how can the ordered crystalline core sustain a persistent mean rotation opposite to the direction favored by the external magnetic drive? Second, what changes within the cluster as the driving frequency is increased such that coherent core–surface rotation is eventually restored? Addressing these questions requires examining how structural defects, microscopic bond deformations, and interaction torques evolve across the cluster. We therefore turn next to the internal dynamics of the defect-rich interface.

\begin{figure*}[!t]
\centering
 \subfloat[]{\includegraphics[width=0.48\textwidth]{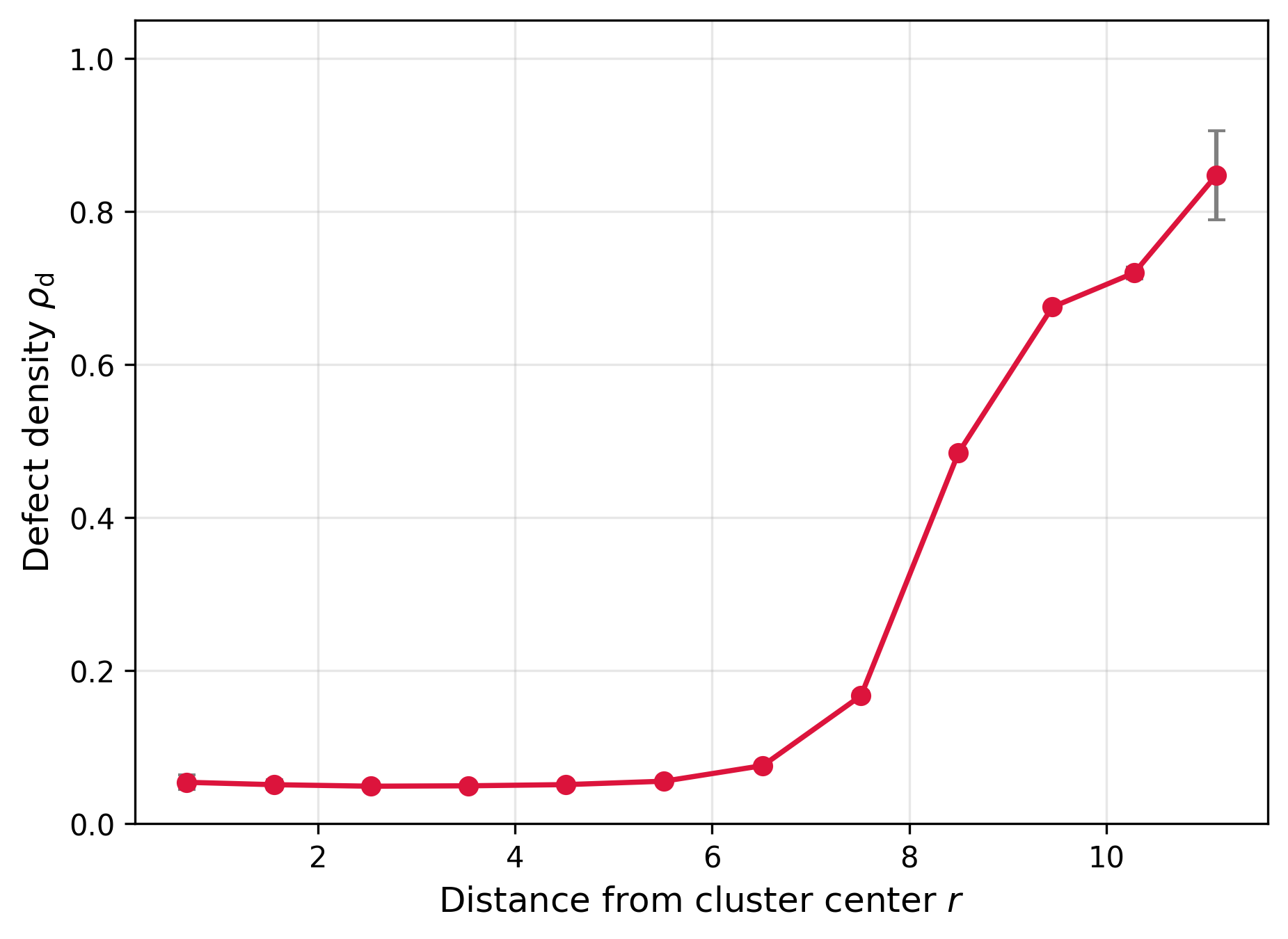}} \hfill
 \subfloat[]{\includegraphics[width=0.48\textwidth]{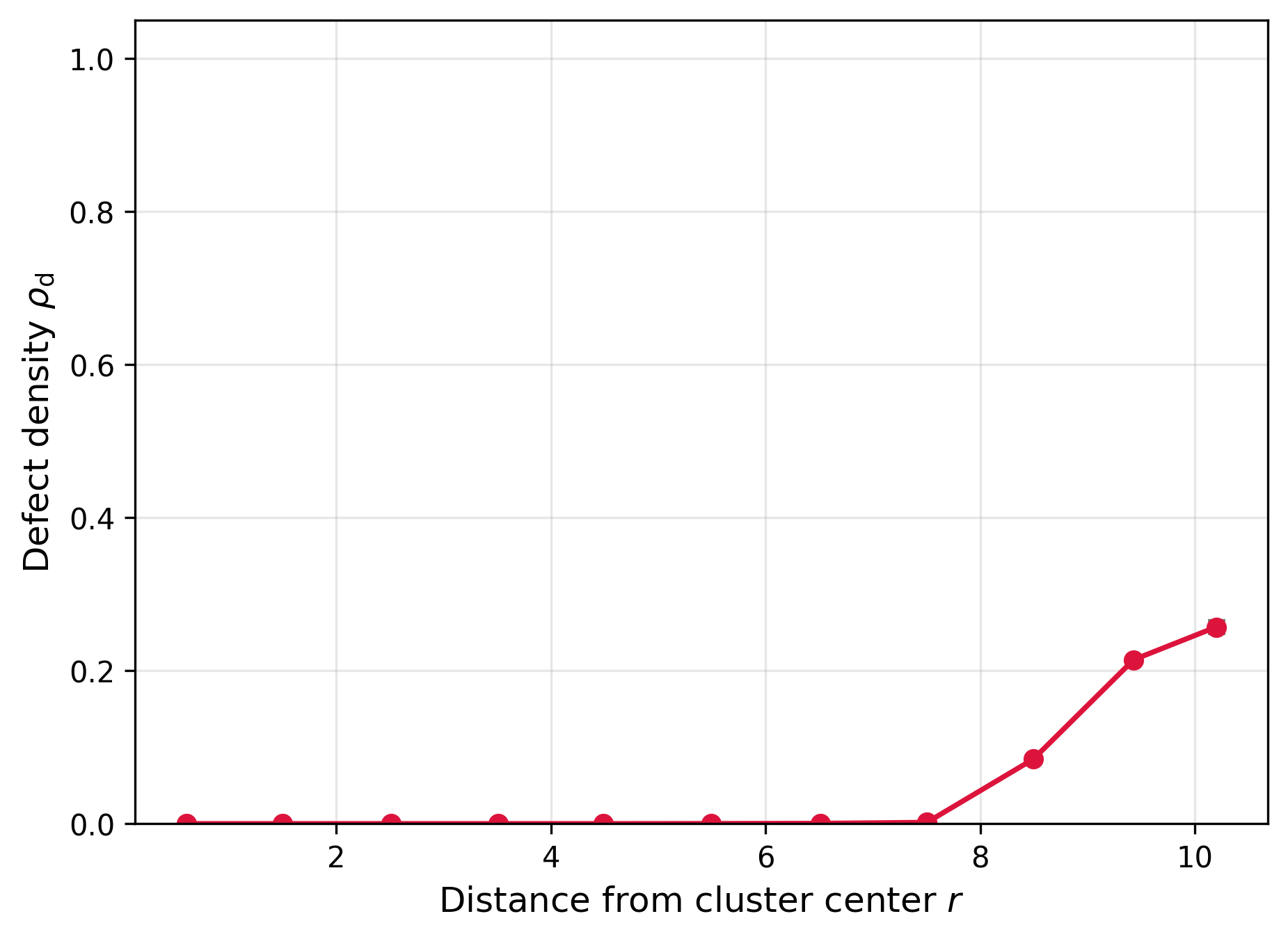}}
\caption{
Radial variation of the defect density in the anti-synchronized and synchronized phases.(a) Anti-synchronized state, $\omega_{\rm{f}}=860$, $B=1200$. (b) Synchronized state, $\omega_{\rm{f}}=1374$, $B=1200$
}
\label{fig:defect_density}
\end{figure*}

\subsection{Defect Dynamics and lattice reorientation} 

Structural defects and localized particle rearrangements are known to play an important role in accommodating deformation in colloidal solids, particularly through dislocation activity and spatially heterogeneous structural relaxation \cite{schall,schall1}. Motivated by these observations, we examine whether the contrasting rotational responses of the crystalline core and the outer boundary are accompanied by corresponding differences in the spatial distribution and dynamics of structural disorder. To examine this connection, we first characterize the spatial distribution of defects across the cluster. To quantify the spatial organization of disorder, the cluster is divided into successive surface-following layers based on the distance from cluster center.To quantify the rotational dynamics of the crystalline lattice in each layer, we compute the hexatic angular velocity $\Omega_{\psi_6}(r,t)$:
\begin{equation} 
    \Omega_{\psi_6}(r,t) = \frac{1}{N_r} \sum_{i=1}^{N_r} \frac{d\theta_{6,i}(t)}{dt},
\end{equation}
\noindent where $\theta_{6,i}(t)$ denotes the instantaneous hexatic orientation angle of particle $i$ at radial distance $r$. While $\Omega_{\psi_6} = \Omega_{\rm{cl}}$ for a perfectly rigid crystal, the two quantities diverge significantly in the anti-synchronized due to large bond oscillation and nucleation of new orientational domains in the lattice. For each layer, we calculate the local defect density defined as the fraction of particles satisfying $|\psi_6|<0.9$, which identifies particles with reduced hexagonal order. Figure~\ref{fig:defect_density} shows the defect density as a function of the distance from the cluster boundary for representative anti-synchronized and synchronized states. In both cases, defects are concentrated predominantly near the surface, as expected for a finite crystalline cluster. A clear difference nevertheless emerges in the spatial extent of the defect-rich region. In the anti-synchronized state [Fig.~\ref{fig:defect_density}(a)], the defect density is comparatively large near the surface and it decreases relatively gradually on moving inward from the surface, with an appreciable defect population remaining even in internal layers and a small residual contribution extending farther into the cluster. In the synchronized state [Fig.~\ref{fig:defect_density}(b)], by contrast, the defect density remains comparatively low at the surface and becomes essentially negligible beyond the 3 outermost layers. The anti-synchronized state is therefore characterized not simply by the presence of surface defects, but by a broader defect-rich interfacial region separating the crystalline interior from the driven boundary.

\begin{figure*}[!t]
\centering
 \subfloat[]{\includegraphics[width=0.48\textwidth]{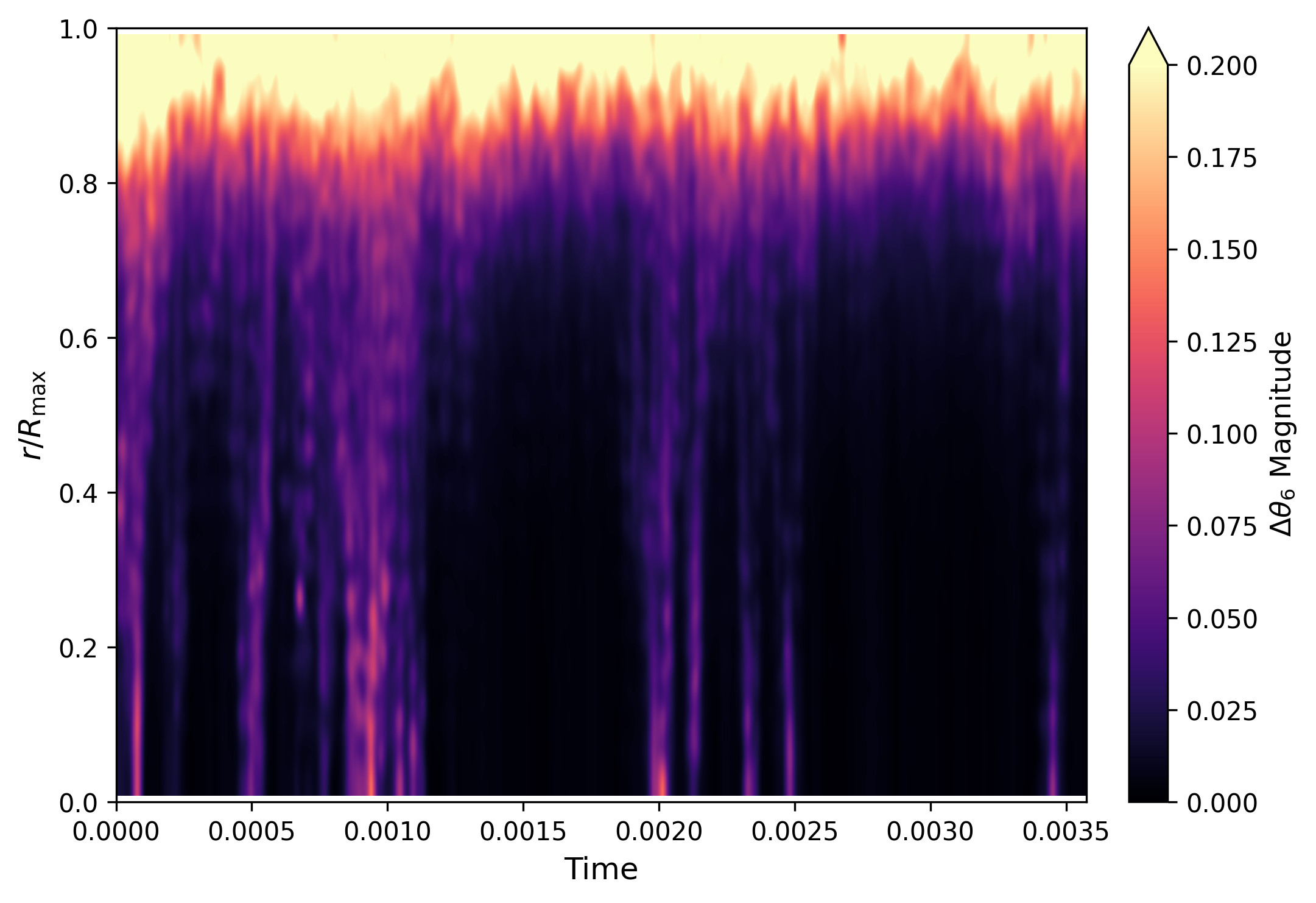}\label{fig:kymo_anti}} \hfill
 \subfloat[]{\includegraphics[width=0.48\textwidth]{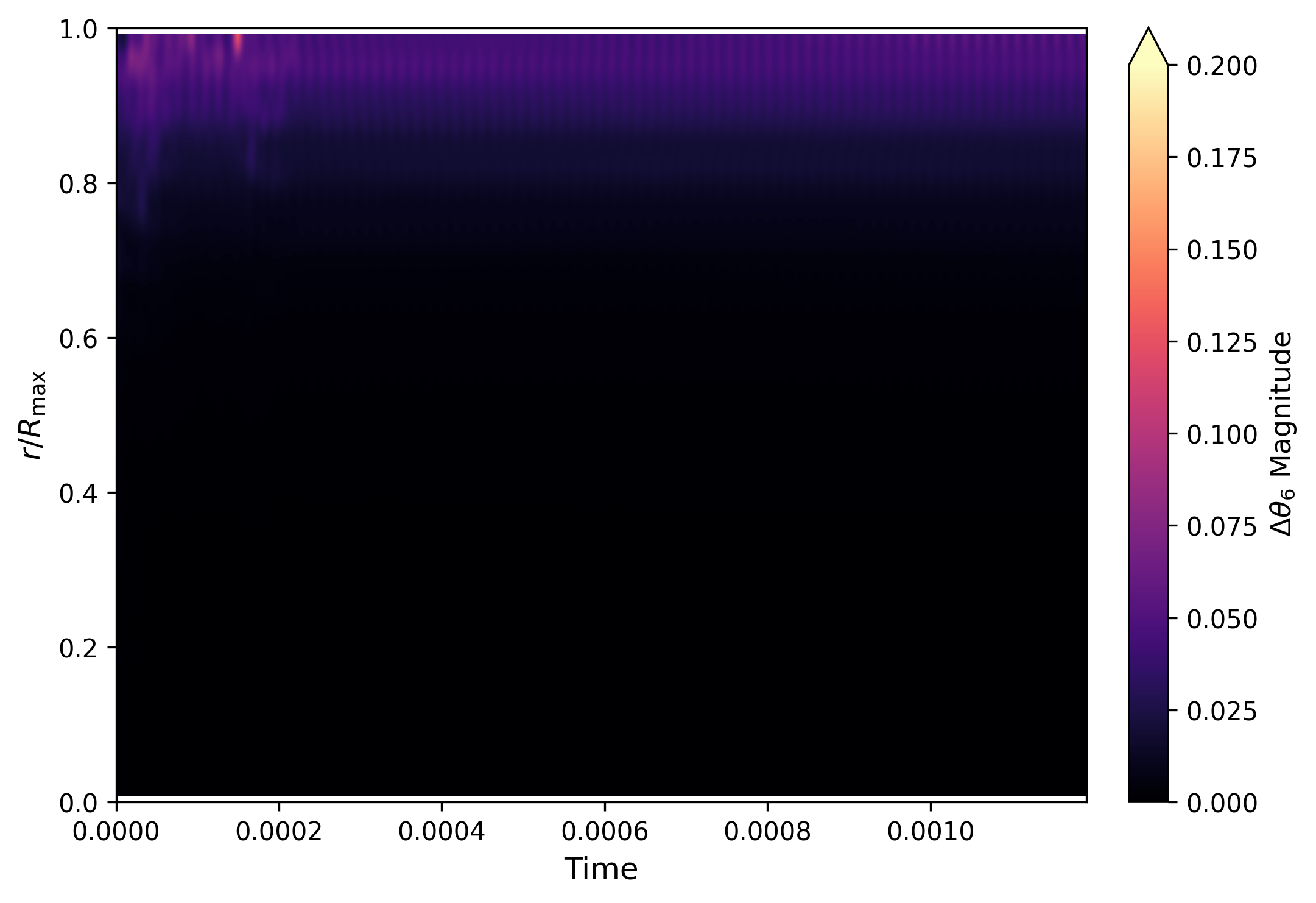}\label{fig:kymo_synch}}
\caption{
Spatiotemporal evolution of the lattice orientation in the anti-synchronized and synchronized phases. (a) Anti-synchronized state, $\omega_{\rm{f}}=860$, $B=1200$. (b) Synchronized state, $\omega_{\rm{f}}=1374$, $B=1200$.
}
\label{fig:nucleation}
\end{figure*}

To determine whether this difference is accompanied by distinct orientational dynamics, we examine the spatiotemporal evolution of the local sixfold bond orientation. Figure~\ref{fig:nucleation} shows the magnitude of the local orientational change, $|\Delta\theta_6|$, as a function of normalized radial position $r/R_{\max}$ and time. In the anti-synchronized state [Fig.~\ref{fig:nucleation}(a)], the strongest orientational fluctuations occur near the outer boundary, but the activity is not restricted to this region. Intermittent bands of enhanced $|\Delta\theta_6|$ extend from the surface toward the interior, and appreciable orientational fluctuations are observed over a broad range of radial positions. The resulting spatiotemporal pattern indicates that structural disturbances generated in the defect-rich outer region are accompanied by reorientation dynamics extending substantially into the crystalline interior. The corresponding behavior in the synchronized state is markedly different [Fig.~\ref{fig:nucleation}(b)].  The interior develops a nearly uniform region of very small $|\Delta\theta_6|$, whereas appreciable orientational activity remains concentrated primarily near the outermost portion of the cluster. Thus, although the surface remains structurally active in both dynamical regimes, the synchronized crystalline interior is comparatively insensitive to these surface fluctuations. The principal distinction between the two states is therefore not the complete disappearance of boundary disorder upon synchronization, but rather the degree to which surface-induced structural fluctuations extend into the interior.

\begin{figure*}[!t]
\centering
 \subfloat[]{\includegraphics[width=0.48\textwidth]{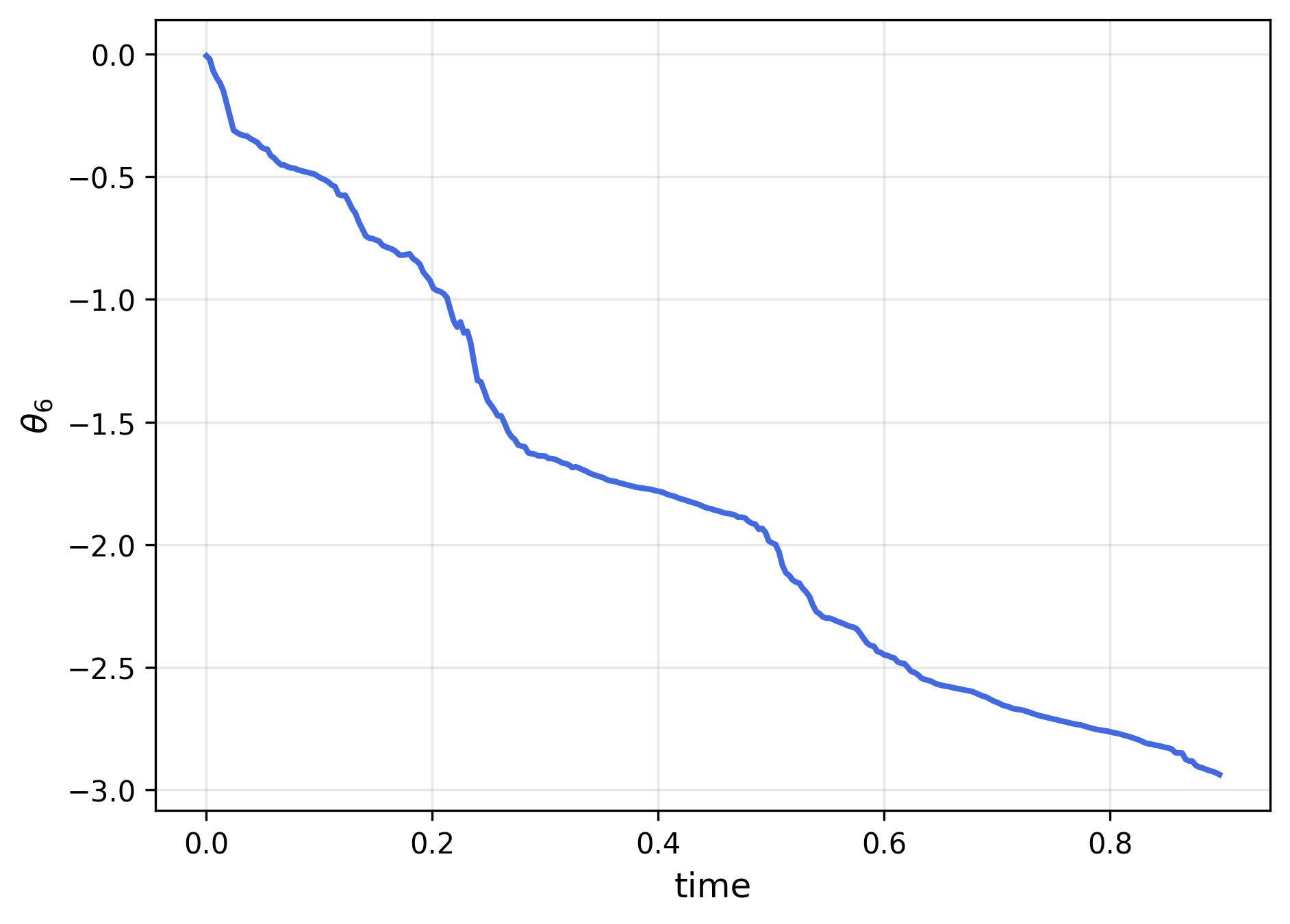}} \hfill
 \subfloat[]{\includegraphics[width=0.48\textwidth]{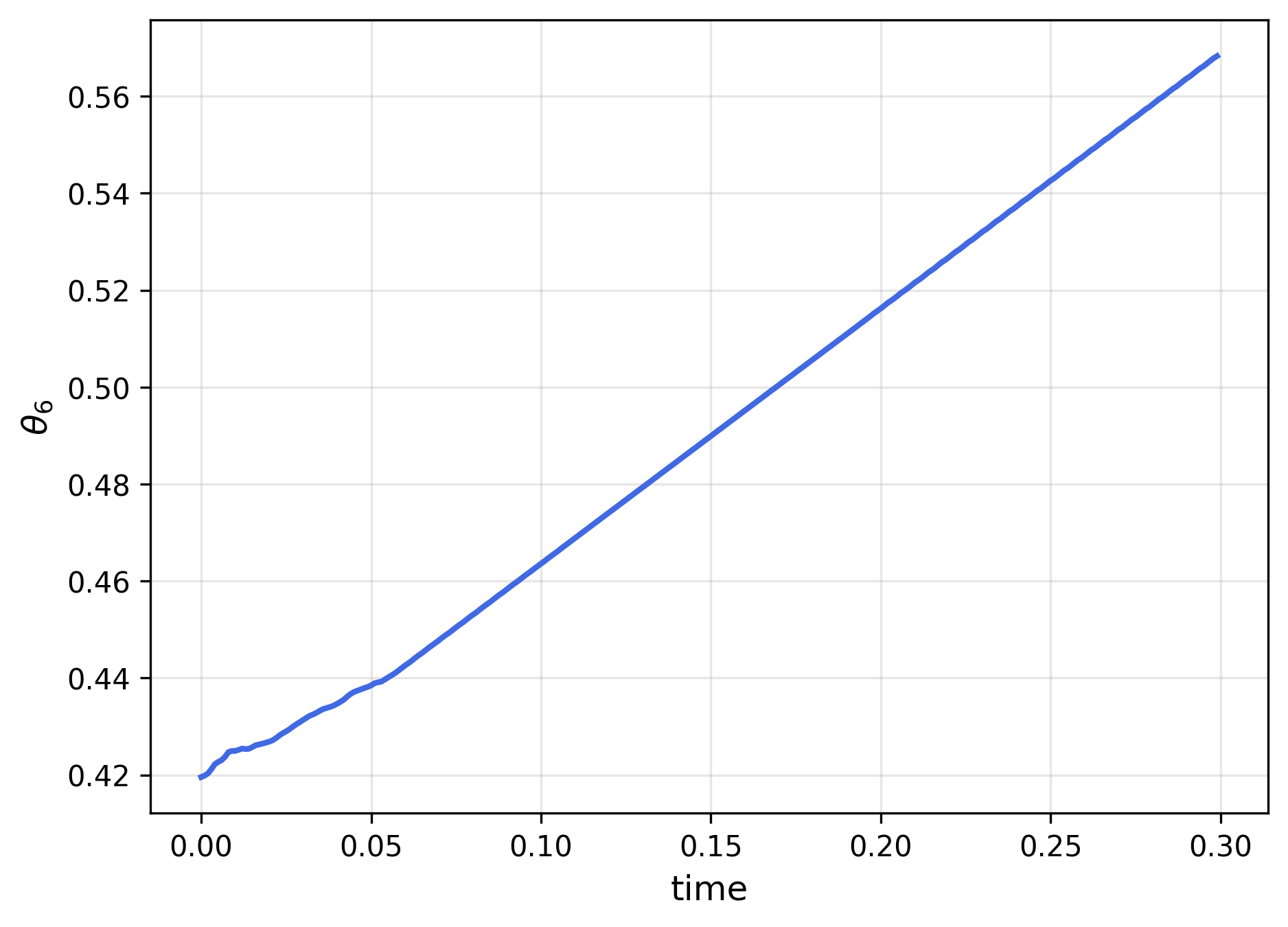}}
\caption{
Temporal evolution of the lattice orientation in the anti-synchronized and synchronized phases.(a) Anti-synchronized state, $\omega_{\rm{f}}=860$, $B=1200$. (b) Synchronized state, $\omega_{\rm{f}}=1374$, $B=1200$.
}
\label{fig:lattice_orientation}
\end{figure*}

This difference is also reflected in the temporal evolution of the global crystalline orientation. Figure~\ref{fig:lattice_orientation}(a) shows the sixfold lattice orientation $\theta_6(t)$ in the anti-synchronized state. The lattice orientation evolves systematically in the direction associated with the counter-rotating core, but its evolution is strongly nonuniform: intervals of relatively slow change alternate with more rapid reorientation events. The varying slope within each branch instead demonstrates that the crystalline orientation undergoes an intermittent rather than uniform rotational evolution. In the synchronized state [Fig.~\ref{fig:lattice_orientation}(b)], the behavior is qualitatively different. The lattice orientation varies almost linearly with time, indicating an approximately constant rate of crystalline rotation. This smooth evolution is consistent with the coherent rotational dynamics inferred from the angular-velocity time series and its persistent autocorrelation. The comparison between Figs.~\ref{fig:lattice_orientation}(a) and \ref{fig:lattice_orientation}(b) therefore reveals a direct structural distinction between the two dynamical regimes: synchronization is accompanied by nearly uniform rotation of the crystalline lattice, whereas anti-synchronization is associated with strongly modulated and intermittent lattice reorientation.

 Taken together, Figs.~\ref{fig:defect_density}–\ref{fig:lattice_orientation} establish a close connection between boundary disorder and the anomalous rotational dynamics of the crystalline interior. In the anti-synchronized state, the defect-rich interfacial region extends farther into the cluster, orientational fluctuations penetrate more deeply from the surface, and the global crystalline orientation evolves intermittently. In the synchronized state, structural fluctuations are largely confined to the outer boundary and the interior rotates much more uniformly. These observations suggest that the defect-rich interface provides a dynamically compliant region through which the surface and crystalline core can partially decouple. Such structural compliance is expected to modify the transmission of interaction forces and torques between the field-driven surface and the interior, a possibility that we examine quantitatively below through the bond and layer-resolved torque dynamics.

\begin{figure*}[!t]
    \centering
    \subfloat[\label{fig:cc synched}]{%
        \includegraphics[width=0.32\textwidth]{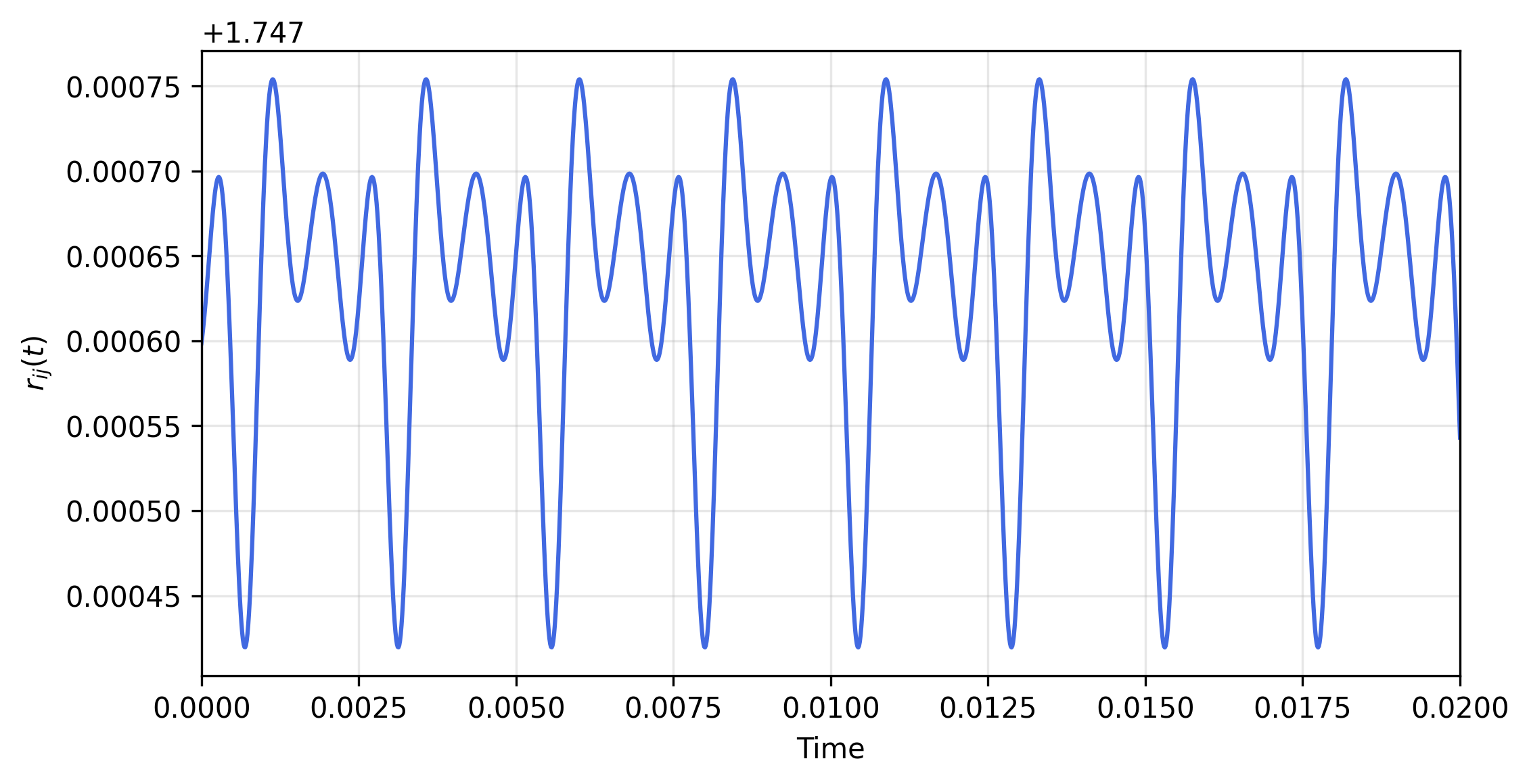}%
    }\hfill
    \subfloat[\label{fig:cc fft}]{%
        \includegraphics[width=0.32\textwidth]{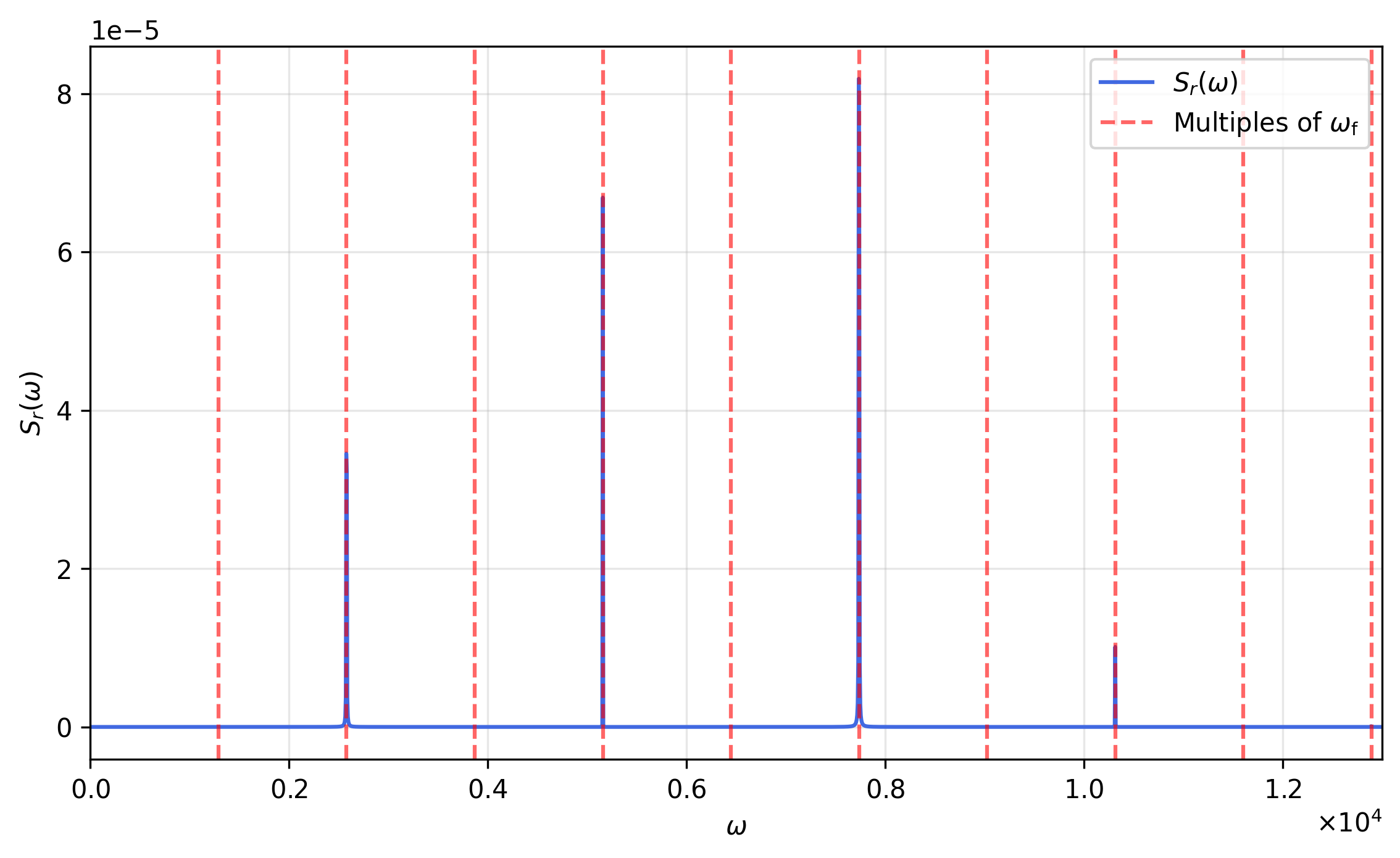}%
    }\hfill
    \subfloat[\label{fig:cc dom freq}]{%
        \includegraphics[width=0.32\textwidth]{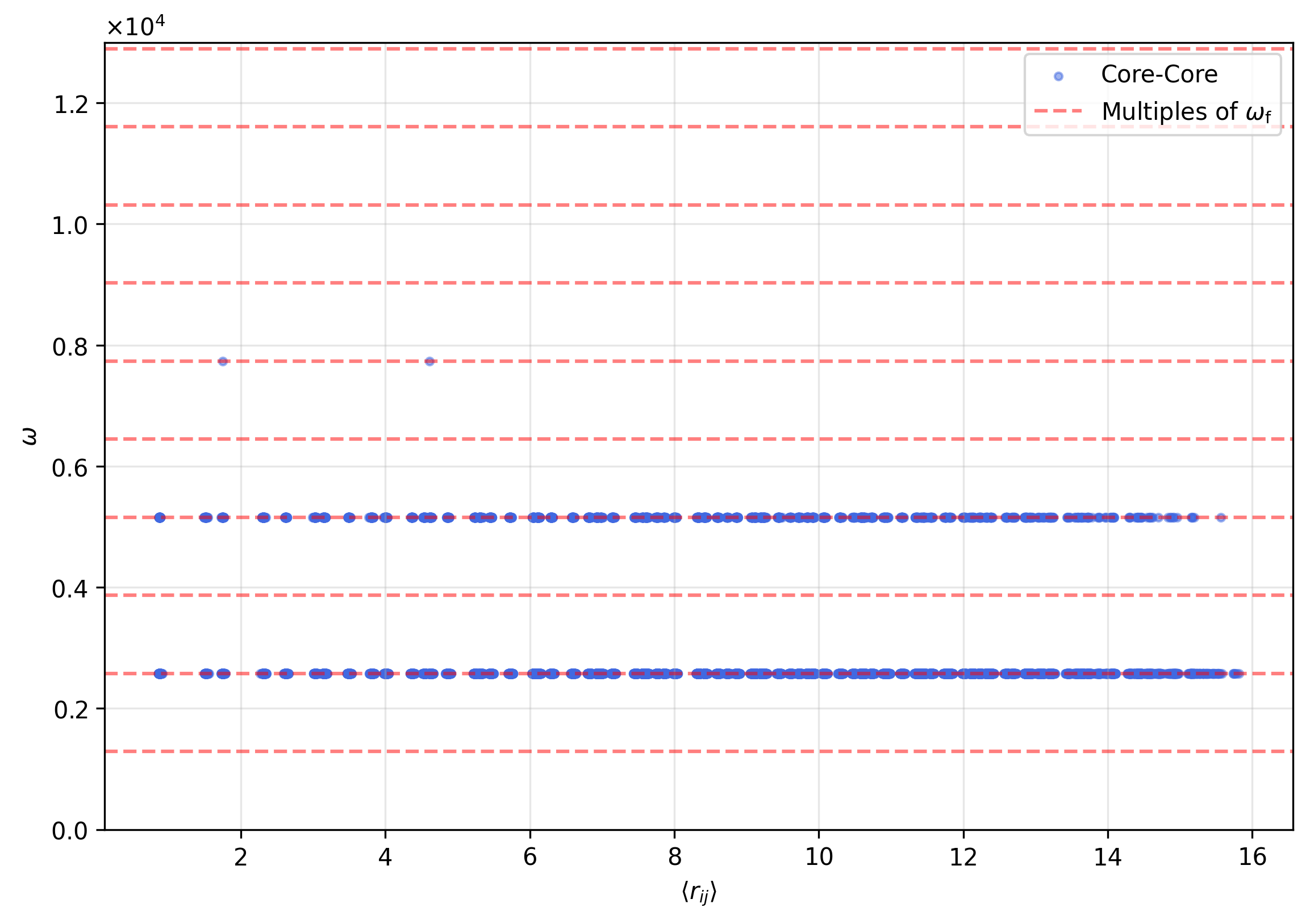}%
    }
    \caption{Temporal evolution of representative interparticle bond lengths in
the synched cluster.The plots are from a cluster at $\omega_{\rm f}=1374$, $B=1200$.(a) Time evolution of a representative core-core bond length, (b) Power spectrum of the core-core bond length time evolution, (c) Dominant frequency distribution of core-core bonds.
}
    \label{fig:bond_time_synch}
\end{figure*}

\subsection{Microscopic bond dynamics}

The structural analysis presented above indicates that the anti-synchronized state is accompanied by enhanced deformation and orientational fluctuations within the particle network. To determine how these differences manifest at the microscopic level, we examine the temporal dynamics of individual interparticle bonds. This analysis is particularly relevant because the dipolar interaction depends sensitively on both the instantaneous interparticle separation and the orientation of the bond relative to the external field. Consequently, the manner in which individual bonds deform provides a microscopic measure of how the particle network responds to and transmits the periodic magnetic forcing. For a pair of particles $i$ and $j$, we define the instantaneous bond length as

\begin{equation}
r_{ij}(t)
=
\left|
\mathbf r_i(t)-\mathbf r_j(t)
\right|.
\label{eq:bond_length}
\end{equation}

\noindent To identify the characteristic timescales associated with these microscopic rearrangements, we calculate the Fourier spectrum of the bond-length fluctuations,

\begin{equation}
\delta r_{ij}(t)
=
r_{ij}(t)-\langle r_{ij}\rangle,
\label{eq:bond_fluct}
\end{equation}
\noindent  and define

\begin{equation}
\tilde r_{ij}(\omega)
=
\int dt\,
\delta r_{ij}(t)e^{-i\omega t}.
\end{equation}
\noindent The corresponding power spectrum is

\begin{equation}
S_r(\omega)
=
\left|
\tilde r_{ij}(\omega)
\right|^2.
\end{equation}

We first consider the synchronized state. Figure~\ref{fig:bond_time_synch}(a) shows the time evolution of a representative core--core bond at $\omega_{\rm f}=1374$ and $B=1200$. The bond length undergoes regular oscillations with a well-defined periodic structure, demonstrating that microscopic deformation persists even though the cluster rotates coherently at the macroscopic level. The corresponding Fourier spectrum [Figure~\ref{fig:bond_time_synch}(b)] reveals a pronounced hierarchy of discrete peaks, with dominant contributions at $2\omega_{\rm f}$, $4\omega_{\rm f}$, and $6\omega_{\rm f}$.  The prominence of even harmonics is consistent with the twofold angular symmetry of the dipolar interaction, while the appearance of higher harmonics reflects the nonlinear response of the deformable interaction network to the periodic magnetic forcing. Thus, synchronization does not imply a rigid particle network; rather, the bonds undergo an organized periodic deformation that remains strongly frequency-locked to the external drive. The collective nature of this microscopic response is demonstrated more clearly in Figure~\ref{fig:bond_time_synch}(c), where the dominant oscillation frequency is shown for individual core--core bonds as a function of their mean bond length. The dominant frequencies cluster around discrete multiples of the driving frequency, particularly $2\omega_{\rm f}$, $4\omega_{\rm f}$, and $6\omega_{\rm f}$, rather than varying continuously from bond to bond. The occurrence of the same characteristic frequencies across bonds with different mean separations indicates that the harmonic response is not a peculiarity of a single representative pair. Instead, it reflects an organized response of the interaction network as a whole. The synchronized cluster can therefore sustain microscopic bond deformation while maintaining a coherent frequency-locked structure across its crystalline interior.

\begin{figure*}[!t]
    \centering
    \subfloat[\label{fig:cc anti time}]{%
        \includegraphics[width=0.32\textwidth]{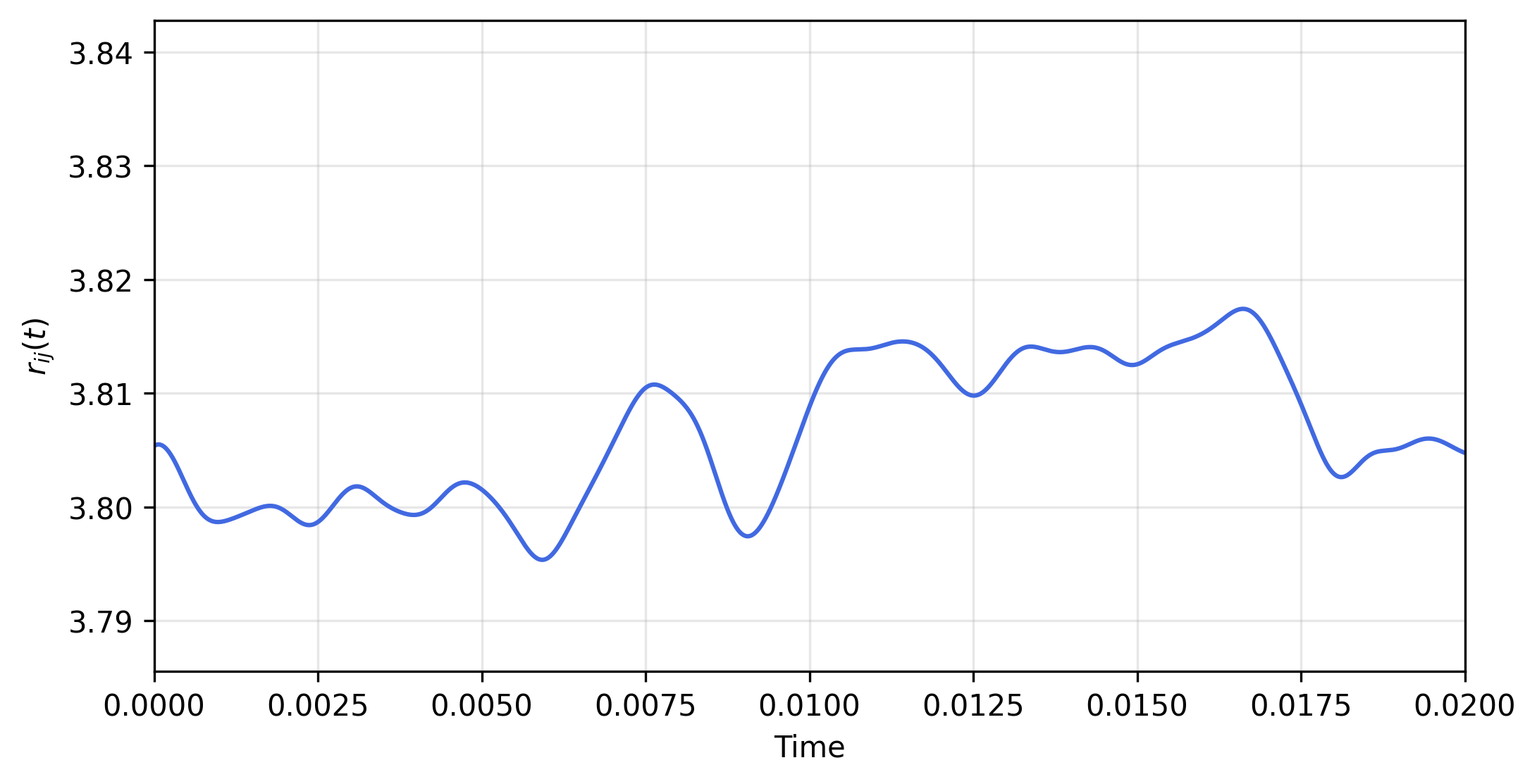}%
    }\hfill
    \subfloat[\label{fig:cc anti fft}]{%
        \includegraphics[width=0.32\textwidth]{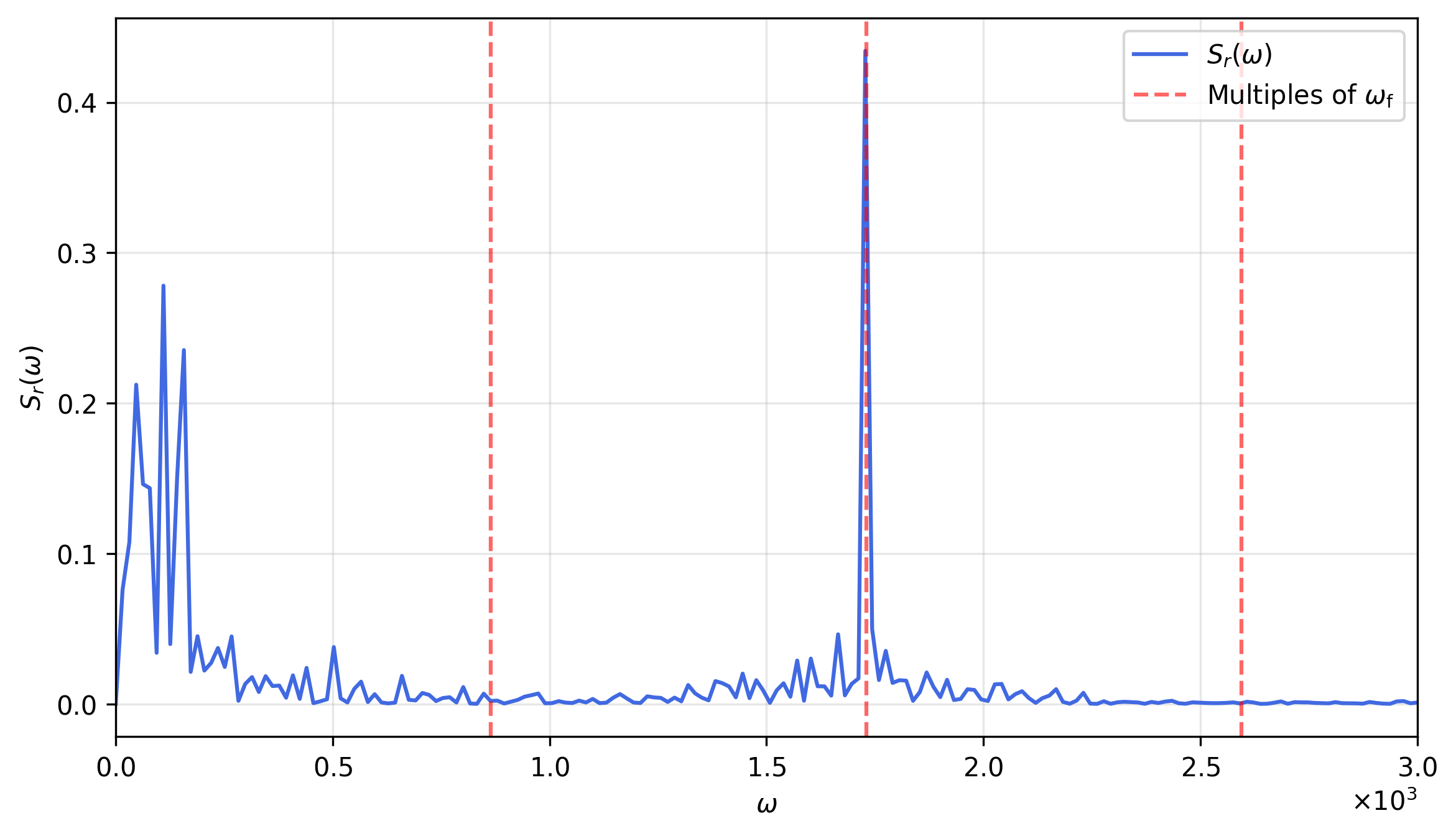}%
    }\hfill
    \subfloat[\label{fig:cc anti dom freq}]{%
        \includegraphics[width=0.32\textwidth]{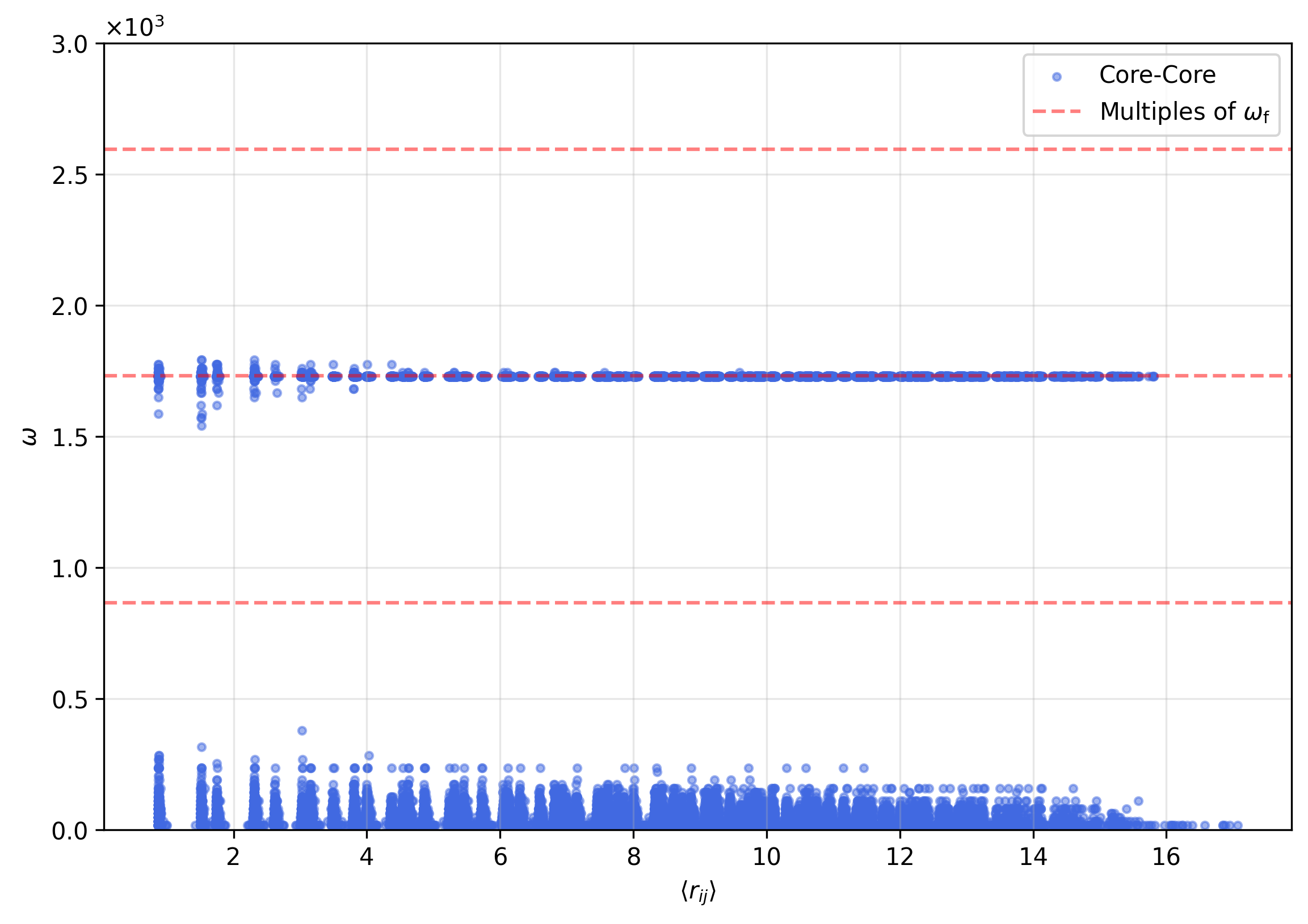}%
    }
    \caption{The plots are from a cluster at $\omega_{\rm f}=865$, $B=1200$.(a)Time evolution of a representative core-core bond length, (b) Power spectrum of the core-core bond length time evolution, (c) Dominant frequency distribution of core-core bonds.}
    \label{fig:bond_time_antisyn}
\end{figure*}

The bond dynamics change markedly in the anti-synchronized state. Figure~\ref{fig:bond_time_antisyn}(a) shows a representative core--core bond at $\omega_{\rm f}=865$ and $B=1200$. Compared with the synchronized state, the bond-length oscillations are substantially less regular and exhibit noticeable cycle-to-cycle variations in both amplitude and waveform. The corresponding power spectrum [Fig.~\ref{fig:bond_time_antisyn}(b)] retains a prominent contribution near $2\omega_{\rm f}$, showing that the bond dynamics remain coupled to the rotating magnetic field. However, the well-resolved sequence of higher harmonics characteristic of the synchronized state is strongly diminished. At the same time, substantial spectral weight appears at frequencies below $2\omega_{\rm f}$, demonstrating the emergence of slower modes of bond deformation. The bond-resolved analysis in Fig.~\ref{fig:bond_time_antisyn}(c) further emphasizes this difference. In contrast to the synchronized state, where the dominant frequencies are concentrated around a discrete hierarchy of field harmonics, the  anti-synchronized state displays a much broader distribution of dominant bond frequencies, with spectral weight distributed between the field-coupled response near \(2\omega_{\rm f}\) and a broad range of substantially lower frequencies.  A population of bonds continues to respond near $2\omega_{\rm f}$, while many others exhibit dominant dynamics at substantially lower frequencies. The microscopic response of the core is therefore no longer characterized by a single organized hierarchy of harmonics. Instead, the particle network supports both a field-driven component and a broad spectrum of slower bond rearrangements.

\begin{figure*}[!t]
    \centering
    \subfloat[]{\includegraphics[width=0.48\textwidth]{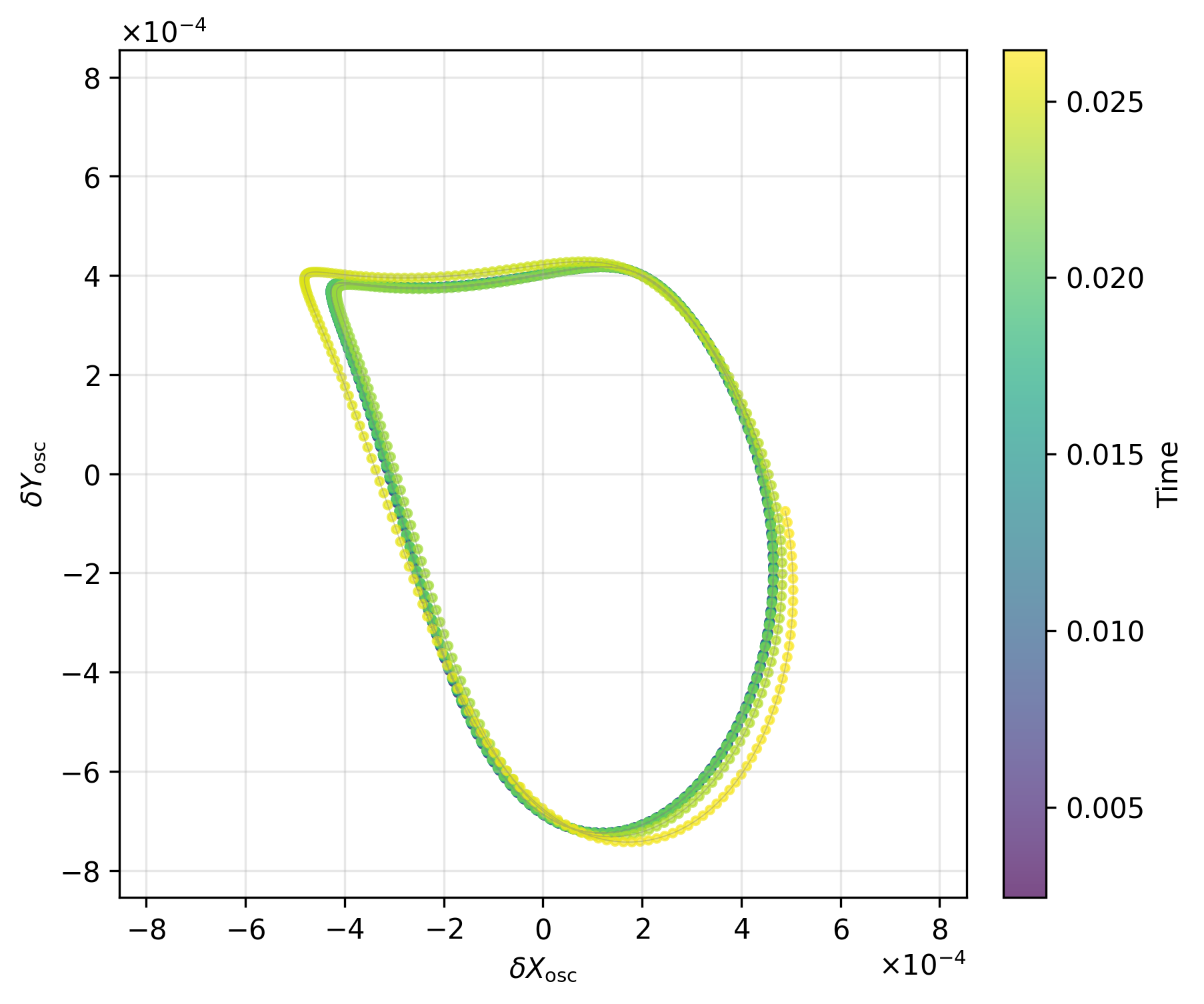}\label{fig:synch_2w}}
    \hfill
    \subfloat[]{\includegraphics[width=0.48\textwidth]{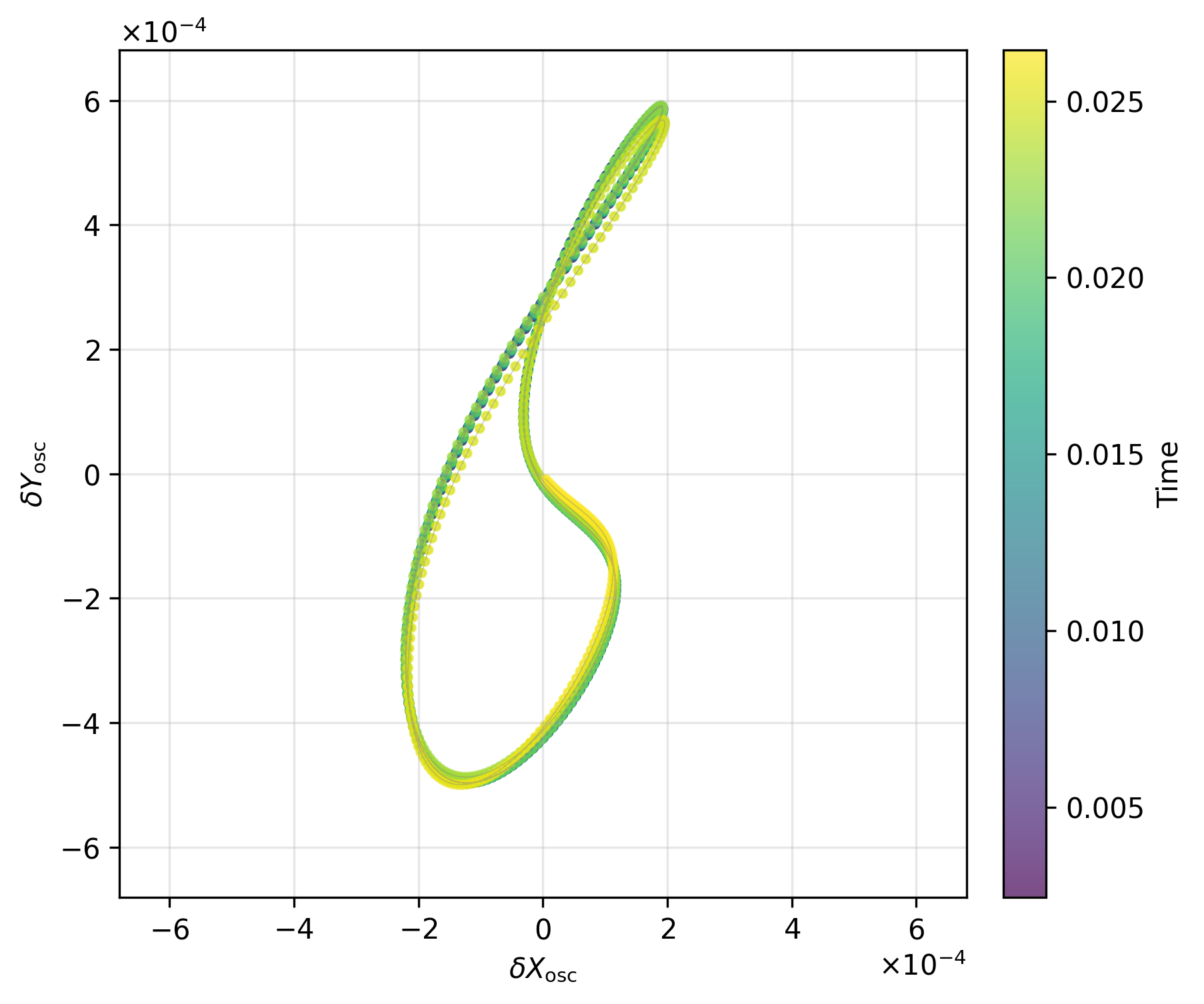}\label{fig:synch_4w}}
    \\[-1ex]
    \subfloat[]{\includegraphics[width=0.48\textwidth]{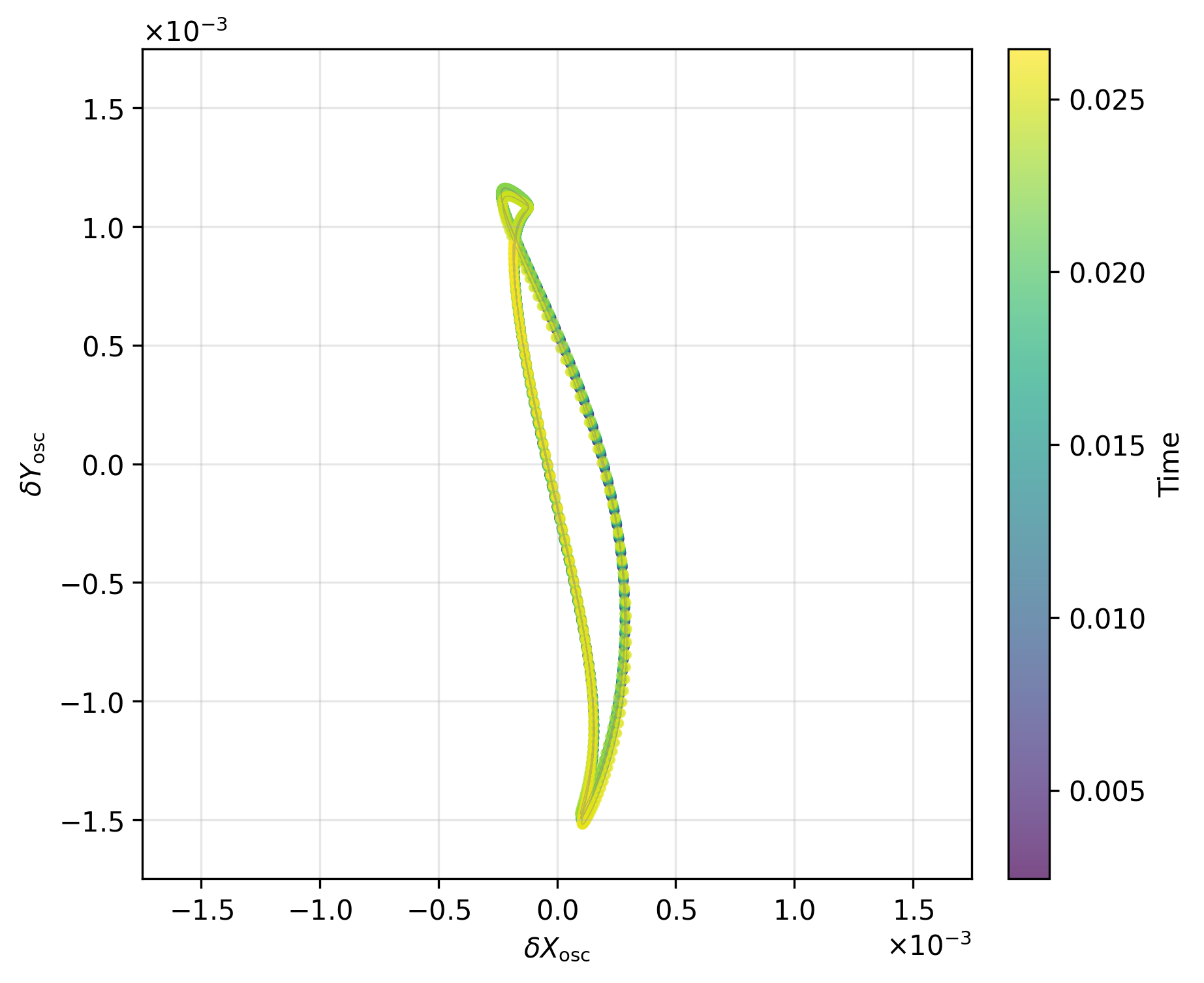}\label{fig:synch_6w}}
    \hfill
    \subfloat[]{\includegraphics[width=0.48\textwidth]{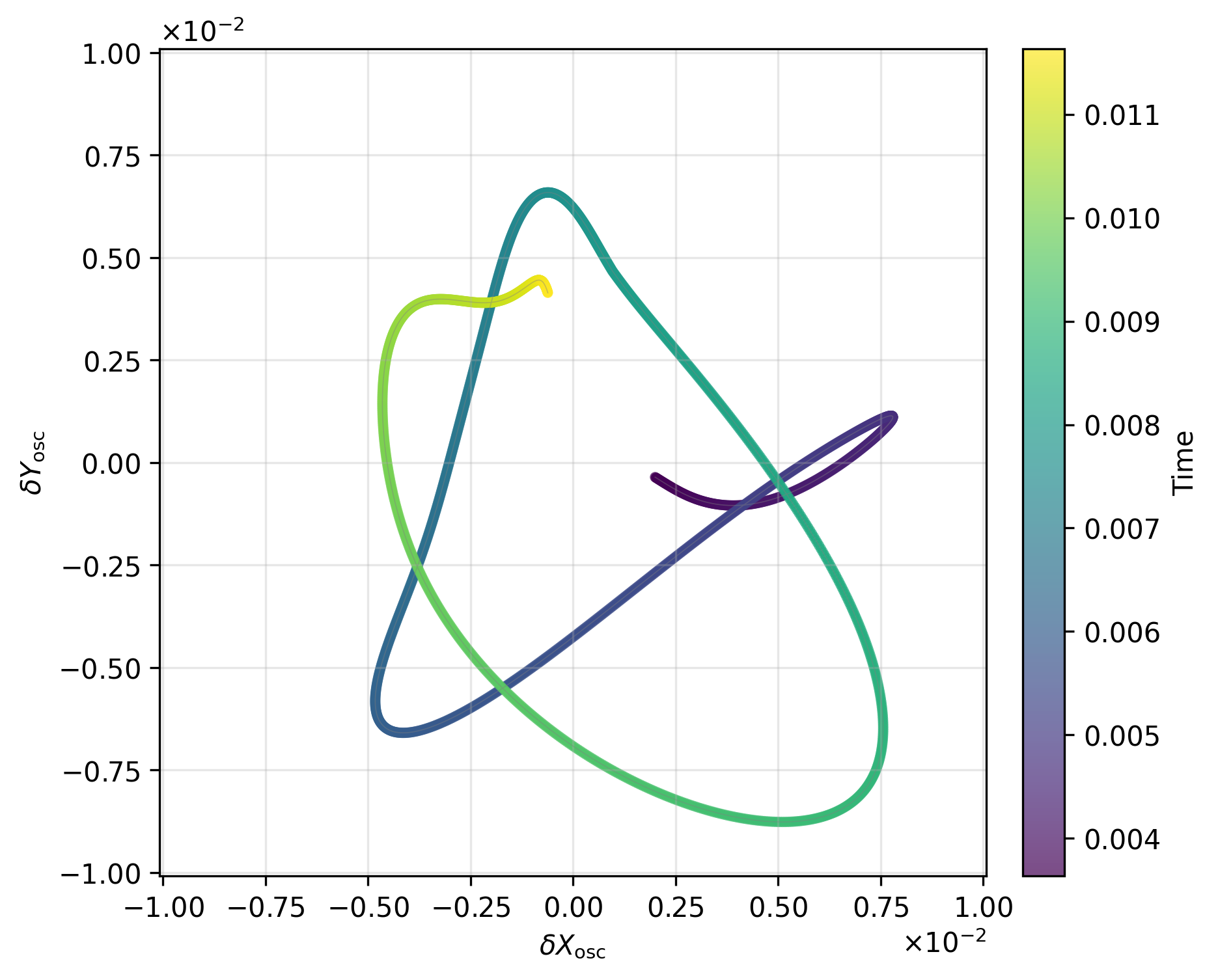}\label{fig:anti_synch_2w}}
    \caption{Lissajous representations of microscopic bond oscillations. (a)--(c) Representative bonds in the synchronized state ($\omega_{\rm f} = 1290$, $B = 1200$) exhibiting dominant harmonics at $2\omega_{\rm f}$, $4\omega_{\rm f}$, and $6\omega_{\rm f}$, respectively. (d) A representative bond in the anti-synchronized state ($\omega_{\rm f} = 865$, $B = 1200$) oscillating primarily at $2\omega_{\rm f}$.}
    \label{fig:Lissajous}
\end{figure*}

The qualitative difference between these two microscopic responses becomes particularly transparent when the bond motion is represented in a local coordinate frame.  To visualize the geometry of these microscopic bond oscillations, we resolve the instantaneous bond displacement into components $\delta x_{\rm osc}$ and $\delta y_{\rm osc}$ in a local bond-based coordinate frame, with the time-averaged bond orientation defining the reference direction. Plotting $\delta y_{\rm osc}$ against $\delta x_{\rm osc}$ then provides a Lissajous representation of the bond deformation over successive oscillation cycles. Figure~\ref{fig:Lissajous} shows Lissajous representations of the oscillatory bond displacement, $(\delta x_{\rm osc},\delta y_{\rm osc})$, for representative bonds. In the synchronized state [Figs.~\ref{fig:Lissajous}(a)-\ref{fig:Lissajous}(c)], bonds associated with the dominant $2\omega_{\rm f}$, $4\omega_{\rm f}$, and $6\omega_{\rm f}$ components trace well-defined, reproducible closed trajectories. The different geometries of these loops reflect the distinct harmonic components of the periodic bond deformation. Importantly, successive portions of the trajectory largely retrace the same paths, demonstrating that the microscopic deformation is organized into repeatable cycles. A qualitatively different trajectory is observed in the anti-synchronized state [Fig.~\ref{fig:Lissajous}(d)]. Although the selected bond retains a dominant component near $2\omega_{\rm f}$, its trajectory is considerably broader and does not collapse onto the same sharply defined periodic orbit observed in the synchronized state. Successive portions of the trajectory explore different regions of the local displacement space, indicating that the field-driven oscillation is superimposed on slower variations of the bond configuration. The anti-synchronized bond motion therefore contains a periodic component inherited from the external drive together with slower structural dynamics that continuously modify the local deformation cycle. In addition to this loss of a sharply repeatable trajectory, the characteristic displacement scale in the anti-synchronized example is substantially larger than that of the synchronized bonds. This difference indicates a pronounced enhancement of microscopic bond deformation in the anti-synchronized state and such enhanced deformation is consistent with the broader defect-rich interface and stronger orientational fluctuations identified in the above section. The contrast between reproducible periodic bond trajectories and spatially heterogeneous, less repeatable deformation is reminiscent of the distinction between reversible and irreversible microscopic responses in periodically driven many-body systems \cite{corte,keim}. Although the underlying interactions and driving mechanisms differ, these studies highlight how changes in microscopic rearrangement dynamics can accompany transitions between distinct collective responses. In the present system, the additional low-frequency components and broader bond trajectories indicate that anti-synchronized rotation is associated with a less coherent, more strongly deforming particle network.
  
These microscopic observations point to a qualitative reorganization of the interaction network between the synchronized and anti-synchronized states. In the synchronized phase, individual bonds undergo coherent and repeatable periodic deformation, producing a well-defined hierarchy of harmonics across the crystalline core. In the anti-synchronized phase, the field-coupled response near $2\omega_{\rm f}$ persists, but the higher-harmonic organization is substantially weakened and accompanied by pronounced low-frequency dynamics and enhanced cycle-to-cycle variability. The distinction between the two states therefore lies not only in the magnitude of bond deformation, but also in its temporal organization: synchronization is associated with a coherently driven bond network, whereas anti-synchronization involves a more dynamically heterogeneous microscopic response. Because the dipolar and steric interaction forces depend directly on the instantaneous bond geometry, such a reorganization of the bond dynamics is expected to influence the transmission of mechanical torque through the cluster. To determine whether this microscopic difference is connected to the counter-rotation of the crystalline interior, we next examine the layer-resolved dipolar and steric torques.

\subsection{Torque transfer and core–surface decoupling}

\begin{figure*}[!t]
    \centering
    \subfloat[]{\includegraphics[width=0.48\textwidth]{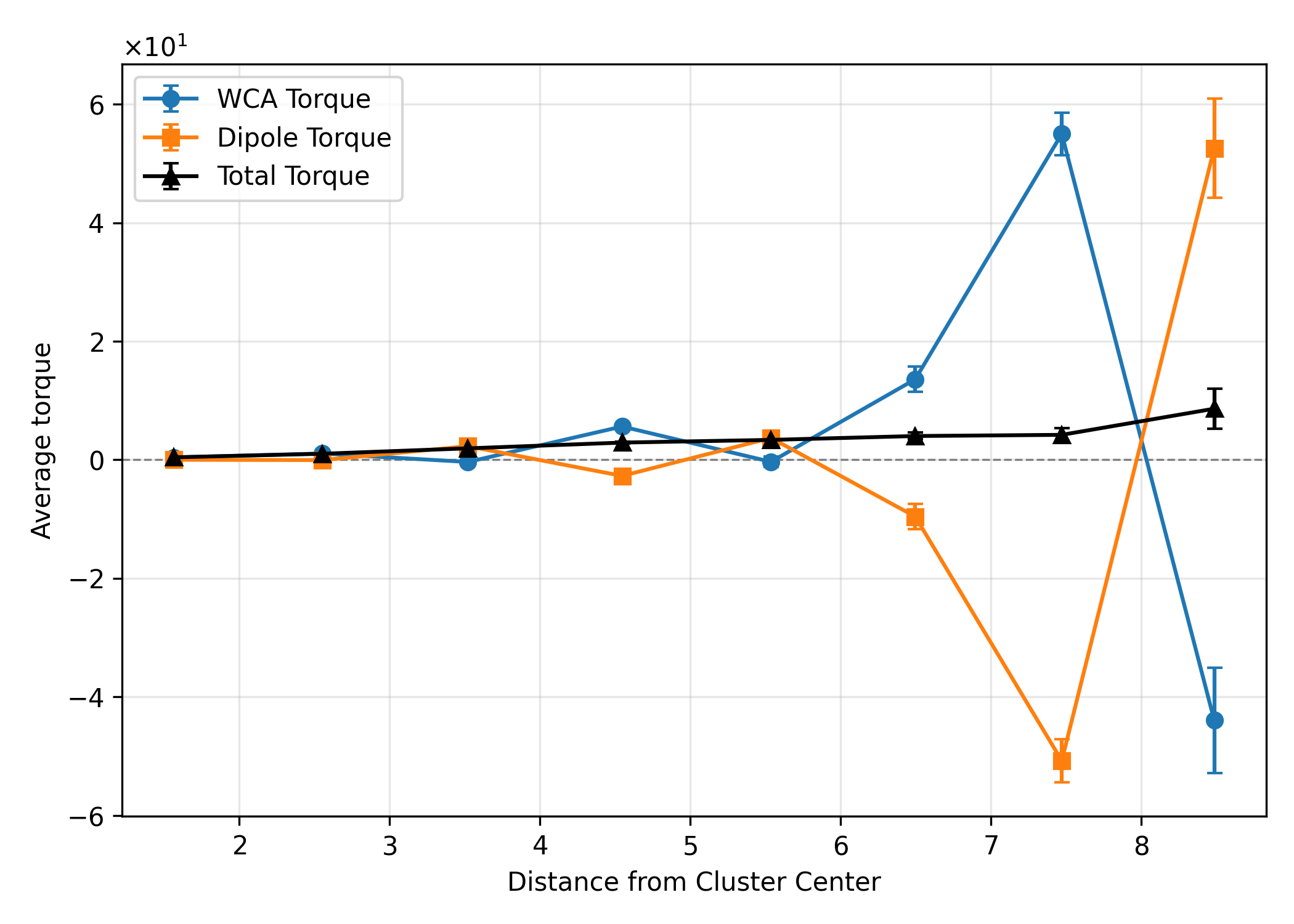}\label{fig:synch_torque}}
    \hfill
    \subfloat[]{\includegraphics[width=0.48\textwidth]{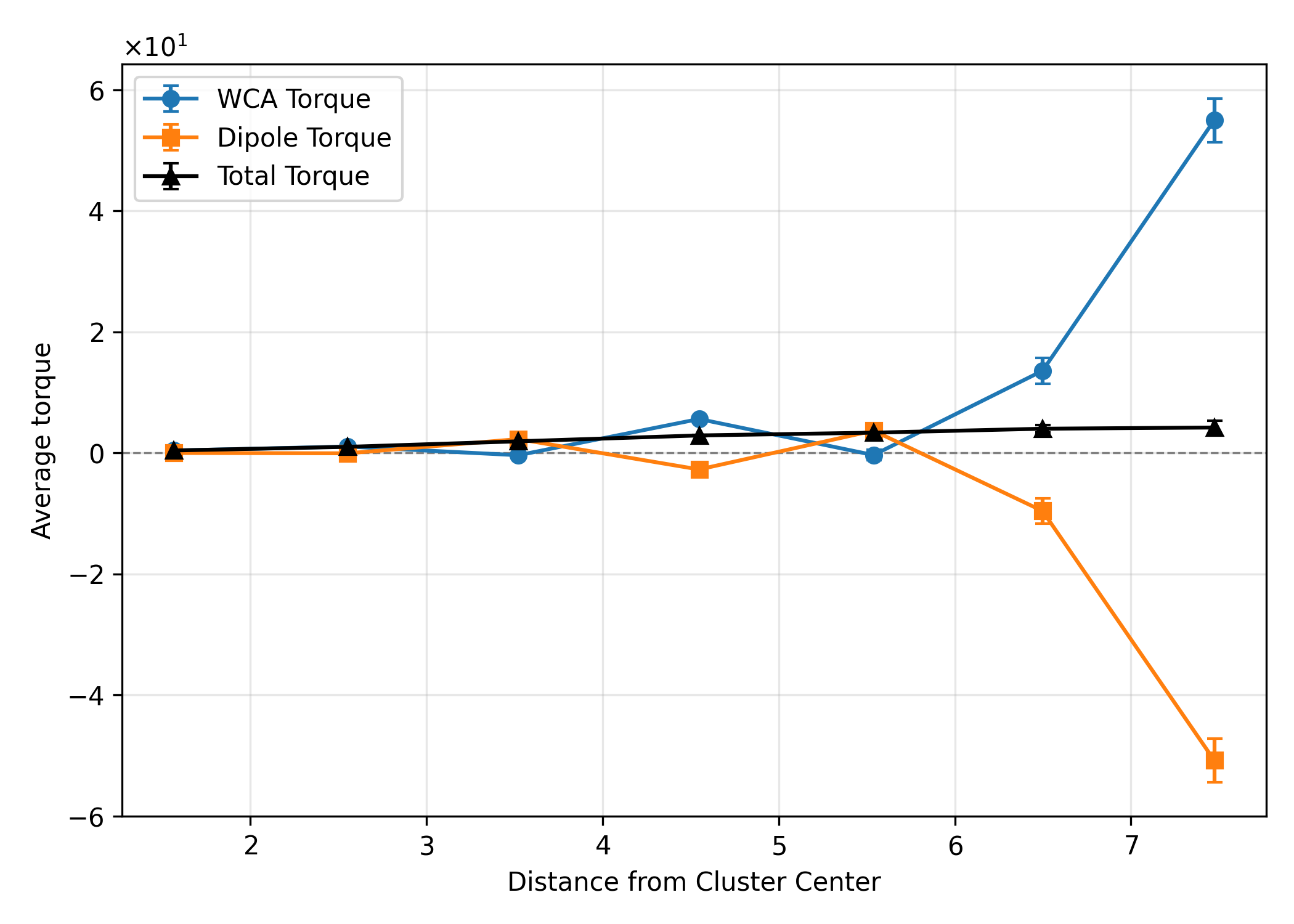}\label{fig:synch_core_torque}}
    \\[-1ex]
    \subfloat[]{\includegraphics[width=0.48\textwidth]{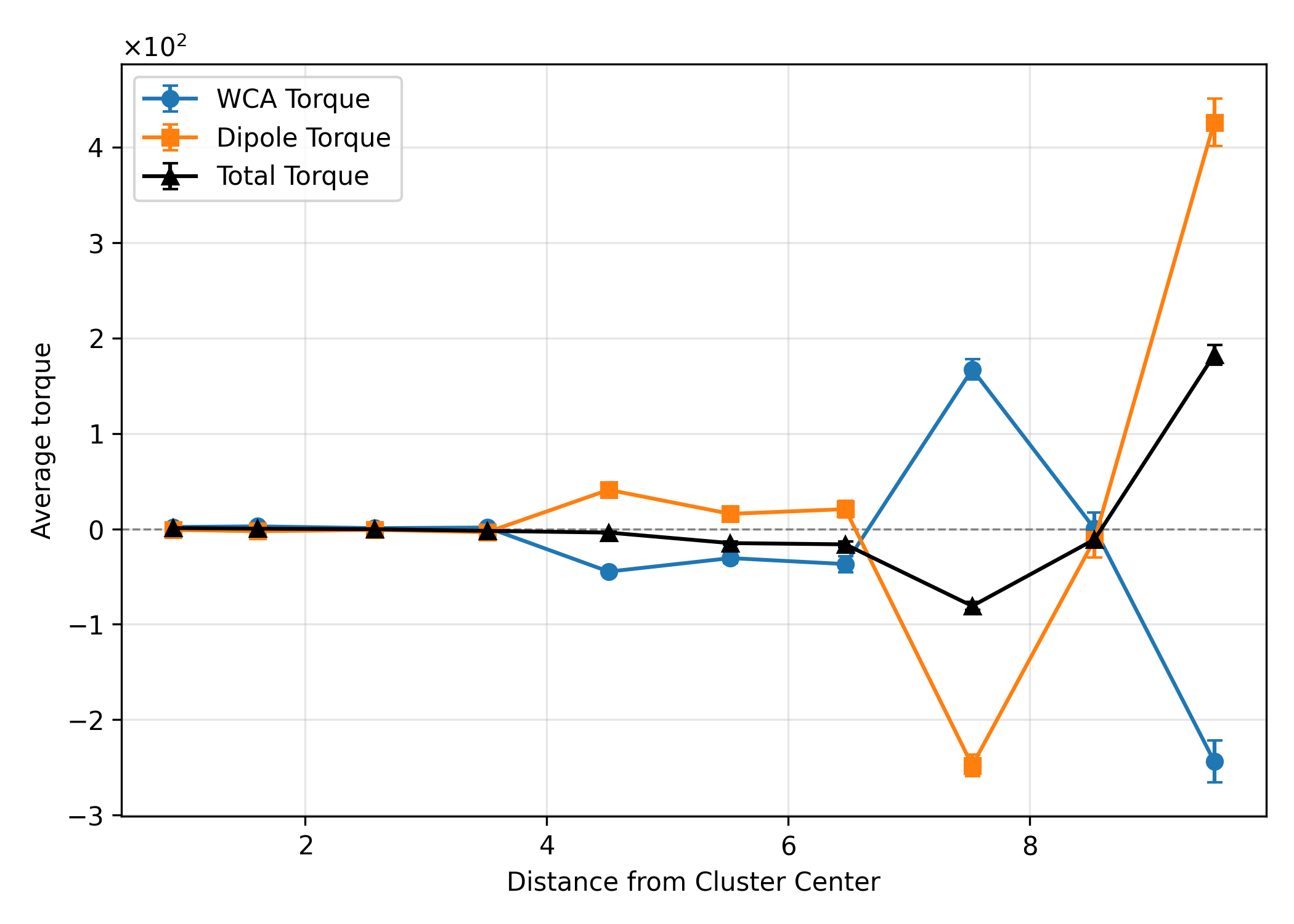}\label{fig:antisyn_torque}}
    \hfill
    \subfloat[]{\includegraphics[width=0.48\textwidth]{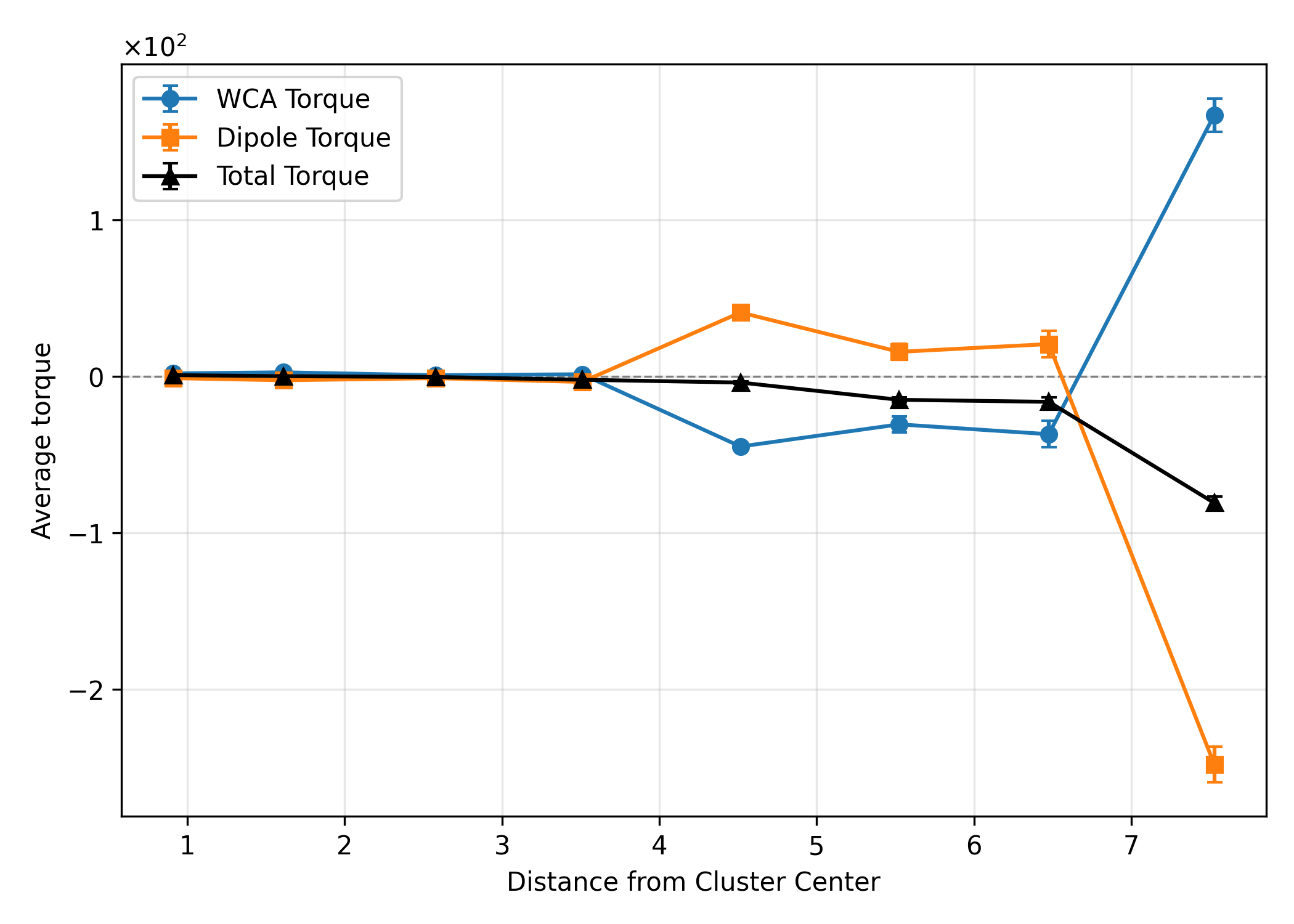}\label{fig:antisyn_core_torque}}
    \caption{Layer-resolved orbital torque in the synchronized and anti-synchronized states at $B=1200$.Here, the core boundary is defined as the first radial layer inward from the surface where the dipolar torque becomes negative. (a) Dipolar, WCA, and total interaction torque across the synchronized cluster at $\omega_f=1100$; (b) corresponding torque profile restricted to the crystalline core. (c,d) Same quantities for the anti-synchronized state at $\omega_f=889$. The reversal of the total interaction torque in the outer core distinguishes the anti-synchronized state from the synchronized state. The data points are layered torques averaged over 1000 frames of a single simulation.The error bars indicate the standard deviation of the torques for each layer in these 1000 frames.}
    \label{fig:torque}
\end{figure*}

The microscopic bond dynamics discussed above indicate that the synchronized and anti-synchronized states differ substantially in the manner in which deformation is accommodated within the interaction network. We now examine how this difference affects the transmission of mechanical torque through the cluster. To this end, we calculate the orbital torque generated separately by the dipolar and short-range WCA forces and resolve these contributions as a function of radial position within the cluster. This decomposition is particularly useful because the dipolar interaction provides the field-dependent anisotropic forcing, whereas the short-range repulsive interaction transmits mechanical stresses generated by deformation of the particle network. It is important to distinguish the magnetic torque acting directly on the orientation of an individual particle from the orbital torque responsible for motion of the particle about the cluster center. The external rotating field acts directly on the particle dipoles, but the collective orbital rotation of the cluster is generated through the interaction forces transmitted between particles. Consequently, counter-rotation of the crystalline core does not require the particle dipoles themselves to rotate persistently against the magnetic torque. Instead, the direction of orbital motion is determined by the net interaction torque transmitted to the core.

For each particle $i$, we decompose the orbital interaction torque about the instantaneous cluster center into dipolar and steric contributions,

\begin{equation}
 \tau_i^{\rm int} = \tau_i^{\rm dip}+\tau_i^{\rm WCA}
\end{equation} 

\noindent where

$$ \tau_i^\alpha= \left[ (\mathbf r_i-\mathbf R_{\rm cm}) \times\mathbf F_i^\alpha \right]_z, \qquad \alpha=\mathrm{dip,WCA}. $$
\noindent The corresponding torques are averaged over particles belonging to successive radial layers, allowing the transmission of interaction torque from the surface toward the crystalline interior to be resolved spatially.
Figures~\ref{fig:torque}(a) and \ref{fig:torque}(b) show the layer-resolved torque distribution in the synchronized state. The interaction torques are relatively small in the inner region of the cluster and increase strongly toward the periphery, demonstrating that the mechanically important torque transfer occurs predominantly within the outer layers. The dipolar and WCA contributions generally act in opposite directions, indicating that the rotational response results from a balance between magnetic interactions and steric stresses generated within the deformable particle network. This competition is particularly clear when only the crystalline core is considered [Fig.~\ref{fig:torque}(b)]. Near the outer part of the core, the dipolar torque becomes negative, whereas the WCA contribution is strongly positive. The positive steric contribution more than compensates the opposing dipolar contribution, leaving a positive residual total torque. The positive residual interaction torque in the outer core is consistent with the observed direction of its mean orbital rotation in the synchronized state. The WCA interaction should consequently not be viewed merely as an excluded-volume constraint: through contact forces generated by the deforming particle network, it provides an important channel for transmitting orbital torque across neighboring layers.

The torque balance changes qualitatively in the anti-synchronized state [Figs.~\ref{fig:torque}(c) and \ref{fig:torque}(d)]. As in the synchronized state, the largest interaction torques occur toward the cluster periphery and the dipolar and WCA contributions oppose one another over much of the outer region. However, their relative magnitudes are now substantially different. Most importantly, in the outer part of the crystalline core [Fig.~\ref{fig:torque}(d)], the dipolar contribution becomes strongly negative similar to the synchronization phase, but the compensating positive WCA torque is no longer sufficient to overcome it. The resulting total interaction torque therefore becomes negative. This reversal of the net torque provides a direct mechanical connection to the counter-rotation observed in the anti-synchronized phase. Although the surface remains driven in the direction of the rotating magnetic field, the interaction torque transmitted to the outer core has the opposite sign. The crystalline interior can therefore acquire a negative mean orbital angular velocity even though the externally driven boundary continues to rotate positively. Anti-synchronization thus emerges not because the entire cluster responds oppositely to the magnetic field, but because the internal transmission of torque permits the surface and core to sustain opposite rotational responses.

The change in torque balance is consistent with the structural and microscopic differences identified in the preceding sections. In the anti-synchronized state, the broader defect-rich interfacial region and the enhanced, less coherently organized bond deformation make the outer particle network mechanically more compliant. Such deformation modifies the instantaneous contact geometry and therefore the steric forces through which mechanical stresses are transmitted between neighboring layers. The layer-resolved torque profiles show that, under these conditions, the positive WCA contribution in the outer core is insufficient to compensate the negative dipolar interaction torque. By contrast, in the synchronized state the more coherently organized bond network supports a stronger positive steric contribution, allowing the net torque on the outer core to remain positive. These results suggest the following microscopic picture of anti-synchronization. The rotating magnetic field drives the particle orientations and produces strong interaction stresses near the cluster surface. In the synchronized state, the mechanically coherent particle network efficiently transmits these stresses through steric contacts, and the resulting interaction torque on the core remains positive. In the anti-synchronized state, enhanced deformation and structural compliance of the interfacial region modify this transmission. The compensating steric torque is weakened relative to the negative dipolar contribution in the outer core, causing the net interaction torque to reverse sign. The surface can therefore continue to follow the external drive while the crystalline interior experiences an interaction torque in the opposite direction, producing the observed core–surface rotational decoupling and counter-rotation.

\subsection{Transition from Anti-Synchronized to Synchronized Rotation}

\begin{figure}[!htbp]
    \centering
        \includegraphics[width=0.85\linewidth]{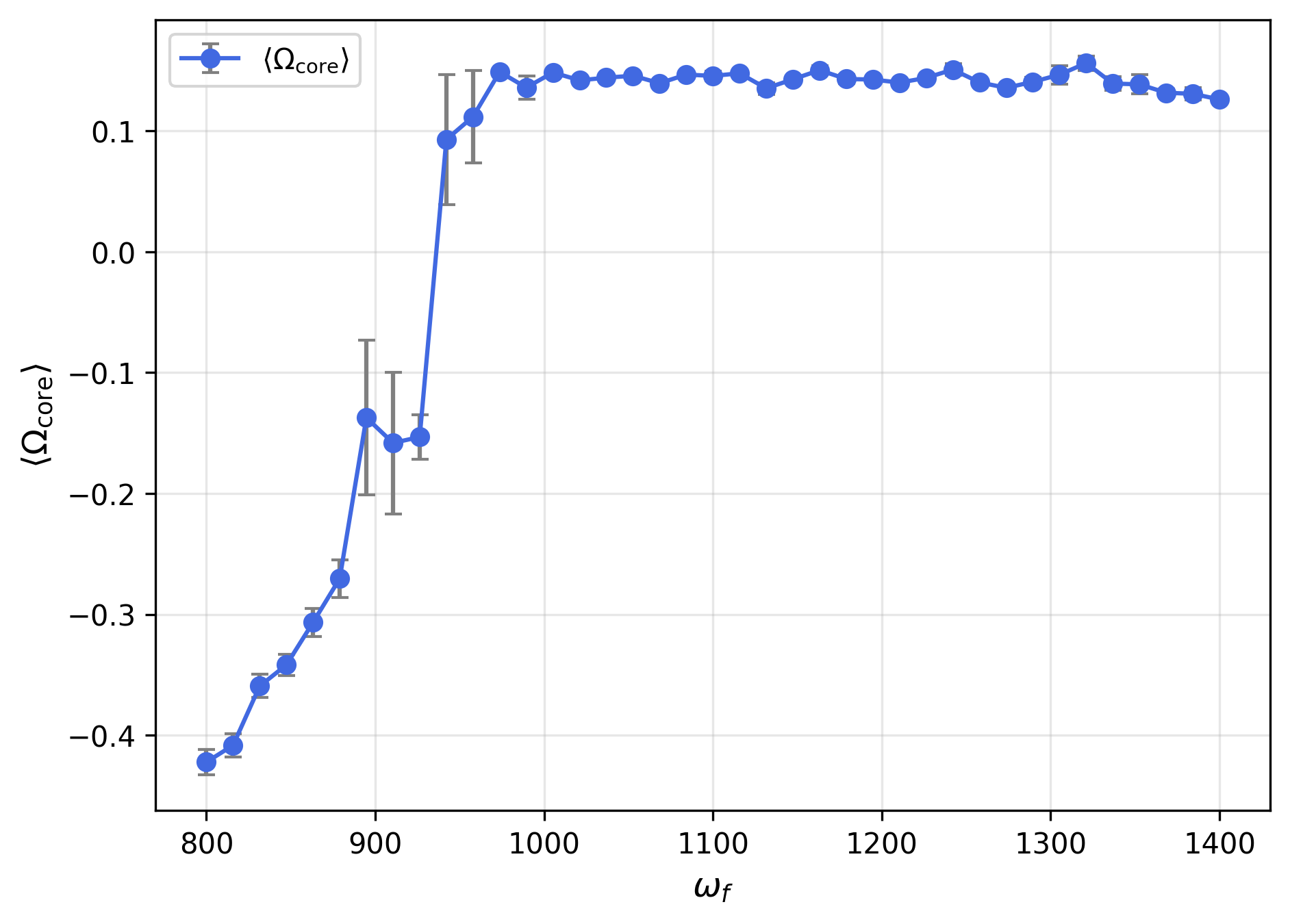}

    \caption{  Mean angular velocity of the core as a function of the driving frequency, characterising the transition between anti-synchronized and synchronized rotation.Each data point is averaged over 5 simulations of random initial conditions. The error bars shows the standard deviation of the mean angular velocities from these 5 simulations.}
    \label{fig:mean}
\end{figure}

The torque analysis above establishes that the anti-synchronized and synchronized regimes correspond to dynamically distinct states characterized by different internal torque balances. We now examine how the system switches between these states as the driving frequency is varied. We use the instantaneous angular velocity of the crystalline core, $\Omega_{\rm core}(t)$, as a dynamical order parameter and characterize its mean value, probability distribution, higher statistical moments, and dependence on the direction of the frequency sweep. Figure~\ref{fig:mean} shows the mean core angular velocity as a function of the driving frequency. In the anti-synchronized regime, $\langle\Omega_{\rm core}\rangle<0$, corresponding to persistent counter-rotation of the crystalline interior. With increasing driving frequency, the magnitude of the counter-rotation decreases before $\langle\Omega_{\rm core}\rangle$ undergoes a rapid sign reversal and becomes positive. At still higher frequencies, the mean angular velocity remains positive, corresponding to the synchronized state. The sign of $\langle\Omega_{\rm core}\rangle$ therefore provides a convenient dynamical marker distinguishing the two rotational states. The mean value alone, however, does not establish whether the reversal occurs continuously or through competition between distinct dynamical states. We therefore examine the full probability distribution $P(\Omega_{\rm core})$.

\begin{figure}[!htbp]
    \centering
    \subfloat[]{%
        \includegraphics[width=0.85\linewidth]{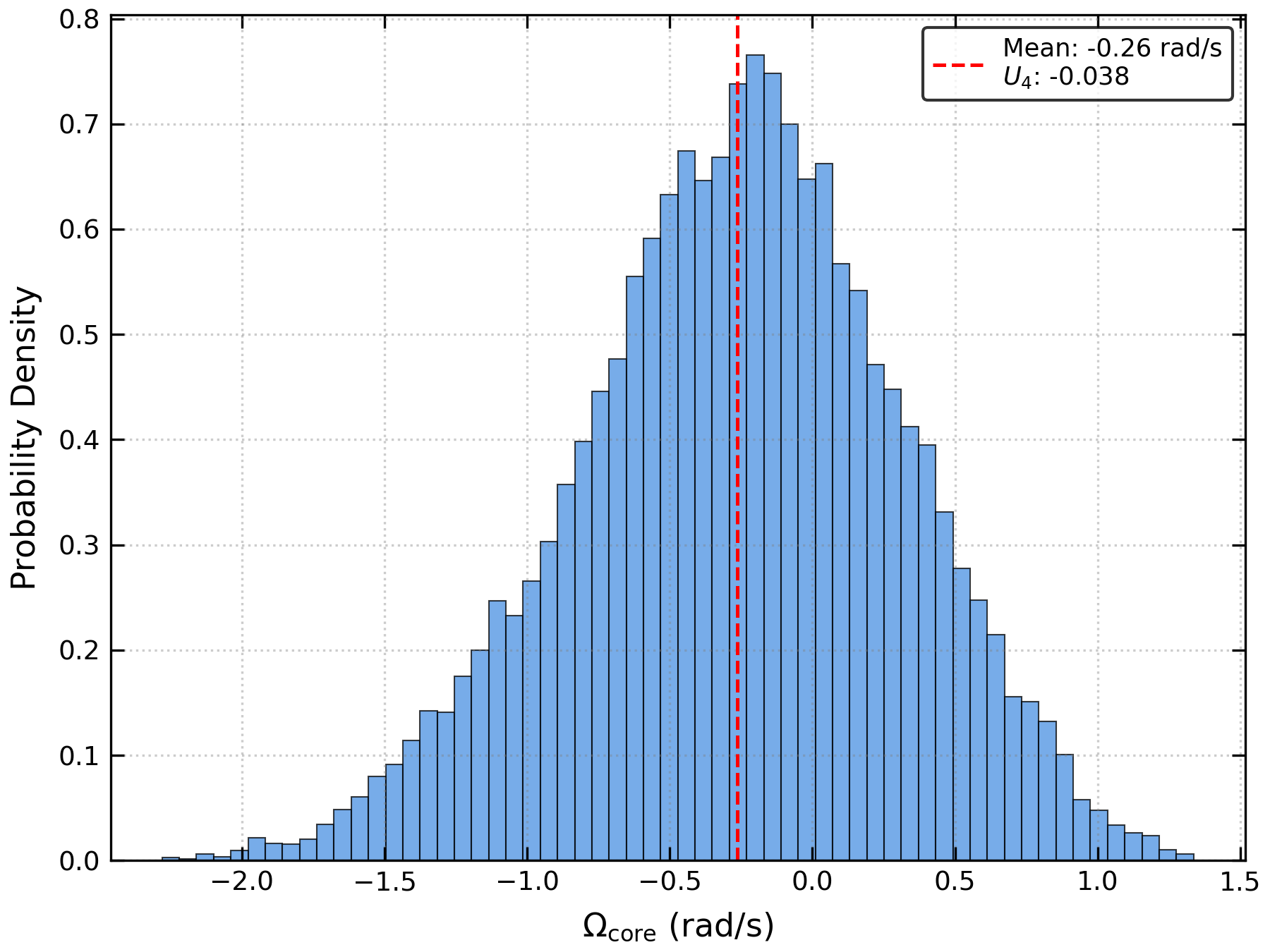}%
    }
    \vspace{0.5em}

    \subfloat[]{%
        \includegraphics[width=0.85\linewidth]{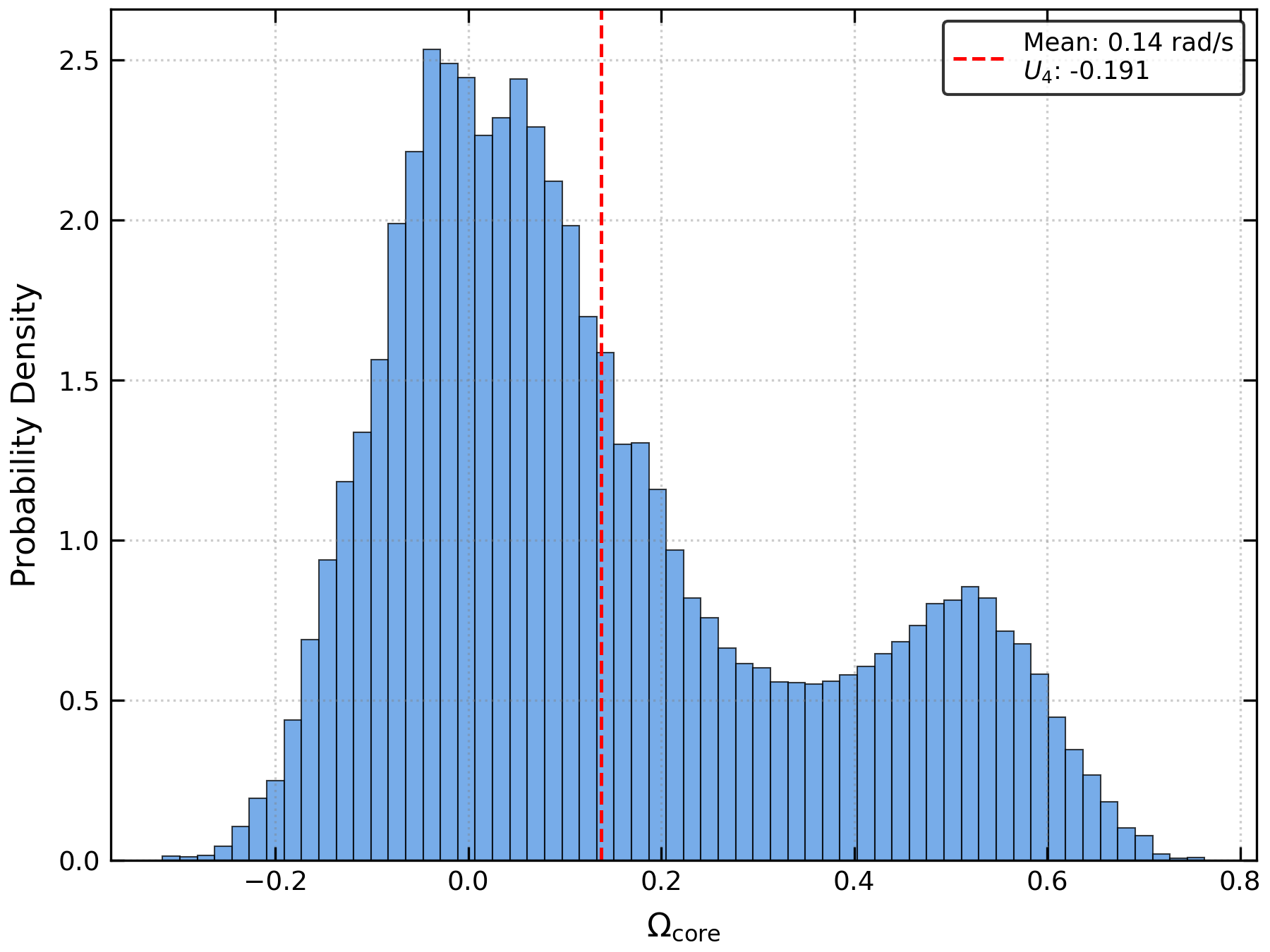}%
    }
    \vspace{0.5em}

    \subfloat[]{
        \includegraphics[width=0.85\linewidth]{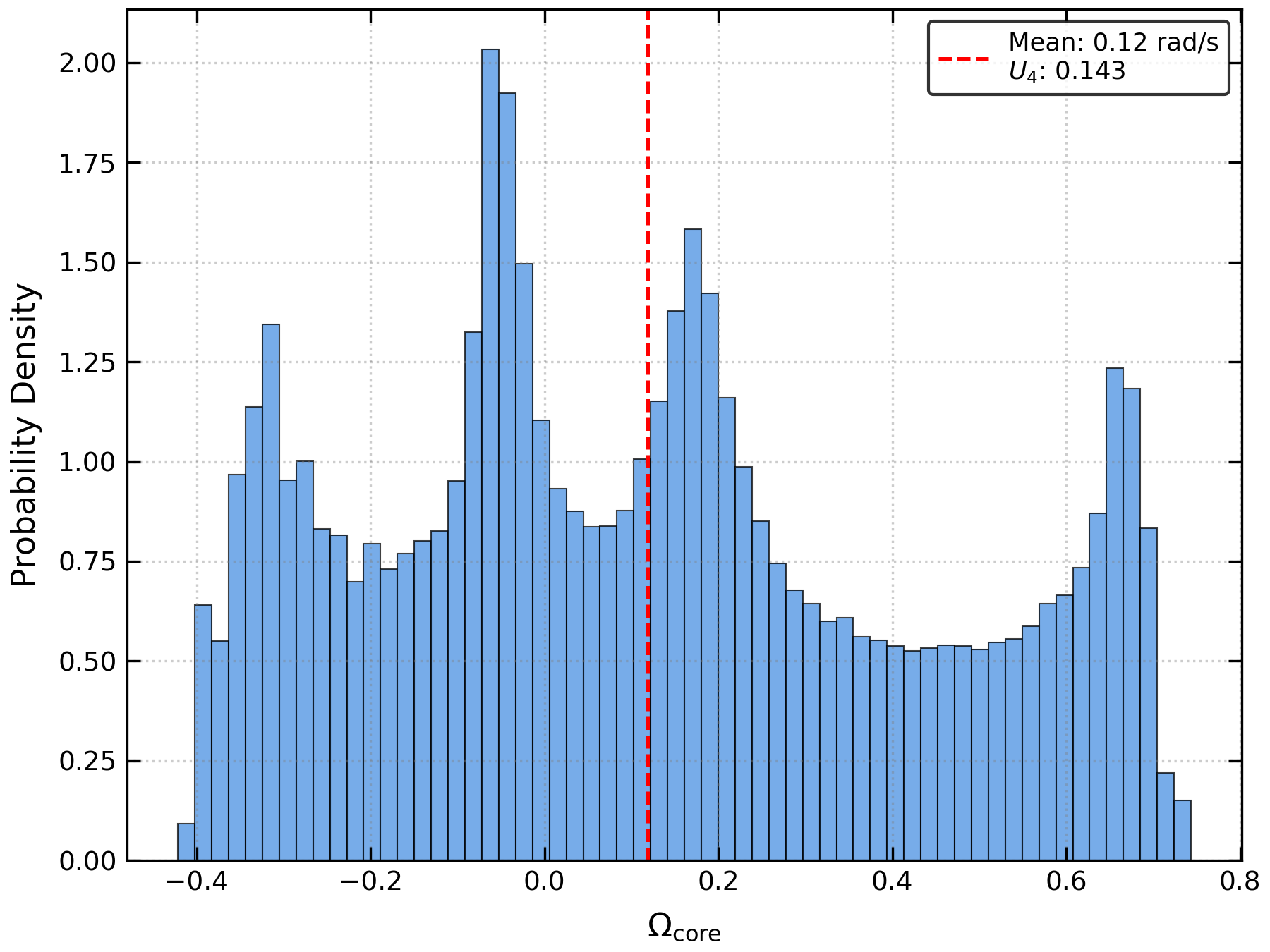}%
    }

    \caption{Probability distributions of the instantaneous cluster angular velocity
$P(\Omega_{\rm core})$ for driving frequencies below, near, and above the transition.(a) Cluster at 863. (b) Cluster at 946. (c) Cluster at 1400.
}
    \label{fig:omega_distributions}
\end{figure}

Representative angular-velocity distributions below, near, and well above the transition are shown in Fig.~\ref{fig:omega_distributions}. At $\omega_{\rm f}\simeq863$ [Fig.~\ref{fig:omega_distributions}(a)], the distribution is broad but predominantly centered at negative angular velocities, with $\langle\Omega_{\rm core}\rangle<0$, characteristic of the anti-synchronized state. Near the transition, at $\omega_{\rm f}\simeq946$ [Fig.~\ref{fig:omega_distributions}(b)], the distribution develops two distinct populations at negative and positive angular velocities. The simultaneous statistical weight associated with counter-rotating and co-rotating motion indicates competition between the two dynamical states rather than a simple continuous passage through a narrowly distributed state with $\Omega_{\rm core}=0$. At substantially higher driving frequency [Fig.~\ref{fig:omega_distributions}(c)], the distribution is confined predominantly to positive angular velocities, consistent with synchronized rotation. The distribution nevertheless retains considerable internal structure, indicating that synchronization of the mean rotational direction does not imply the absence of microscopic temporal modulation. This observation is consistent with the harmonic angular-velocity and bond dynamics discussed in the preceding sections.

\begin{figure}[!htbp]
\centering
\includegraphics[width=\linewidth]{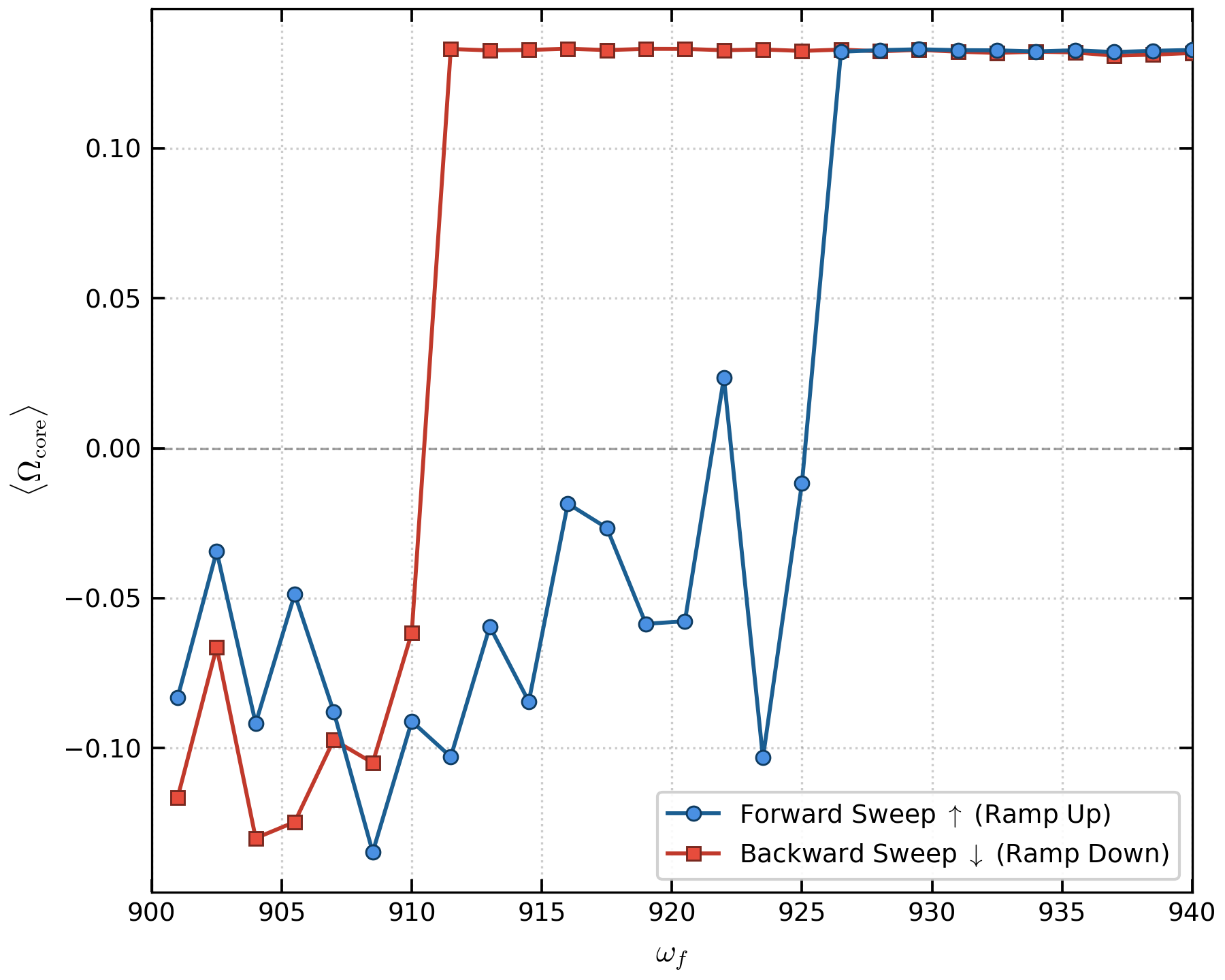}
\caption{
Hysteretic response of the cluster angular velocity under increasing and decreasing driving frequency.B=1200,$\nu=7.00$. The driving frequency was varied quasi-statically in discrete steps of $\Delta \omega_{\rm f} = 1.5$, held for $2 \times 10^5$ time steps at each plateau. The first $30\%$ of each frequency plateau was discarded as transient relaxation, and $\langle \Omega_{\rm core} \rangle$ was averaged over the remaining steady state, excluding the outer 3 surface layers.}
\label{fig:hysteresis}
\end{figure}

More direct evidence for bistability is provided by the response to increasing and decreasing driving frequency. Figure~\ref{fig:hysteresis} shows $\langle\Omega_{\rm core}\rangle$ obtained from frequency sweeps performed in opposite directions. The two protocols follow different rotational branches over a finite interval of driving frequencies. During the upward sweep, the system remains on the anti-synchronized branch to substantially higher frequency before switching to positive rotation, whereas during the downward sweep the synchronized branch persists to lower frequency before the system returns to counter-rotation. The resulting hysteresis demonstrates that, within this interval, the observed rotational state depends on the dynamical history of the cluster.

The three observations—the rapid reversal of the mean core angular velocity, the emergence of competing negative- and positive-angular-velocity populations near the transition, and the hysteretic response under opposite frequency sweeps—provide mutually consistent evidence for bistability between the anti-synchronized and synchronized rotational states. Similar phenomenology, particularly the appearance of competing states and hysteretic switching, has been extensively investigated in the broader context of first-order transitions and driven systems\cite{binder,sethna}. In the present system, these features arise in the collective orbital dynamics of a periodically driven colloidal cluster, indicating a discontinuous, first-order-like nonequilibrium dynamical transition between anti-synchronized and synchronized rotation. From the microscopic perspective developed in the preceding sections, the transition can be viewed as switching between two distinct modes of internal torque transmission. On the anti-synchronized branch, the negative dipolar contribution dominates the torque balance in the outer core, producing a negative net interaction torque and counter-rotation. On the synchronized branch, the steric contribution compensates the opposing dipolar torque, restoring a positive net torque and coherent rotation of the core and surface. The bistability observed near the transition therefore reflects competition not merely between opposite signs of $\Omega_{\rm core}$, but between two distinct collective mechanical states of the driven cluster.

\section{Conclusions}

In this work, we have investigated the collective rotational dynamics of magnetic colloidal particles subjected to a rotating magnetic field. We identify distinct disordered, anti-synchronized, synchronized, and slipping dynamical regimes. The most striking behavior is the anti-synchronized state, in which the structurally ordered crystalline core develops a persistent mean rotation opposite to the applied field, while the defect-rich outer region continues to rotate in the direction of the drive. Thus, a single mechanically connected  cluster can sustain spatially separated regions of opposite rotational response. Our analysis reveals that anti-synchronization is accompanied by a qualitative reorganization of the internal particle network. Compared with the synchronized state, the anti-synchronized regime exhibits a broader defect-rich interface, orientational fluctuations extending farther into the interior, and less coherently organized bond dynamics with substantial low-frequency contributions. The layer-resolved torque analysis provides a microscopic explanation for the resulting counter-rotation. In the synchronized state, steric torques compensate the opposing dipolar contribution in the outer core, allowing the surface-generated forcing to be transmitted through the cluster. In the anti-synchronized state, this compensation becomes insufficient, leaving the outer core under a negative net interaction torque and enabling rotational decoupling between the field-following surface and the crystalline interior. Anti-synchronization therefore emerges from the interplay between microscopic deformation, interfacial compliance, and spatially heterogeneous torque transmission. Increasing the driving frequency restores coherent core–surface rotation through a transition characterized by competing positive- and negative-angular-velocity states and hysteresis, consistent with a discontinuous, first-order-like nonequilibrium dynamical transition. The transition therefore represents not simply a reversal of the mean angular velocity, but a reorganization of the mechanism by which torque is transmitted through the cluster. More broadly, our results demonstrate that a periodically driven colloidal assembly can develop a collective response locally opposite to the external forcing through internal mechanical reorganization, suggesting a route for controlling counter-rotating and dynamically decoupled states in driven soft and magnetic matter.

\appendix
\section{overdamped equation of motion for a dimer}

The dipole dipole potential for a dipole pair,with particles having parallel dipole moments is given by:
\begin{equation}
    U = \frac{\mu^2}{r_{ij}^3}
    \left(1 - 3\cos^2(\psi_{ij})\right).
\end{equation}

The dipole dipole force is given by:
\begin{equation}
    \mathbf{F}_{ij}^{\rm{dd}}
    = -\nabla_i U
    = \frac{3\mu^2}{r_{ij}^4}
    \left[
        \left(1 - 3\cos^2(\psi_{ij})\right)\mathbf{e}_{r_{ij}}
        + \sin(2\psi_{ij})\mathbf{e}_{\theta_{ij}}
    \right].
\end{equation}

Here
\begin{equation}
    \psi_{ij}
    = \phi_i - \theta_{ij}
    = \omega_{\rm f} t - \delta - \theta_{ij},
\end{equation}
where $\delta$ is the phase lag between magnetic field and particle orientation.

The overdamped equation for a magnetic dipole in an external magnetic field is given by:
\begin{equation}
    \dot{\mathbf{r}}_i
    =
    \mathbf{F}_{ij}^{\rm{dd}}
    + \mathbf{F}_{ij}^{\rm{WCA}}
    + \mathbf{F}^{ext}_{i}.
    \label{eq:r_i_dot}
\end{equation}

Where $F^{ext}$ is the force due to external magnetic field, and is 0 if the field is uniform, which is our case. For the orientation of the dipole the overdamped equation is given by:
\begin{equation}
    3\dot{\phi}_i
    =
    -\frac{\partial U}{\partial \phi_i}
    + \mu B \sin(\omega_{\rm f} t - \phi_i).
    \label{eq:phi_i_dot}
\end{equation}
\begin{equation}
    \dot{\phi}_i
    =
    -\frac{\mu^2}{r_{ij}^3}
    \sin(2\psi_{ij})
    + \frac{\mu B}{3} \sin(\omega_{\rm f} t - \phi_i).
    \label{eq:phi_i_dot_expanded}
\end{equation}

In case of a dimer the bond length is given by
\begin{equation}
    \mathbf{r}_{ij}
    =
    \mathbf{r}_i - \mathbf{r}_j.
\end{equation}

For the overdamped equation
\begin{equation}
    \dot{\mathbf{r}}_{ij}
    =
    \dot{\mathbf{r}}_i - \dot{\mathbf{r}}_j.
\end{equation}

\begin{equation}
    \dot{\mathbf{r}}_{ij}
    =
    \mathbf{F}_{ij}^{\rm{dd}}
    -\mathbf{F}_{ji}^{\rm{dd}}
    + \mathbf{F}_{ij}^{\rm{WCA}}
    - \mathbf{F}_{ji}^{\rm{WCA}}
    + \mathbf{F}^{ext}_{i}.
\end{equation}

By Newton's third law
\[
    \mathbf{F}_{ij}^{\rm{dd}} = -\mathbf{F}_{ji}^{\rm{dd}},
    \qquad
    \mathbf{F}_{ij}^{\rm{WCA}} = -\mathbf{F}_{ji}^{\rm{WCA}}.
\]

giving
\begin{equation}
    \frac{\dot{\mathbf{r}}_{ij}}{2}
    =
    \mathbf{F}_{ij}^{\rm{dd}}
    + \mathbf{F}_{ij}^{\rm{WCA}}.
    \label{eq:r_ij_vector_dot}
\end{equation}

The WCA force acts purely in $\mathbf{e}_{r_{ij}}$ direction, giving us
\begin{equation}
    \frac{\dot{r}_{ij}}{2}
    =
    \frac{3\mu^2}{r_{ij}^4}
    \left(1 - 3\cos^2(\psi_{ij})\right)
    + F^{WCA}_{ij}.
    \label{eq:r_ij_dot}
\end{equation}

\begin{equation}
    \dot{\theta}_{ij}
    =
    \frac{6\mu^2}{r_{ij}^5}
    \sin(2\psi_{ij}).
    \label{eq:theta_ij_dot}
\end{equation}
\section{Characteristic Dipolar Time and length scales}
To estimate the characteristic scales associated with the dipole--dipole interaction, we consider the relative dynamics of a pair of magnetic particles and neglect the short-range WCA interaction. The latter primarily provides excluded-volume repulsion and therefore does not enter the characteristic scale of the dipolar dynamics considered here. For a pair separated by a distance $r$ and with bond angle $\theta$, the radial and angular equations of motion due to the dipolar interaction are

\begin{equation}
\dot{r}
=
\frac{6\mu^2}{r^4}
\left[
1-3\cos^2(\omega_{\rm f}t-\theta-\delta)
\right],
\end{equation}

\begin{equation}
\dot{\theta}
=
\frac{6\mu^2}{r^5}
\sin\left[2(\omega_{\rm f} t-\theta-\delta)\right].
\end{equation}

The characteristic timescale of the externally imposed rotation is

\begin{equation}
t_s = \frac{1}{\omega_{\rm f}}.
\end{equation}

Introducing the dimensionless variables
\begin{equation}
t' = \frac{t}{t_s},
\qquad
r' = \frac{r}{r_s},
\end{equation}
and choosing $r_s$ such that the prefactor of the dimensionless radial equation is unity gives

\begin{equation}
r_s^5 = \frac{6\mu^2}{\omega_{\rm f}},
\end{equation}

and hence

\begin{equation}
r_s =
\left(
\frac{6\mu^2}{\omega_{\rm f}}
\right)^{1/5}
\end{equation}

Thus, the characteristic length scale of the dipolar dynamics decreases with increasing driving frequency as

\begin{equation}
r_s \sim \omega_{\rm f}^{-1/5}.
\end{equation}

This provides a simple timescale argument for the observed reduction in the amplitude of bond-length oscillations with increasing driving frequency. As $\omega_{\rm f}$ increases, the direction of the dipolar interaction changes on a progressively shorter timescale, leaving less time for the bond to respond to the instantaneous interaction.

\begin{acknowledgements}
The authors acknowledge financial support from the Department of Atomic Energy, India through the plan Project (RIN4001-SPS).
\end{acknowledgements}

\end{document}